\documentclass[longauth]{aa}  
\usepackage{graphicx}
\usepackage{longtable}
\usepackage{multirow}
\usepackage{lscape}
\usepackage{pdflscape}
\usepackage{caption}
\usepackage{rotating}
\usepackage{natbib}
\usepackage{fixltx2e}
\usepackage{float}
\usepackage[above,below]{placeins}
\usepackage{amsmath}
\usepackage{mathtools}
\usepackage{txfonts}

\usepackage{txfonts}

\newcommand{\um}{$\mu$m}                                 
\newcommand{\lsim}{\;\lower.6ex\hbox{$\sim$}\kern-7.75pt\raise.65ex\hbox{$<$}\;}
\newcommand{\gsim}{\;\lower.6ex\hbox{$\sim$}\kern-7.75pt\raise.65ex\hbox{$>$}\;}
\newcommand{\gl}{\;\lower.6ex\hbox{$<$}\kern-7.75pt\raise.65ex\hbox {$>$}\;}
\DeclarePairedDelimiter\abs{\lvert}{\rvert}
\newcommand{\cutex}{{\textit{\sc{CuTEx}}}}
\newcommand{\higal}{Hi-GAL}
\newcommand{\amin}{$^{\prime}$}                   
\newcommand{\asec}{$^{\prime \prime}$}
\newcommand{\adeg}{$^{\circ}$}

\begin{document}
\title{Hi-GAL, the Herschel\thanks{{\it Herschel} is an ESA space observatory with science instruments provided by European-led Principal Investigator consortia and with important participation from NASA.} infrared Galactic Plane Survey: photometric maps and compact source catalogues. }

   \subtitle{First data release for Inner Milky Way: +68\adeg $\geq l \geq-70$\adeg.}

\author{S. Molinari\inst{\ref{iaps}} \and
E. Schisano\inst{\ref{iaps}}\and
D. Elia\inst{\ref{iaps}}\and
M. Pestalozzi\inst{\ref{iaps}}\and
A. Traficante\inst{\ref{manchester}}\and
S. Pezzuto\inst{\ref{iaps}}\and
B. M. Swinyard\inst{\ref{ral}}\and
A. Noriega-Crespo\inst{\ref{stsci}}\and
J. Bally\inst{\ref{colourado}}\and
T. J. T. Moore\inst{\ref{moores}}\and
R. Plume\inst{\ref{calgary}}\and
A. Zavagno\inst{\ref{lam}}\and
A. M. di Giorgio\inst{\ref{iaps}}\and
S. J. Liu\inst{\ref{iaps}}\and
G. L. Pilbratt\inst{\ref{estec}}\and
J. C. Mottram\inst{\ref{leiden}}\and
D. Russeil\inst{\ref{lam}}\and
L. Piazzo\inst{\ref{dietroma}}\and
M. Veneziani\inst{\ref{ipac}}\and
M. Benedettini\inst{\ref{iaps}}\and
L. Calzoletti\inst{\ref{esac}, \ref{asdc}}\and
F. Faustini\inst{\ref{asdc}}\and
P. Natoli\inst{\ref{ferrara}}\and
F. Piacentini\inst{\ref{roma1}}\and
M. Merello\inst{\ref{iaps}}\and
A. Palmese\inst{\ref{iaps}}\and
R. Del Grande\inst{\ref{iaps}}\and
D. Polychroni\inst{\ref{atene}}\and
K. L. J. Rygl\inst{\ref{ira}}\and
G. Polenta\inst{\ref{asdc}}\and
M. J. Barlow\inst{\ref{ucl}}\and
J.-P. Bernard\inst{\ref{toulouseuniv},\ref{toulousecesr}}\and
P. G. Martin\inst{\ref{cita}}\and
L. Testi\inst{\ref{eso},\ref{arcetri}}\and
B. Ali\inst{\ref{boulder}}\and
P. Andr\'e\inst{\ref{saclay}}\and
M.T. Beltr\'an\inst{\ref{arcetri}}\and
N. Billot\inst{\ref{iram}}\and
C. Brunt\inst{\ref{exeter}}\and
S. Carey\inst{\ref{ssc}}\and
R. Cesaroni\inst{\ref{arcetri}}\and
M. Compi\`{e}gne\inst{\ref{hygeos}}\and
D. Eden\inst{\ref{strasbourg}}\and
Y. Fukui\inst{\ref{nagoya}}\and
P. Garcia-Lario\inst{\ref{esac}}\and
M. G. Hoare\inst{\ref{leeds}}\and
M. Huang\inst{\ref{beijing}}\and
G. Joncas\inst{\ref{laval}}\and
T. L. Lim\inst{\ref{ral},\ref{esac}}\and
S. D. Lord\inst{\ref{seti}}\and
S. Martinavarro-Armengol\inst{\ref{ucl}}\and
F. Motte\inst{\ref{saclay}}\and
R. Paladini\inst{\ref{ipac}}\and
D. Paradis\inst{\ref{toulouseuniv},\ref{toulousecesr}}\and
N. Peretto\inst{\ref{cardiff}}\and
T. Robitaille\inst{\ref{mpia}}\and
P. Schilke\inst{\ref{colonia}}\and
N. Schneider\inst{\ref{bordeaux}}\and
B. Schulz\inst{\ref{ipac}}\and
B. Sibthorpe\inst{\ref{sron}}\and
F. Strafella\inst{\ref{lecce}}\and
M. A. Thompson\inst{\ref{herts}}\and
G. Umana\inst{\ref{catania}}\and
D. Ward-Thompson\inst{\ref{lancashire}}\and
F. Wyrowski\inst{\ref{mpifr}}
}

\institute{INAF-Istituto di Astrofisica e Planetologia Spaziale, Via Fosso del Cavaliere 100, I-00133 Roma, Italy \email{molinari@iaps.inaf.it} \label{iaps} 
\and 
Jodrell Bank centre for Astrophysics, School of Physics and Astronomy, University of Manchester, Manchester, M13 9PL, UK\label{manchester}
\and
STFC, Rutherford Appleton Labs, Didcot, UK \label{ral} 
\and 
Space Telescope Science Institute, 3700 San Martin Dr., Baltimore, 21218 MD, USA\label{stsci}
\and
centre for Astrophysics and Space Astronomy (CASA), Department of Astrophysical and Planetary Sciences, University of colourado, Boulder, USA\label{colourado} 
\and
Astrophysics Research Institute, Liverpool John Moores University, Ic2 Liverpool Science Park, 146 Brownlow Hill, Liverpool L3 5RF, UK\label{moores}
\and
Department of Physics \& Astronomy, University of Calgary, Canada\label{calgary}
\and
LAM, Universit\'{e} de Provence, Marseille, France\label{lam}
\and
ESA, Directorate of Science, Scientific Support Office, European Space Research and Technology Centre (ESTEC/SCI-S), Keplerlaan 1, NL-2201 AZ Noordwijk, The Netherlands\label{estec}
\and
Leiden Observatory, Leiden University, PO Box 9513, 2300 RA, Leiden, The Netherlands\label{leiden}
\and
DIET, Universit\`{a} di Roma $^{\prime \prime}$La Sapienza$^{\prime \prime}$, Roma, Italy\label{dietroma}
\and
Infrared Processing Analysis centre, California Institute of Technology, Pasadena, CA 91125, USA\label{ipac}
\and
Herschel Science centre, ESA/ESAC, P.O. Box 78, Villanueva de la Ca\~{n}ada, E-28691 Madrid, Spain\label{esac}
\and
ASI Science Data centre, I-00044 Frascati (Roma), Italy\label{asdc}
\and
Dipartimento di Fisica e Scienze della Terra, Universit\`{a} degli Studi di Ferrara e Sezione INFN di Ferrara, via Saragat 1, I-44100 Ferrara, Italy\label{ferrara}
\and
Dipartimento di Fisica, Universit\`{a} di Roma  $^{\prime \prime}$La Sapienza$^{\prime \prime}$, P.le Aldo Moro 2, Roma, Italy\label{roma1}
\and
Departement of Astrophysics, Astronomy and Mechanics, Faculty of Physics, University of Athens, Panepistimiopolis, 15784 Zografos, Athens, Greece \label{atene}
\and
INAF - Istituto di Radioastronomia, Via P. Gobetti 101, I-40129, Bologna, Italy\label{ira}
\and
Department of Physics and Astronomy, University College London, London, UK\label{ucl}
\and
Universit\'e de Toulouse, UPS, CESR, and CNRS, UMR5187, Toulouse, France \label{toulouseuniv}
\and
CNRS, IRAP, 9 Av. Colonel Roche, BP 44346, 31028 Toulouse Cedex 4, France\label{toulousecesr}
\and
Canadian Institute for Theoretical Astrophysics, University of Toronto, Toronto, Canada\label{cita}
\and
European Southern Observatory, Garching bei M\"unchen, Germany\label{eso}
\and
INAF - Osservatorio Astrofisico di Arcetri, Firenze, Italy\label{arcetri}
\and
Space Science Institute, Boulder, CO, USA\label{boulder}
\and 
Laboratoire AIM, CEA/DSM - INSU/CNRS - Universit\'e Paris Diderot, IRFU/SAp CEA-Saclay, 91191 Gif-sur-Yvette, France\label{saclay}
\and
Institut de RadioAstronomie Millim\'etrique (IRAM), Granada, Spain\label{iram}
\and
School of Physics, University of Exeter, Stocker Road, Exeter, EX4 4QL, UK\label{exeter}
\and
Spitzer Science centre, California Institute of Technology, Pasadena, CA\label{ssc}
\and
HYGEOS, Euratechnologies, 165 Avenue de Bretagne, 59000 Lille, France\label{hygeos}
\and
Universit\'e de Strasbourg, CNRS, UMR 7550, 11 rue de l'Universit\'e, F-67000 Strasbourg, France\label{strasbourg}
\and
Department of Astrophysics, Nagoya University, Nagoya, Japan\label{nagoya}
\and
School of Physics and Astronomy, University of Leeds, Leeds, UK\label{leeds}
\and
National Astronomical Observatories, Chinese Academy of Sciences, Beijing, China\label{beijing}
\and
Departement de Physique, Universit\'e Laval, Qu\'ebec, Canada\label{laval}
\and
The SETI Institute, 189 Bernardo Avenue Suite 100, Mountain View, CA 94043, USA\label{seti}
\and
School of Physics and Astronomy, Cardiff University, Cardiff, UK\label{cardiff}
\and
Max-Planck-Institut f\"ur Astronomie, K\"onigstuhl 17, D-69117 Heidelberg, Germany\label{mpia}
\and
Physikalisches Institut der Universit\"at zu K\"oln, Z\"ulpicher Str. 77, 50937 K\"oln, Germany\label{colonia}
\and
Univ. Bordeaux, LAB, UMR 5804, CNRS, 33270, Floirac, France\label{bordeaux}
\and
SRON Netherlands Institute for Space Research, Zernike Building, PO Box 800, NL-9700 AV Groningen, the Netherlands\label{sron}
\and
Dipartimento di Matematica e Fisica, Universit\`a del Salento, Lecce, Italy\label{lecce}
\and
Centre for Astrophysics Research, Science and Technology Research Institute, University of Hertfordshire, Hatfield AL10 9AB, UK\label{herts}
\and
INAF-Osservatorio Astrofisico di Catania, Catania, Italy\label{catania}
\and
Jeremiah Horrocks Institute, University of Central Lancashire, Preston PR1 2HE, UK\label{lancashire}
\and
Max-Planck-Institut f\"ur Radioastronomie (MPIfR), Bonn, Germany\label{mpifr}
}

   \date{Received ; accepted}

  \abstract
   {}
  {We present the first public release of high-quality data products (DR1) from Hi-GAL, the \textit{Herschel} infrared Galactic Plane Survey. Hi-GAL is the keystone of a suite of continuum Galactic Plane surveys from the near-IR to the radio, and covers five wavebands at 70, 160, 250, 350 and 500\um, encompassing the peak of the spectral energy distribution of cold dust for 8$\lesssim T \lesssim$50K. This first Hi-GAL data release covers the inner Milky Way in the longitude range 68\adeg $\gtrsim \ell \gtrsim -70$\adeg\ in a $|b| \leq1$\adeg\ latitude strip.}
  {Photometric maps have been produced with the ROMAGAL pipeline, 
 that optimally capitalizes on the excellent sensitivity and stability of the bolometer arrays of the {\em Herschel} PACS and SPIRE photometric cameras, to deliver images of exquisite quality and dynamical range, absolutely calibrated with {\em Planck} and {\em IRAS}, and recovering extended emission at all wavelengths and all spatial scales, from the point-spread function to the size of an entire 2\adeg $\times$ 2\adeg\ "tile" that is the unit observing block of the survey. The compact source catalogues have been generated with the CuTEx algorithm, specifically developed to optimize source detection and extraction in the extreme conditions of intense and spatially varying background that are found in the Galactic Plane in the thermal infrared. 
 }
  {Hi-GAL DR1 images are cirrus-noise-limited, reaching the 1$\sigma$-rms predicted by the Herschel Time Estimators for parallel-mode observations at 60\asec\,s$^{-1}$ scanning speed only in relatively low cirrus emission regions. Hi-GAL DR1 images will be accessible via a dedicated web-based image cutout service. The DR1 Compact Source Catalogues are delivered as single-band photometric lists containing, in addition to source position, peak and integrated flux and source sizes, a variety of parameters useful to assess the quality and reliability of the extracted sources; caveats and hints to help this assessment are provided. 
Flux completeness limits in all bands are determined from extensive synthetic source experiments and greatly depend on the specific line of sight along the Galactic Plane, due to the greatly varying background as a function of Galactic longitude. Hi-GAL DR1 catalogues contain 123210, 308509, 280685, 160972 and 85460 compact sources in the five bands, respectively.
   }
{}

   \keywords{ISM: dust - Galaxy: disk - Infrared: ISM - star: formation - Methods: data analysis - Techniques: photometric}

	\authorrunning{Molinari et al.}
	\titlerunning{Hi-GAL Data Release 1.}
   \maketitle
%

\section{Introduction}
\label{intro}

The Milky Way Galaxy, our home, is a complex ecosystem in which a cyclical transformation process brings diffuse baryonic matter into dense, unstable condensations to form stars.  The stars produce radiant energy for billions of years before releasing chemically enriched material back into the Interstellar medium (ISM) in their final stages of evolution. 

Although considerable progress has been made in the last two decades in 
understanding the evolution of isolated dense molecular clumps 
toward the onset of gravitational collapse and the formation of stars and 
planetary systems, a lot remains hidden. We do not know the relative 
importance of gravity, turbulence or the perturbation from spiral arms in 
assembling the diffuse and mostly atomic Galactic ISM into dense, molecular,  
filamentary structures and compact clumps. We do not know how turbulence, 
gravity and magnetic fields interact on different spatial scales to bring 
a diffuse cloud to the verge of star formation. We still do not have a comprehensive quantitative understanding of the relative 
importance of external triggers in the process, although available evidence suggests that triggering is not a major pathway for star formation \citep{thompson:2012, kendrew:2012}. We do not know how the 
relative roles played by these different agents changes from extreme 
environments like the Galactic Centre to the quiet neighbourhoods of the 
Galaxy beyond the solar circle.

Today, for the first time, it is possible to engage with this ambitious challenge, thanks to a new suite of cutting-edge Milky Way surveys that provide homogenous coverage of the entire Galactic Plane and that have already started to transform the view of our Galaxy as a global star-formation engine (see \citealt{Molinari+2014} for a recent review). 

The UKIDSS Galactic Plane Survey \citep{Lucas+2008} on the 4m UK Infrared 
Telescope on Hawaii covered the three near-IR photometric bands (J, H and 
K) to 18$^{th}$ magnitude, producing catalogues of over a billion stars. The 
unprecedented depth (15$^{th}$ mag) and resolution (2\asec) of the NASA 
{\em Spitzer} satellite's GLIMPSE survey was the first to deliver a new 
global view of the Galaxy at wavelengths of 3.6, 4.5 5.8, and 8.0\um\ 
\citep{Benjamin+2005}, until then only partially accessible from the 
ground and with imaging capabilities limited to resolutions of a few arcminutes, 
at best. The resulting catalogue of 49 million sources is dominated by stars 
and, to a lesser extent, by pre-Main Sequence young stellar objects (YSOs), with the 8.0-\um\ channel 
also showing strong extended emission that probes the interaction between 
the UV radiation from hot stars and molecular clouds. The {\em Spitzer}-MIPSGAL survey at 24\um\ 
\citep{Carey+2009} enables much deeper penetration into the dense 
molecular clouds to reveal the presence of nascent intermediate and 
high-mass stars. Such surveys, that were limited to the inner third of the 
Milky Way Galactic Plane (GP), were complemented by GLIMPSE360, that 
used {\em Spitzer} in its "warm mission" to complete the coverage of 
the entire GP at 3.6 and 4.5\um, and by the {\em WISE} 
satellite \citep{Wright+2010} that, as part of its all-sky survey, is 
covering the entire GP (although at lower resolution than {\em Spitzer})
between 3 and 25\um.

At far-infrared and millimetre wavelengths, AKARI surveyed the entire sky between 65\um\ and 160\um\ in 2006--2007.  Its spatial resolution of between 1\amin\ and 1\amin\!.5 \citep{Doi+2015} represented an improvement of a factor $\sim$3 over that of IRAS, although still a factor $\sim$5 larger than \textit{Herschel}. The \textit{Planck} satellite \citep{Planck+2011-1st} also surveyed the entire sky at wavelengths between 350\um\  and 1cm,  but with a resolution >5\amin, insufficient to resolve the complexity of the thermal dust emission internal to star-forming clouds. 

Only ground-based facilities can, at the moment, achieve resolutions below 1\amin\ in the millimetre regime The ATLASGAL survey \citep{Schuller2009} 
has used the 12-m APEX telescope in Chile to map the portion of the GP 
at longitudes between roughly +60\adeg\ and $-60$\adeg\ at 870\um, the JPS 
survey \citep{Moore+2015}, using the JCMT antenna in Hawaii, gives deeper coverage at somewhat higher resolution in the northern part of this same region at 850\um, while the Bolocam GPS covers the 
1$^{st}$ quadrant at 1.1mm \citep{Aguirre2011}. These (sub-)millimetre surveys provide a 
census of the cold and compact dust condensations that harbour 
star-formation; however, mass estimates require assumptions about 
dust temperatures that the single-band survey data themselves cannot 
constrain. 

Radio-wavelength continuum observations provide extinction-free views of 
bremsstrahlung radiation from ultra-compact HII (UCHII) regions and the ionised ISM in general. 
The 1\asec\!.5 resolution, 6-cm CORNISH survey used the Very Large Array telescope to map the 
$\ell = +10$\adeg\ to $+65$\adeg\ section of the GP at resolutions of $\sim$1\asec\ to $\sim$10\asec
\citep{Purcell2013}.  The 
CORNISH-South extension of the project, carried out with the ATCA array 
will complement this information for the corresponding region of the 4$^{th}$ quadrant, 
augmented with imaging in radio recombination lines.

This suite of continuum GP surveys sees its ideal complement 
in a family of spectroscopic surveys of molecular and atomic emission lines. Kinematic 
information on the same dense clouds traced by the thermal emission from cool dust 
can also be traced using
molecular-line emission. The Galactic Ring Survey (GRS; \citealt{Jackson+2006}), at 
46\asec\ resolution, uses the FCRAO 14-m antenna to map the $^{13}$CO (J=1--0) 
transition in the range 15\adeg $\lesssim \ell \lesssim$56\adeg. The JCMT COHRS survey \citep{Dempsey+2013} covers essentially the same longitude range as the GRS, but in the CO (J=3--2) line and at a spatial resolution of 14\asec.

Further extensions to the GRS, in the 1$^{st}$ and 2$^{nd}$ quadrants, toward the Galactic 
Anticentre, also in $^{12}$CO (J=1--0), have been carried out with the FCRAO \citep{Heyer1998, Brunt+2015}. The International Galactic Plane Survey 
(IGPS) has combined three interferometric 21-cm HI surveys at 45--60\asec\ 
resolution, the combination giving an ideal tool to study the transformation of atomic into 
molecular gas in the spiral arms (e.g., \citealt{McClure+2001}).

The coverage of the 3$^{rd}$ and 4$^{th}$ quadrants in molecular lines is more sparse and less systematic. Together with targeted-source line surveys like MALT90 \citep{Jackson+2013}, unbiased coverage of the Plane is limited to the NANTEN survey (e.g. \citealt{Mizuno+2004}), now being improved with the NANTEN2/NASCO project that, however, still has limited ($\sim4$\amin) spatial resolution. Recent unbiased surveys with the Mopra antenna in Australia \citep{Burton+2013, Jones2012} are starting to fill the gap with the data quality of the CO surveys in the northern portion of the GP. The SEDIGISM survey is currently in execution to map the 4$^{th}$ quadrant between $\ell = +18$\adeg\ and $\ell = -60$\adeg\ in $^{13}$CO and 
C$^{18}$O (J=2--1) with the APEX telescope.

The Methanol Multi-Beam survey (e.g., \citealt{Green+2012}) is searching the Plane for 6.7-GHz methanol maser emission using the Parkes and ATCA telescopes. Methanol maser emission is characteristic of the early formation stage of massive stars; its association with cool dense clumps is a signpost for ongoing formation of massive stars and associated protoclusters in such objects. A more complete compilation of GP Surveys from the near-IR to the radio is provided in the review of \cite{Molinari+2014}.

The {\em Herschel} infrared Galactic Plane Survey (Hi-GAL, \citealt{Molinari10, Molinari2010b}), carried out with the {\em Herschel} Space Observatory \citep{pilbratt10}, is the keystone in the arch of GP continuum 
surveys. With a full Plane coverage of the thermal far-IR and 
submillimetre continuum in five bands between 70\um\ and 500\um, ideally covering 
the peak of the spectral energy distribution (SED) of dust in the 
temperature range 8\,K$\leq T \leq 50$\,K, Hi-GAL delivers a complete census of 
structures containing cold dust, from the Central Molecular Zone to the outskirts of 
the Galaxy, enabling self-consistent determination of dust 
temperatures and masses. Thanks to its space-borne platform, the {\em 
Herschel} cameras do not suffer from the rapid atmospheric variabilities 
that limit ground-based submillimetre facilities.  This allows full 
exploitation of the excellent sensitivity and stability of the infrared 
bolometric arrays to deliver exquisite-quality images that recover 
extended emission from dust on all spatial scales.  The ability of {\em Herschel} 
to recover multi-wavelength extended emission from the 
diffuse ISM, through dense filamentary structures, down to compact and 
point-like sources \citep{Molinari2010b, andre10} are and will remain 
unparalleled in the coming decades.


Hi-GAL is delivering a transformational view of the complete evolutionary path that brings cold and diffuse interstellar material to condense into clouds and filaments that then fragment into protocluster-forming dense clumps. More than 50 papers have been published by the \higal\ consortium to date, based on \higal\ images and preliminary source catalogues, from studies of the diffuse ISM (e.g. \citealt{bernard10, paradis10, comp10, Traficante+2014, Elia+2014}) to dense, large-scale filaments \citep{Molinari2010b, Schisano+2014, Wang+2015}, dust in HII regions (e.g. \citealt{Paladini+2012, Tibbs+2012}, clumps and massive star formation (e.g. \citealt{elia10, Bally2010_W43, Elia+2013, Battersby+2011, Mottram+2012, Wilcock+2012, Veneziani+2013, Beltran+2013, Strafella+2015, Traficante+2015}), Galactic Central Molecular Zone studies \citep{Molinari+2011, longmore12}, triggered star formation \citep{zavagno:2010} and finally dust around post-Main Sequence objects \citep{Umana+2012, Mart+2015}. More papers are in preparation in the \higal\ Consortium. Although basic \higal\ data have always been open for public access through the Herschel Science Archive, we are now providing access for the larger Community to the high-quality data products (maps and source catalogs) used internally by the Hi-GAL consortium.

In this paper we present the first public release of Hi-GAL data products 
(DR1). DR1 is limited to the inner Milky Way in the longitude range 
+68\adeg $\geq \ell \geq -70$\adeg\ and latitude range 1\adeg $\geq b \geq 
-$1\adeg , and consists of calibrated and astrometrically registered 
images at 70, 160, 250, 350 and 500\um, plus compact-source catalogues, delivered via an image cutout service provided by the ASI Science Data Center and accessible from the VIALACTEA project portal at \url{http://vialactea.iaps.inaf.it}.  We present and discuss the production
methods and characterization of the images and catalogues considered according to their 
band-specific properties. A full systematic analysis of the physical 
properties of dense, star-forming and potentially star-forming 
condensations (reconstructed from the band-merged Hi-GAL 
photometric catalogues with augmented SED coverage from ancillary surveys 
from the mid-IR to the millimetre), will be presented in \cite{Elia+2016}. 
A first systematic analysis of far-IR properties of post-Main Sequence 
objects based on the Hi-GAL Catalogues is presented in \cite{Mart+2015}.

\section{Observations}
\label{observations}

The motivations and observing strategy adopted for the Hi-GAL Survey are 
described in detail in \cite{Molinari10}. The complete survey was 
assembled in three instalments of observing time granted in 
Open Time competition in each of the three calls issued during the 
{\em Herschel} project lifetime. Due to a clerical inconsistency in
determining the duration time of the observations, a longitude range 
of about 6\adeg\ in extent in the outer Galaxy could not be executed in 
the observing time formally granted for the complete Plane coverage, and 
Director's Discretionary Time was additionally granted to obtain the 
360\adeg -wide coverage. The total observing time amounted to slightly in 
excess of 900 hours, making the full Hi-GAL survey the largest observing program carried out by {\em Herschel}.

The Hi-GAL observations were acquired by subdividing the surveyed area into 
square tiles of $\sim$ 2\adeg .2 in size, to obtain complete coverage of a 
$|b| \leq 1$ strip of the Galactic Plane at 70, 160, 250, 350 and 500\um\ 
simultaneously. Each tile was observed with the PACS \citep{poglitsch10} and SPIRE \citep{griffin10} cameras in parallel mode
(pMode), specifically designed to optimise data acquisition 
for large-area multi-wavelength surveys. In pMode the PACS and SPIRE 
cameras are used simultaneously, effectively making {\em Herschel} into a five-band 
imaging camera spanning a decade in wavelength. Since the field of view of 
the PACS and SPIRE cameras are offset by $\sim 20$\arcmin\ in the plane of 
the sky, slight oversizing of the individual observing tiles was 
needed to make sure that a 2\adeg x2\adeg\ area was covered in all five 
photometric bands.

As the bolometers that constitute the elemental pixels of the PACS and SPIRE 
arrays are differential detectors known to be affected by slow thermal 
drifts with typical $1/f$ frequency behaviour, each tile was observed in two 
independent passes with nearly orthogonal scanning directions.  Individual Astronomical Observation Requests (AORs) were concatenated in the {\em Herschel} Observation Planning Tool (HSpot) so that the two scanning passes were executed 
immediately one after the other for each tile. This strategy was chosen so 
that a given position in the sky was observed by as many pixels as 
possible and in different scanning directions, producing the  
degree of redundancy needed to beat 
down the correlated and uncorrelated $1/f$ noise of single detectors, 
thereby allowing recovery of all the emission at the largest possible 
spatial scales. The approach was also designed to perfectly couple to the data 
processing and map-making pipeline specifically developed for the \higal\ 
project (see \S\ref{maps}).

The satellite scan speed in pMode was set to its maximum value of 60 
\arcsec\ per second, with a detector sampling rate of 40\,Hz for PACS and 10\,Hz for 
SPIRE. The spatial sampling is therefore 1.5\arcsec\ and 
6.0\arcsec\ for PACS and SPIRE, respectively, enough to Nyquist sample all 
the nominal diffraction-limit beams ($\simeq$[6.0, 12.0, 18.0, 24.0, 
35.0]\arcsec\ at [70,160, 250, 350, 500] \um, respectively). However, due to 
the limited transmission bandwidth, the PACS data were co-added on-board 
{\em Herschel}, with a compression of 8 and 4 consecutive frames at 70 and 
160 \um, producing an effective spatial sampling of 12\arcsec\ and 6\arcsec\ at 
70 and 160 \um\ respectively. Therefore, in pMode, the PACS beams are not 
Nyquist sampled and the resulting point-spread functions (PSFs) are elongated along the scan direction 
with a measured size of 5.8\arcsec$\times$12.1\arcsec and 
11.4\arcsec$\times$13.4\arcsec\ at 70 \um\ and 160 \um, respectively (Lutz 
2012\footnote{\url{http://herschel.esac.esa.int/twiki/pub/Public/PacsCalibrationWeb/bolopsf\_20.pdf}}).

Table \ref{obs_log} summarizes a few details of the observations. Column 1 is an assigned field name for each tile; cols. 2-5 report the approximate coordinates of the tile centre; cols. 6 and 7 indicate the date of the observation for each tile, both in standard format and in OD number (Observation Day, starting from date of launch);  cols. 8-10 report the start time (UT) of the tile in the nominal and orthogonal scan direction (see below), together with the associated observation identification (OBSID) number uniquely attached to each scan observation.

SPIRE was used in ``bright-source'' mode in the three tiles of the survey 
closest to the Galactic Centre (roughly centred at longitudes +2\adeg, 
0\adeg\ and $-$2\adeg). This was done to avoid the widespread 
saturation and non-linearities in the detector response otherwise likely to occur on 
the extraordinarily strong background emission in that region.  In this 
observing mode, the limited 12-bit dynamical range of the 
Analog-to-Digital converters in the detector chains is centred around 
higher-than-nominal current values. In this way saturation is avoided at the 
cost of greatly decreased sensitivity.  In ``bright-source'' mode SPIRE is much less capable of 
detecting intermediate and low-flux compact sources
(see fig. \ref{glondistr_all}, last three panels).

\begin{table*}
\centering
\tiny
\caption{The log of the observations.}
\begin{tabular}{lcccrcc|lc|c}
\hline\hline
\multicolumn{1}{c}{Field}&RA&Declination&$\ell$&\multicolumn{1}{c}{b}&Date&OD&\multicolumn{2}{c|}{Nominal}&Ortho.\\
&hh:mm:ss&dd:pp:ss&&&&&\multicolumn{1}{c}{Start}&OBSID&Start\\
\hline
290 & 11:05:14.266 & $-$60:57:38.79 & 290.400 & $-$0.700 & 2010-08-15 & 459 & 15:42:32 & 1342203081(+1) & 18:21:52\\
292 & 11:22:31.063 & $-$61:42:12.77 & 292.600 & $-$0.629 & 2010-08-14 & 458 & 20:53:20 & 1342203065($-$1) & 18:01:48\\
294 & 11:40:34.588 & $-$62:18:00.17 & 294.800 & $-$0.552 & 2010-08-14 & 458 & 26:24:29 & 1342203067($-$1) & 23:32:57\\
297 & 11:59:16.713 & $-$62:44:25.30 & 297.000 & $-$0.471 & 2010-08-15 & 458 & 07:55:37 & 1342203069($-$1) & 05:04:05\\
299\tablefootmark{a} & 12:18:26.910 & $-$63:00:59.70 & 299.200 & $-$0.384 & 2009-09-03 & 112 & 03:21:32 & 1342183075(+1) & 06:26:26\\
301 & 12:37:52.625 & $-$63:07:23.91 & 301.400 & $-$0.292 & 2010-08-15 & 459 & 24:06:14 & 1342203084($-$1) & 21:14:42\\
303\tablefootmark{b} & 12:57:20.191 & $-$63:03:29.76 & 303.600 & $-$0.194 & 2010-01-08 & 239 & 04:03:19 & 1342189081(+1) & 07:15:22\\
305\tablefootmark{b} & 13:16:35.719 & $-$62:49:20.56 & 305.800 & $-$0.091 & 2010-01-08 & 239 & 13:40:19 & 1342189084($-$1) & 10:27:43\\
308 & 13:35:27.781 & $-$62:26:15.89 & 308.000 &    0.000 & 2010-08-16 & 459 & 05:38:17 & 1342203086($-$1) & 02:46:45\\
310 & 13:53:56.541 & $-$61:59:11.46 & 310.200 &    0.000 & 2010-08-20 & 464 & 22:27:35 & 1342203279($-$1) & 19:36:03\\
312\tablefootmark{b} & 14:11:47.708 & $-$61:23:08.81 & 312.400 &    0.000 & 2010-01-09 & 240 & 04:06:20 & 1342189110($-$1) & 00:53:44\\
314 & 14:28:53.915 & $-$60:38:41.13 & 314.600 &    0.000 & 2010-08-21 & 464 & 01:07:08 & 1342203280(+1) & 03:46:28\\
316 & 14:45:10.082 & $-$59:46:25.81 & 316.800 &    0.000 & 2010-08-21 & 464 & 06:37:45 & 1342203282(+1) & 09:17:05\\
319 & 15:00:33.335 & $-$58:47:02.32 & 319.000 &    0.000 & 2010-08-21 & 465 & 17:57:13 & 1342203289(+1) & 20:36:33\\
321 & 15:15:02.740 & $-$57:41:10.22 & 321.200 &    0.000 & 2010-08-21 & 465 & 26:19:33 & 1342203292($-$1) & 23:28:01\\
323\tablefootmark{b} & 15:28:38.841 & $-$56:29:28.17 & 323.400 &    0.000 & 2010-01-29 & 261 & 24:48:03 & 1342189879($-$1) & 21:35:27\\
325 & 15:41:23.367 & $-$55:12:32.49 & 325.600 &    0.000 & 2010-08-22 & 465 & 04:59:07 & 1342203293(+1) & 07:38:27\\
327 & 15:53:18.797 & $-$53:50:57.02 & 327.800 &    0.000 & 2010-09-03 & 478 & 24:11:19 & 1342204043($-$1) & 21:19:47\\
330 & 16:04:28.099 & $-$52:25:12.66 & 330.000 &    0.000 & 2010-09-04 & 478 & 05:42:27 & 1342204045($-$1) & 02:50:55\\
332 & 16:14:54.520 & $-$50:55:47.08 & 332.200 &    0.000 & 2010-09-04 & 478 & 08:21:47 & 1342204046(+1) & 11:01:07\\
334 & 16:24:41.348 & $-$49:23:05.26 & 334.400 &    0.000 & 2010-09-04 & 479 & 24:20:00 & 1342204055($-$1) & 21:28:28\\
336 & 16:33:51.865 & $-$47:47:29.17 & 336.600 &    0.000 & 2010-09-05 & 479 & 05:51:08 & 1342204057($-$1) & 02:59:36\\
338 & 16:42:29.198 & $-$46:09:18.44 & 338.800 &    0.000 & 2010-09-05 & 479 & 11:22:16 & 1342204059($-$1) & 08:30:44\\
341 & 16:50:36.309 & $-$44:28:50.42 & 341.000 &    0.000 & 2010-09-06 & 480 & 13:55:21 & 1342204095($-$1) & 11:03:49\\
343 & 16:58:15.976 & $-$42:46:20.19 & 343.200 &    0.000 & 2010-09-06 & 480 & 08:24:15 & 1342204093($-$1) & 05:32:43\\
345 & 17:05:30.743 & $-$41:02:01.36 & 345.400 &    0.000 & 2010-09-06 & 480 & 02:53:09 & 1342204091($-$1) & 00:01:37\\
347 & 17:12:22.972 & $-$39:16:05.62 & 347.600 &    0.000 & 2010-09-06 & 481 & 25:27:08 & 1342204101($-$1) & 22:35:36\\
349 & 17:18:54.803 & $-$37:28:43.53 & 349.800 &    0.000 & 2011-02-20 & 647 & 07:34:31 & 1342214511($-$1) & 04:42:59\\
352 & 17:25:08.192 & $-$35:40:04.45 & 352.000 &    0.000 & 2011-02-20 & 648 & 20:02:47 & 1342214576($-$1) & 17:11:15\\
354 & 17:31:04.932 & $-$33:50:16.46 & 354.200 &    0.000 & 2011-02-24 & 651 & 04:29:59 & 1342214713(+1) & 07:09:19\\
356 & 17:36:46.638 & $-$31:59:27.04 & 356.400 &    0.000 & 2010-09-12 & 486 & 12:19:22 & 1342204369($-$1) & 09:27:50\\
358\tablefootmark{c} & 17:42:14.801 & $-$30:07:42.57 & 358.600 &    0.000 & 2010-09-12 & 486 & 06:48:07 & 1342204367($-$1) & 03:56:26\\
\phantom{00}0\tablefootmark{c} & 17:45:37.199 & $-$28:56:10.23 & \phantom{00}0.000 & 0.000 & 2010-09-07 & 481 & 04:08:11 & 1342204102(+1) & 06:47:44\\
\phantom{00}2\tablefootmark{c} & 17:50:46.049 & $-$27:03:08.22 & \phantom{00}2.200 & 0.000 & 2010-09-07 & 481 & 09:39:03 & 1342204104($-$1) & 12:18:36\\
\phantom{00}4 & 17:55:44.665 & $-$25:09:25.47 & \phantom{00}4.400 & 0.000 & 2011-02-24 & 652 & 13:30:00 & 1342214761(+1) & 16:09:20\\
\phantom{00}6 & 18:00:34.119 & $-$23:15:06.43 & \phantom{00}6.600 & 0.000 & 2011-02-24 & 652 & 19:01:17 & 1342214763(+1) & 21:40:37\\
\phantom{00}8 & 18:05:15.396 & $-$21:20:15.19 & \phantom{00}8.800 & 0.000 & 2011-04-09 & 695 & 02:11:54 & 1342218963(+1) & 04:51:14\\
\phantom{0}11 & 18:09:49.409 & $-$19:24:55.45 & \phantom{0}11.000 & 0.000 & 2011-04-09 & 695 & 07:43:15 & 1342218965(+1) & 10:22:35\\
\phantom{0}13 & 18:14:17.005 & $-$17:29:10.63 & \phantom{0}13.200 & 0.000 & 2011-04-10 & 696 & 13:36:51 & 1342218999(+1) & 16:16:11\\
\phantom{0}15 & 18:18:38.972 & $-$15:33:03.90 & \phantom{0}15.400 & 0.000 & 2011-04-10 & 696 & 10:57:27 & 1342218998($-$1) & 08:05:55\\
\phantom{0}17 & 18:22:56.053 & $-$13:36:38.19 & \phantom{0}17.600 & 0.000 & 2011-04-10 & 696 & 05:26:14 & 1342218996($-$1) & 02:34:42\\
\phantom{0}19 & 18:27:08.946 & $-$11:39:56.26 & \phantom{0}19.800 & 0.000 & 2011-04-15 & 701 & 09:54:54 & 1342218644(+1) & 12:34:14\\
\phantom{0}22 & 18:31:18.313 & \phantom{0}$-$9:43:00.72 & \phantom{0}22.000 & 0.000 & 2011-04-15 & 701 & 07:15:30 & 1342218643($-$1) & 04:23:58\\
\phantom{0}24 & 18:35:24.790 & \phantom{0}$-$7:45:54.02 & \phantom{0}24.200 & 0.000 & 2011-04-15 & 701 & 15:26:36 & 1342218646(+1) & 18:05:56\\
\phantom{0}26 & 18:39:28.986 & \phantom{0}$-$5:48:38.52 & \phantom{0}26.400 & 0.000 & 2011-04-16 & 702 & 14:23:37 & 1342218696(+1) & 17:02:57\\
\phantom{0}28 & 18:43:31.490 & \phantom{0}$-$3:51:16.53 & \phantom{0}28.600 & 0.000 & 2011-04-16 & 702 & 08:52:59 & 1342218694(+1) & 11:32:19\\
\phantom{0}30\tablefootmark{d} & 18:46:05.222 & \phantom{0}$-$2:36:32.90 & \phantom{0}30.000 &  0.000 & 2009-10-24 & 163 & 02:33:54 & 1342186275(+1) & 05:39:08\\
\phantom{0}30filler\tablefootmark{e} & 18:48:28.610 & \phantom{0}$-$1:09:31.80 & \phantom{0}31.563 & +0.130 & 2011-10-24 & 893 & 09:38:00 & 1342231361(+1) & 10:40:30\\
\phantom{0}33 & 18:51:33.726 & \phantom{0}+0:03:38.13 & \phantom{0}33.000 & 0.000 & 2011-04-16 & 702 & 06:13:19 & 1342218693($-$1) & 03:21:47\\
\phantom{0}35 & 18:55:34.588 & \phantom{0}+2:01:06.44 & \phantom{0}35.200 & 0.000 & 2011-04-26 & 712 & 16:39:59 & 1342219631($-$1) & 13:48:27\\
\phantom{0}37 & 18:59:36.031 & \phantom{0}+3:58:32.52 & \phantom{0}37.400 & 0.000 & 2010-10-23 & 528 & 24:48:50 & 1342207027($-$1) & 21:57:18\\
\phantom{0}39 & 19:03:38.623 & \phantom{0}+5:55:54.18 & \phantom{0}39.600 & 0.000 & 2010-10-24 & 528 & 06:19:59 & 1342207029($-$1) & 03:28:27\\
\phantom{0}41 & 19:07:42.942 & \phantom{0}+7:53:09.18 & \phantom{0}41.800 & 0.000 & 2010-10-24 & 528 & 11:51:08 & 1342207031($-$1) & 08:59:36\\
\phantom{0}44 & 19:11:49.580 & \phantom{0}+9:50:15.27 & \phantom{0}44.000 & 0.000 & 2010-10-24 & 529 & 26:42:28 & 1342207053($-$1) & 23:50:56\\
\phantom{0}46 & 19:15:59.148 &   +11:47:10.05 & \phantom{0}46.200 & 0.000 & 2010-10-25 & 529 & 08:13:37 & 1342207055($-$1) & 05:22:05\\
\phantom{0}48 & 19:20:12.280 &   +13:43:51.05 & \phantom{0}48.400 & 0.000 & 2011-11-05 & 905 & 06:33:30 & 1342231859($-$1) & 03:41:58\\
\phantom{0}50 & 19:24:29.643 &   +15:40:15.66 & \phantom{0}50.600 & 0.000 & 2011-11-04 & 904 & 08:55:34 & 1342231851(+1) & 11:34:54\\
\phantom{0}52 & 19:28:32.384 &   +17:38:53.32 & \phantom{0}52.800 & +0.088 & 2011-10-23 & 892 & 10:22:02 & 1342231342($-$1) & 07:30:30\\
\phantom{0}55 & 19:32:35.799 &   +19:37:48.65 & \phantom{0}55.000 & +0.198 & 2011-05-02 & 718 & 13:13:23 & 1342219813($-$1) & 10:21:51\\
\phantom{0}57 & 19:36:46.510 &   +21:36:13.38 & \phantom{0}57.200 & +0.301 & 2011-05-02 & 718 & 18:44:29 & 1342219815($-$1) & 15:52:57\\
\phantom{0}59\tablefootmark{d} & 19:41:44.297 &   +23:01:21.60 & \phantom{0}59.000 & 0.000 & 2009-10-23 & 162 & 10:28:59 & 1342186235(+1) & 13:34:13\\
\phantom{0}59filler\tablefootmark{f} & 19:44:00.000 &   +24:10:00.00 & \phantom{0}60.250 & +0.119 & 2011-11-05 & 905 & 09:14:34 & 1342231860(+1) & 10:16:38\\
\phantom{0}61 & 19:45:33.109 &   +25:31:19.18 & \phantom{0}61.600 & +0.492 & 2011-05-03 & 719 & 15:17:59 & 1342220536($-$1) & 12:26:27\\
\phantom{0}63 & 19:50:10.833 &   +27:27:53.52 & \phantom{0}63.800 & +0.579 & 2011-10-23 & 892 & 04:49:22 & 1342231340($-$1) & 01:57:50\\
\phantom{0}66 & 19:54:59.545 &   +29:23:43.65 & \phantom{0}66.000 & +0.661 & 2011-10-22 & 892 & 23:18:16 & 1342231338($-$1) & 20:26:44\\\hline
\end{tabular}
\tablefoot{\tiny Column Field gives the ID of each tile; next four columns 
report the coordinates (equatorial and Galactic) of the centre of the 
tile; Date is the date of the observation (yyyy-mm-dd); OD is the \textit{Herschel} 
operational day (with OD 1 corresponding to 14 May 2009); Start is the 
start time (hh:mm:ss): if the second observation was begun after midnight, the start time is increased 
by 24 hours; OBSID is the \textit{Herschel} identifier of the 
observation: the OBSID for the orthogonal scan is +1 or $-$1, according to 
the value given in parenthesis. Unless stated differently, all the maps 
have a size of 2$^\circ\times$2$^\circ$ and the observations lasted 9490\,s 
and 10189\,s for the two scans.\\
\tablefoottext{a} Observed during the Performance Verification Phase: duration was 10930\,s for both scans;\\ 
\tablefoottext{b} duration was 11453\,s for both scans;\\ 
\tablefoottext{c} duration was 9499\,s and 10198\,s for the two scans. SPIRE was used in ``bright source'' mode; \\ 
\tablefoottext{d} observed during the Science Demonstration Phase: duration was 10940\,s for both scans;\\ 
\tablefoottext{e} size is 120$\times$35 arcmin$^2$, duration was 3662\,s and 5599\,s for the two scans;\\ 
\tablefoottext{f} size is 130$\times$30 arcmin$^2$, duration was 3654\,s and 5717\,s for the two scans.}
\label{obs_log}
\end{table*}

\onecolumn
\begin{landscape}
\begin{figure*}
\resizebox{\hsize}{!}{\includegraphics{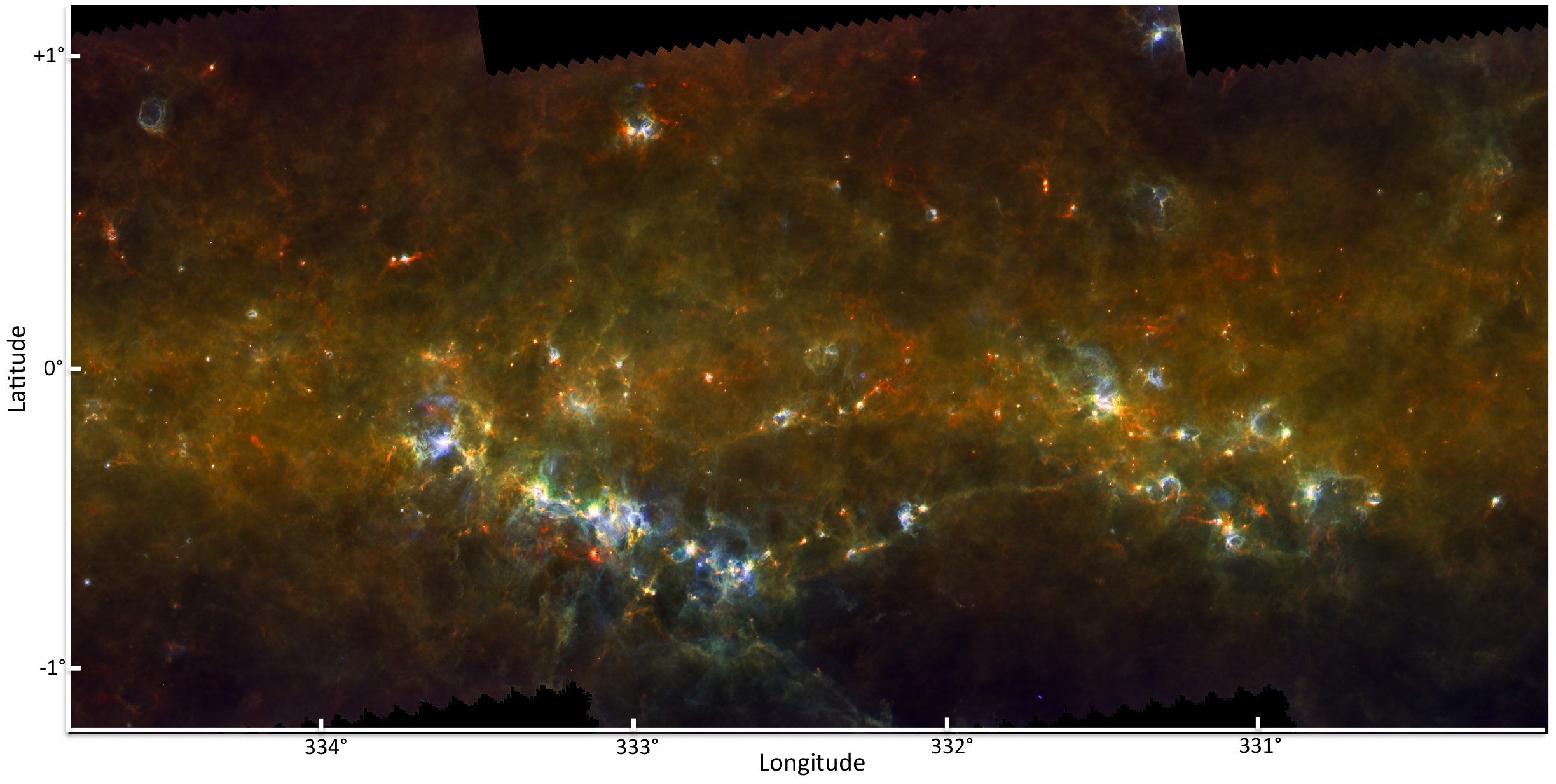}}
\caption{Three-colour image (\textit{blue} 70\um, \textit{green} 160\um, \textit{red} 350\um) of a 3-tile mosaic field around 330\adeg $\lesssim \ell \lesssim$335\adeg.}
\label{image-sample}
\end{figure*}
\end{landscape}
\twocolumn

\FloatBarrier

\section{Production of the Photometric Maps}
\label{maps}

The data reduction was carried out using the ROMAGAL data-processing software
described in detail in \cite{trafi11}. In short, the pipeline uses  
standard {\em Herschel} Interactive Processing Environment (HIPE) \citep{ott10} processing 
up to level 0.5, where the signal from individual detectors is photometrically calibrated and each 
detector has its sky position assigned. Subsequent steps in the data reduction were carried out using a dedicated pipeline written within the \higal\ Consortium. Fast 
and slow detector glitches arising from particle hits onto the detectors 
are identified and the affected portions of the data are flagged in each 
detector's Time Ordered Data (TOD). Slow detector drifts arising from $1/f$ 
noise are estimated and subtracted; for PACS, the drifts are estimated at 
subarray level as each $16 \times 16$ array matrix shares the same readout electronics. 
The core of the map-making implements a Generalised Least-Square (GLS) 
algorithm that is ideally designed to use redundancy to minimise residual 
uncorrelated $1/f$ detector noise by filtering in Fourier space 
\citep{nat01}.  In order to deliver optimal results, the code (i.e.\ each 
GLS-based code) requires that the detector noise properties are 
regularly sampled in time over the entire duration of the observations. 
For this reason we implement a pre-processing stage where the sections of 
the TOD flagged as "bad data" (e.g. due to a glitch removal or signal saturation) are replaced 
with artificial samples in which the data are set to 0, but where the 
noise is added using a "constrained noise realisation" using the noise 
frequency properties estimated from valid data immediately before and 
after the flagged section.


The pixel sizes of the ROMAGAL maps account for the larger-than-nominal 
PACS PSFs and are set to [3.2, 4.5, 6.0, 8.0, 11.5]\arcsec\ at [70, 160, 
250, 350, 500] \um\ respectively.  This choice represents a good compromise 
between the need to sample the PSF as also determined for point-like 
objects in Hi-GAL maps with at least three pixels, while avoiding (in the 
case of PACS 70\um\ and 160\um) excessively small pixels in which the hit statistics 
of the detector sampling are too low, resulting in increased pixel-to-pixel 
noise. For the PACS bands this is due to the fact that the {\em Herschel} scanning strategy in pMode implements an on-board frame co-adding (see \S\ref{observations}), resulting in an effective decrease in sampling rate. The pixel size of the images is therefore such that the beam FWHM is sampled with 3 pixels for the three SPIRE bands, and with 2.66 pixels in the PACS bands. Saturated pixels in the maps are a consequence of signal saturation for all TODs covering the specific pixel, due to the necessary limitations in the dynamical range of DAC converters at the detection stage. A list of locations where saturation is reached is reported in appendix \ref{saturation}.


It is clearly not possible to report in the paper, even in electronic 
form, the complete list of images for all wavelengths and all the tiles of 
Table \ref{obs_log}. We choose here to show only one figure (Fig. 
\ref{image-sample}) as a 3-colour image of a 3-tile mosaic in the longitude range 330\adeg $\lesssim \ell \lesssim$335\adeg  to 
set the framework for the 
subsequent sections (see \S\ref{catalogues} and \S\ref{reliability}) 
describing the properties of the compact-source catalogues. \textit{The 
maps deliver a stunning view of the GP at all Hi-GAL 
wavelengths with a detail that is unattainable from any 
ground-based millimetre-wave facility now and in the foreseeable future.  Extended emission 
with at least two orders of magnitude dynamical range in intensity is retrieved at 
all spatial scales from the most compact objects to the extent of the 
entire tile. We will show in the next sections that compact sources 
within these multiple complex, extended structures have very 
low peak/background contrast ratio (generally below 1).  This makes the 
detection and flux computation of compact sources an extremely complex task, 
where it is, in particular, difficult to identify a figure of merit that can be 
used to unambiguously distinguish reliable from unreliable sources}.

\begin{figure}[h!]
\includegraphics[width=0.5\textwidth]{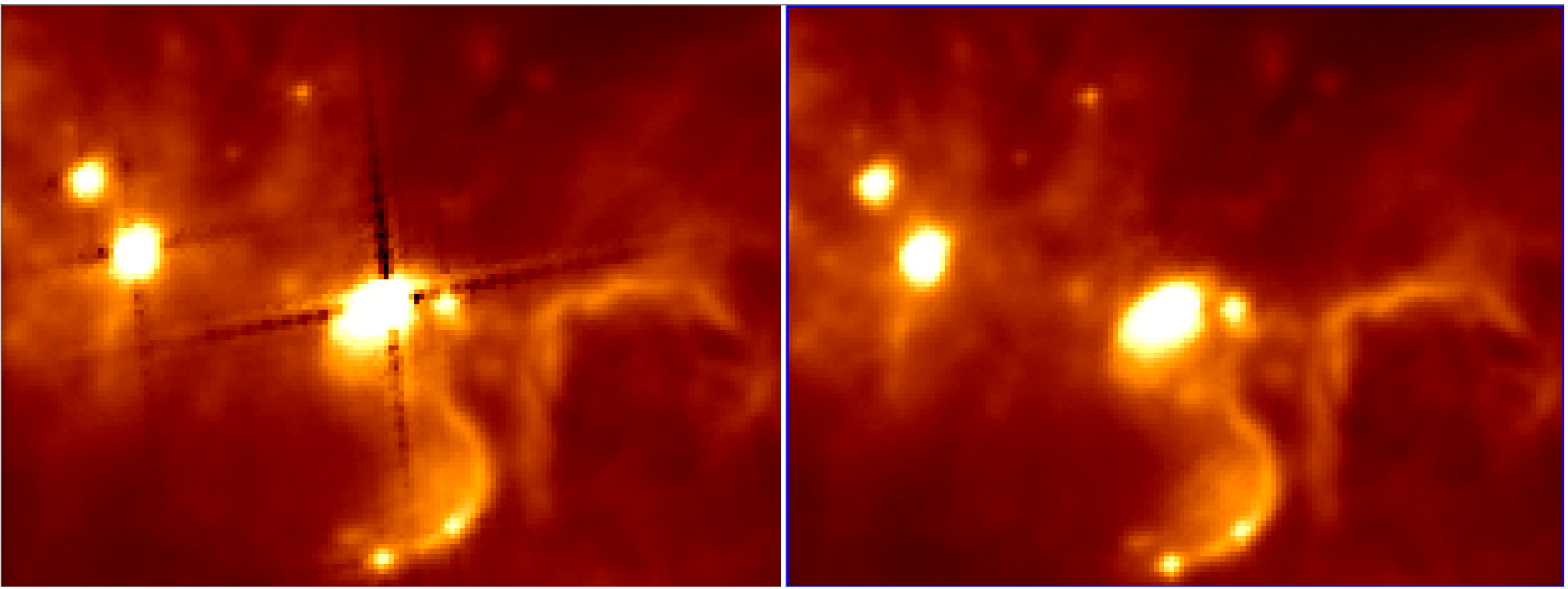}
\caption{\textit{Left panel:} Particulars of a point source as 
reconstructed by the ROMAGAL map-making for PACS at 70\um; the typical 
"cross" feature introduced by the GLS map-maker when performing the noise 
filter deconvolution in Fourier space over strongly varying signal (as is 
the case for point-like sources) is clearly seen (image in log scale). 
The minimum within the negative cross scales as $\sim 2.5$\% of
the peak source flux. 
\textit{Right panel:} 
same as left panel with the same scale and colour stretch, but after 
applying the correction devised by \cite{Piazzo+2012}. The angular extent of the region imaged is $\sim$6\amin $\times$ 4\amin.}
\label{crosses}
\end{figure}

The pipeline is augmented with a module 
specifically developed by the Hi-GAL team to cure the high-frequency 
artefacts that the GLS map-making technique used in ROMAGAL (as in many 
other approaches, like MadMap or Scanamorphos) is known to introduce to the maps, namely 
crosses and stripes corresponding to the brightest sources. The left 
panel in Fig. \ref{crosses} shows a typical example of these features that 
are introduced by the noise filter deconvolution carried out by the GLS 
map-maker in Fourier space when the flux is strongly varying with position, 
as is the case for point-like sources. We find that the 
minimum within a negative cross feature is 
proportional to the peak brightness of the source, and amounts to 
$\sim2.5$\%\ of this value. It is therefore not a strong effect in 
principle, but is can be quite annoying for relatively faint nearby 
objects and for the determination of the surrounding diffuse emission; it 
is also aesthetically undesirable.

To correct for such effects, particularly visible in the PACS 70-\um, and to a lesser extent, in the 160-\um\ images, a weighted post-processing of the GLS maps (WGLS, \citealt{Piazzo+2012})  has been applied to finally obtain images in which these artefacts are removed (right panel in Fig. \ref{crosses}).

\subsection{Noise properties of the Hi-GAL maps}
\label{maps-noise}

\begin{figure}[h!]
\includegraphics[width=0.5\textwidth]{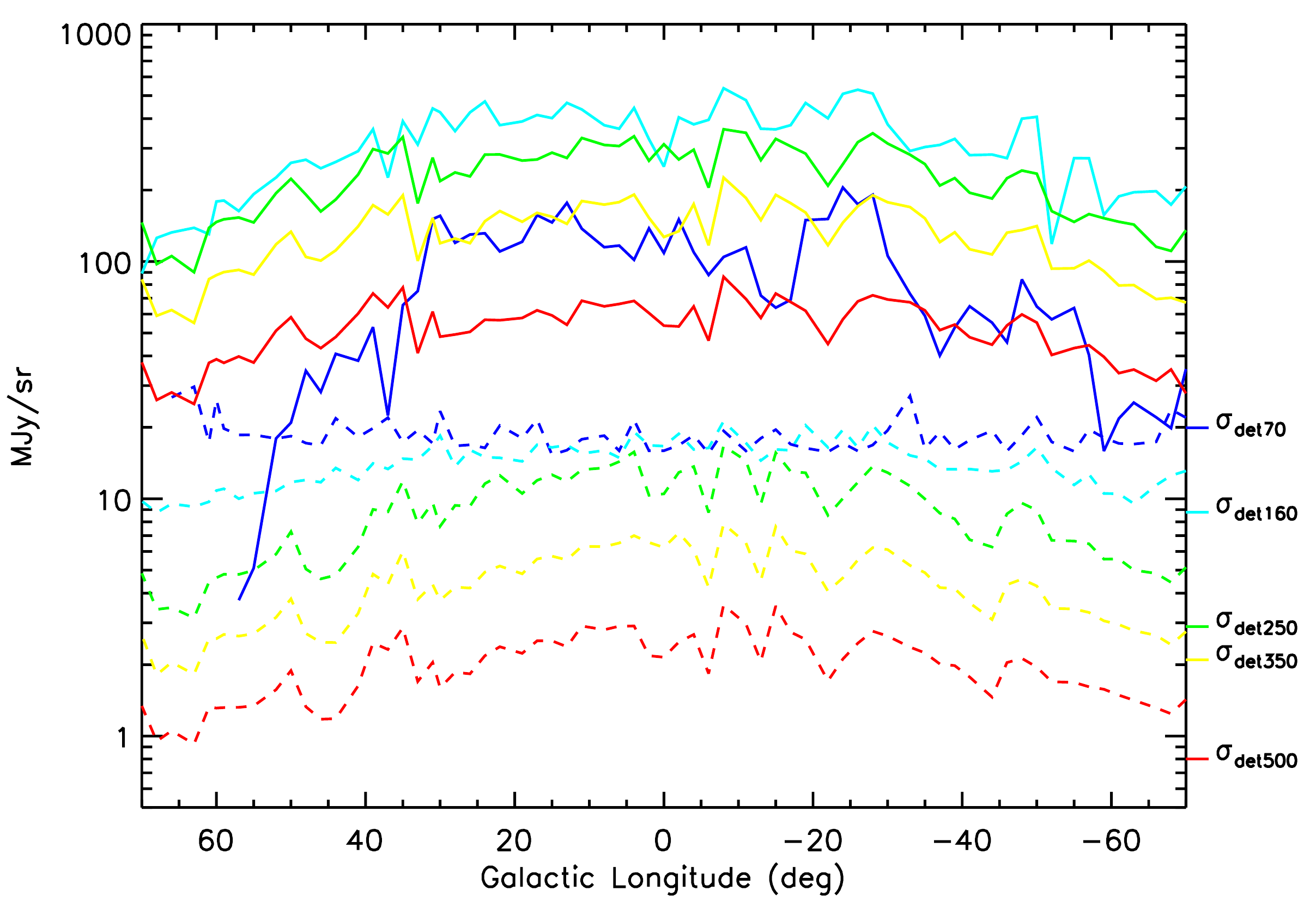}
\caption{Distribution as a function of Galactic longitude of the median brightness (full lines) and its r.m.s. (dashed lines), in regions within each Hi-GAL tile where the brightness levels are below the 10\% percentiles of the brightness distribution for that tile. Hi-GAL bands are colour-coded as: blue for PACS 70\um , cyan for PACS 160\um , green for SPIRE 250\um , yellow for SPIRE 350\um\ and red for SPIRE 500\um .  Ticks on the right margin of the figure mark the values of the theoretical sensitivities predicted by the official PACS/SPIRE time estimator (available in HSpot) for observations in pMode with 60\asec\,s$^{-1}$ scanning speed.}
\label{noisefig}
\end{figure}

To characterize the noise properties of the Hi-GAL DR1 maps, we consider 
all the tiles in each band, locating and analyzing those map regions where the lowest 
signal is found.  This is done by computing the pixel brightness 
distribution and selecting pixels where the brightness is below the 
lowest 10\% percentiles. We subsequently consider, always for each tile 
and each band separately, only those pixels that form connected areas 
with at least 100 pixels each and we therein compute the median of the 
brightness and the mean of its r.m.s.  These quantities are reported in 
Fig. \ref{noisefig} as full and dashed lines, respectively, as function of 
Galactic longitude. 
The figure reports for each band the distribution of the lowest brightness 
levels, and the corresponding r.m.s., found in each tile. The 
coloured ticks on the right margin of the figure represent the 1$\sigma$ 
brightness sensitivities in MJy/sr predicted by the PACS and SPIRE time 
estimator for the Hi-GAL observing strategy, with two independent 
orthogonal scans taken in parallel mode at a scanning speed of 60\asec\,s$^{-1}$ .

Brightness levels are always well above the instrument sensitivities, 
showing that, even in the faintest regions mapped by Hi-GAL, we are limited 
by cirrus brightness and cirrus noise emission by big grains \citep{dbp90} for $\lambda \geq$ 160\um, 
except perhaps at the outskirts of the Hi-GAL DR1 longitude range where 
the minimum signal r.m.s. is close or equal to the predicted detector 
noise. An exception is the 70-\um\ emission, where the brightness of the diffuse cirrus that dominates at longer wavelengths drops significantly \citep{bernard10}.  The 70-\um\ 
brightness levels reach (or cross) the respective r.m.s. values much 
earlier, moving away from the Galactic Centre, than in the other bands. The 
fact that the most intense emission is reached at 160\um, and then 
decreases toward 500\um\ is in excellent agreement with expectations 
for diffuse, optically thin cirrus dust at temperatures 16\,K $\lesssim 
T \lesssim$20\,K, as determined by \cite{paradis10} from detailed modelling 
of \higal\ data in selected regions of the Galactic Plane.

It should be noted that the Hi-GAL ROMAGAL pipeline used for DR1 is successfully delivering the PACS and SPIRE predicted sensitivities with the very bright and complex ISM emission on the Galactic Plane, while preserving in the data processing chain the signal at all spatial scales with no spatial scale filtering.

\subsection{Astrometric corrections}
\label{astrometry}

Although the map-making algorithm was run for each tile using the same projection centre for all bands, the PACS and SPIRE maps are slightly mis-aligned, possibly due to a residual uncalibrated effect in the basic astrometric calibration that is carried out in the HIPE environment. Excellent map alignment is essential to generate products such as column-density maps (e.g. \citealt{Elia+2013}) or to positionally match source counterparts at different wavelengths.

As the images obtained with the same instrument (PACS or SPIRE) are 
internally aligned, we initially align the PACS 70\um\ images to match the 
astrometry of the {\em Spitzer}/MIPSGAL images at 24\um. This has the 
advantage that the two instrument/wavelength combinations deliver the same 
spatial resolution. The astrometric accuracy of the MIPSGAL images with 
respect to higher resolution IRAC and 2MASS is better than $\sim$1\arcsec\ on average \citep{Carey+2009}.

\begin{figure}[h]
\includegraphics[width=0.5\textwidth]{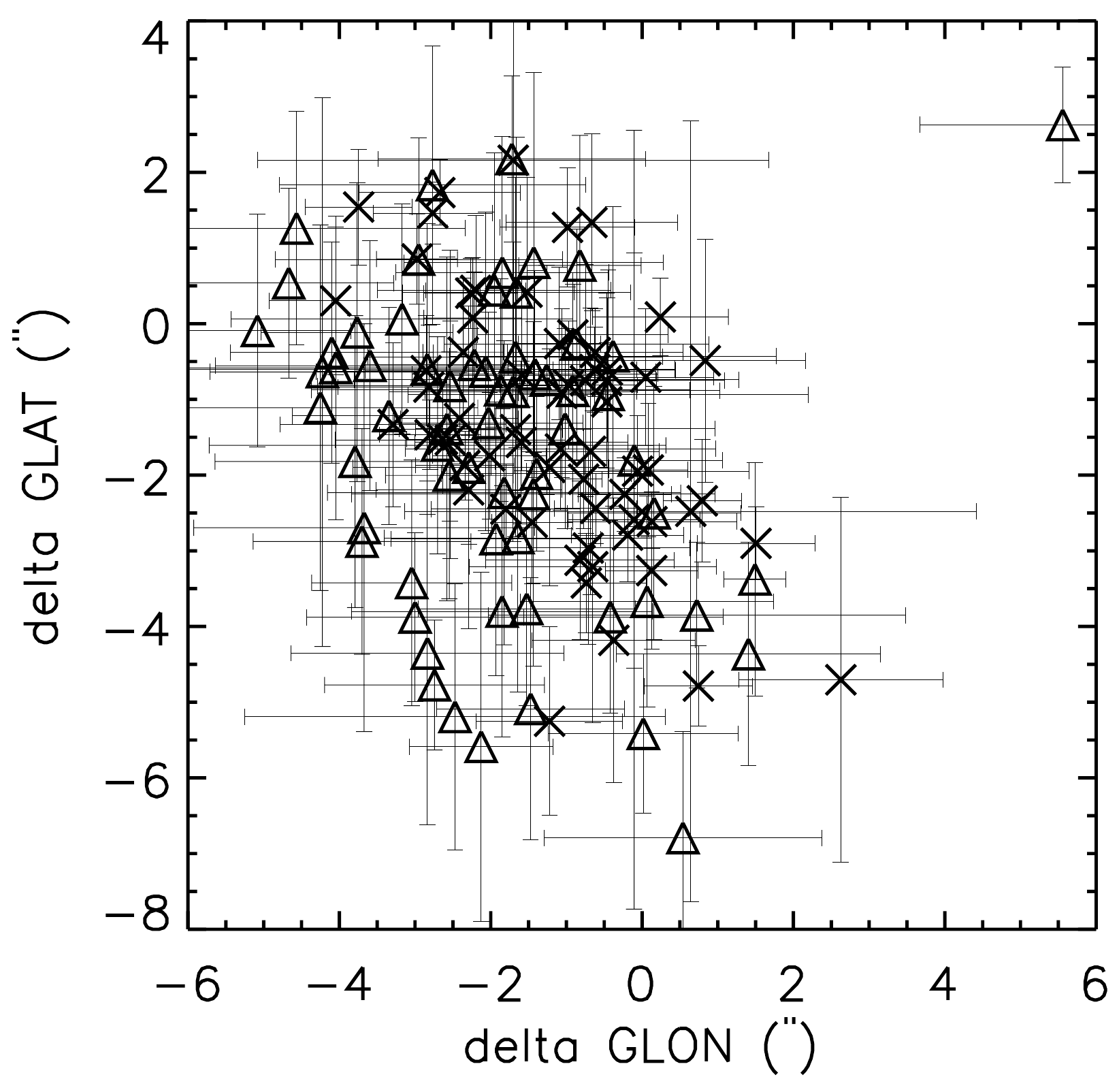}
\includegraphics[width=0.5\textwidth]{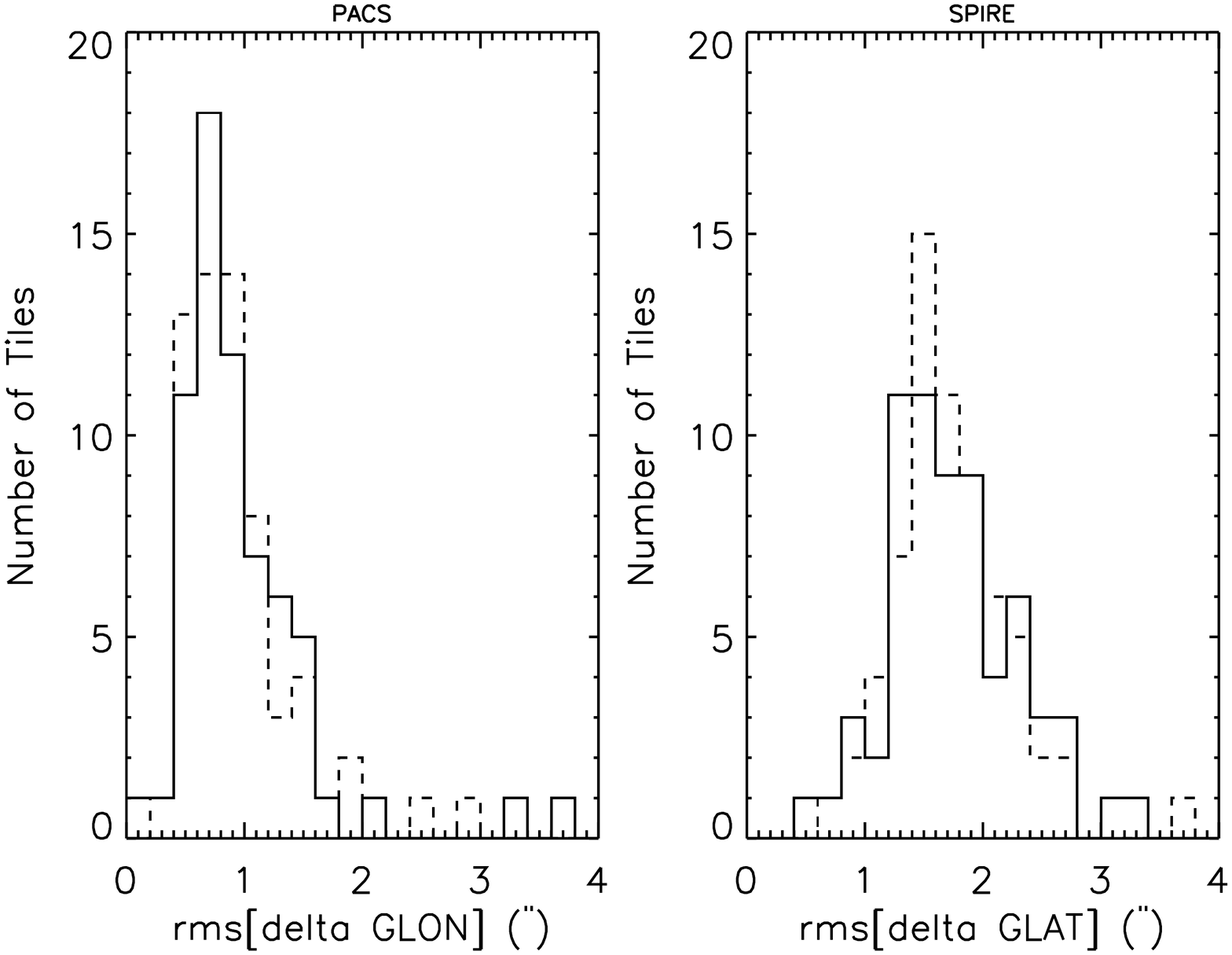}
\caption{\textit{Top Panel:}Astrometry shifts in Galactic longitude ($x$ axis) and latitude ($y$ axis), estimated for each tile in arcseconds. Crosses are for PACS tiles while triangles are for SPIRE tiles. The error bars represent the r.m.s. of the source coordinates used to estimate the offset corrections with respect to their respective mean value. \textit{Bottom Panels:} Histograms of the r.m.s. of the longitude (full lines) and latitude (dashed lines) shifts estimated for PACS (left panel) and SPIRE (right panel). }
\label{astro_plot}
\end{figure}

\emph{For each tile} 
For each tile, we visually select a number of sources across the 
maps (typically more than 6) that appear relatively isolated and compact 
both at 24 and 70\um. The implicit assumption is that the two counterparts 
are the same physical source; This is reasonable as long as we avoid 
selecting sources in relatively crowded star forming regions where sources 
in different evolutionary stages (and hence intrinsically different SED 
shapes) are generally found. We extract the selected sources in both 
images and we determine an average $\overline{[\delta l, \delta b]}$ shift 
to minimize the offsets between the positions of the selected sources in 
the 24-\um\ and 70-\um\ maps. This mean shift correction is then applied to 
the astrometric keywords in the FITS headers of the PACS maps.

The SPIRE maps were aligned by bootstrapping from the aligned PACS images. For 
each tile we selected a number of sources that appear compact and isolated 
both in PACS 160\um\ and SPIRE 250\um. In a similar way to the 
alignment of the 70-\um\ PACS images, we extract the selected sources in 
both maps and compare the source positions in the two bands to determine 
an average shift that minimizes the positional differences. This average 
shift is then applied to correct the astrometric keywords in the FITS 
headers of all SPIRE maps.

The corrections estimated for each tile are shown in Fig. \ref{astro_plot} 
for PACS (cross signs) and SPIRE (triangles) images, taking the {\em 
Spitzer}/MIPSGAL images as a reference. Corrections can be as large as 
6\asec\ in absolute terms, meaning they are particularly significant for 
the PACS 70-\um\ band where they can reach about 2/3 of the image 
reconstructed FWHM beamwidth. 
The outlier point at the top-right of the plot corresponds to the tile centred 
at $\ell$=299\adeg, which was taken during the {\em Herschel} Performance 
Verification Phase. The {\em Herschel} astrometric accuracy evolved throughout the mission, as sources of errors in the star trackers and in general in the pointing reconstruction have been isolated and recovered. One of the major issue up to OD 320 was the ``speed bumps'' that caused large variations in the scanning speed of the telescope. These bumps happened when a tracking star passed over bad pixels of the optical telecope's CCD. This effect was corrected by lowering the operational temperature of the tracking telescopes. In general, the astrometric accuracy up to OD 320 was better than 2 arcsec but outliers at more than 8 arcsec were observed (for a detailed report on the Herschel astrometric accuracy see \citealt{SP+2014}). 

The error bars in Fig. \ref{astro_plot} represent the r.m.s. of the source 
coordinates used to estimate the offset corrections with respect to their 
mean value. The distribution of these values is reported in the lower panels of Fig. \ref{astro_plot}; they are centred around the median values [$\Delta$GLON, $\Delta$GLAT]=[0\asec\!.9, 0\asec\!.8] for the PACS images (lower-left panel of fig. \ref{astro_plot}), and [1\asec\!.7, 1\asec\!.6] for the SPIRE images (lower-right panel), and may be assumed as an estimate of the typical residual uncertainty of the source coordinates. These amount to $\sim$ 10\%\ of the PSF FWHM as estimated from compact sources in the images. It is interesting to note that there are a few outliers in the distributions, particularly apparent for the PACS shifts, but even for their maximum values they are below half of the PACS beam at 70\um.
As mentioned at the beginning of the section, an additional 
average 1\asec\ uncertainty should be added in quadrature to account for 
the MIPSGAL pointing accuracy.

\subsection{Map photometric offset calibration}

Although the PACS and SPIRE images are calibrated internally in Jy/pixel 
and Jy/beam, respectively, their zero point level is not. So in order to 
bring the images to a common calibrated zero level an offset was applied 
to the maps. The photometric offsets of the \higal\ maps were determined 
through a comparison between the \higal\ data and the {\em Planck} and 
IRIS (Improved Reprocessing of the IRAS Survey) all-sky maps, following 
the procedure described in \cite{bernard10}. We 
smoothed the {\em Herschel} maps to the common resolution of the IRIS and
{\em Planck} high frequency maps of 5\arcmin\ and projected them into the 
HEALPix pixelisation scheme \citep{Gorski+2005} following the drizzling 
procedure described in \cite{Paradis+2012}, which preserves the 
photometric accuracy of the input maps. These smoothed \higal\ maps are 
compared with the IRIS and {\em Planck} all-sky maps (hereafter called 
"model").

To make this model, we used the IRIS maps projected into HEALPix taken from the CADE web site 
(http://cade.irap.omp.eu) and the {\em Planck} maps shown in 
\cite{planck+2011}. Since the {\em Herschel}, {\em Planck} and {\em IRAS} 
photometric channels are different, the comparison requires frequency 
interpolation with differential colour correction, and the use 
of a model. We predict the shape of the emission spectrum in each pixel using 
the DustEM\footnote{See http://dustemwrap.irap.omp.eu/ and 
http://www.ias.u-psud.fr/DUSTEM/ } code \citep{comp11}, computed for an 
intensity of the radiation field best matching the dust temperature, 
derived from the combination of the IRIS 100-\um\ and the {\em Planck} 857-GHz 
and 353-GHz maps. The dust temperature assumed is that of \cite{planck+2011} with the standard dust distribution of \cite{comp11}. For a given PACS or SPIRE band, the model 
is normalized to the data at the IRAS or {\em Planck} band at the nearest 
frequency to the considered \textit{Herschel} band, and a predicted 5\arcmin\ 
resolution model image is constructed.These nearest frequencies are the IRAS 60-\um\ and {\em Planck} 857-GHz bands for the PACS 70-\um\ and 160-\um\ bands, respectively, and the {\em Planck} 857-GHz, 857-GHz and 545-GHz 
bands for the SPIRE 250-\um\, 350-\um\ and 500-\um\ bands, respectively. In this 
process, the differential colour correction between IRAS or {\em 
Planck} and the {\em Herschel} band under consideration is also taken into account, 
using the spectral shape predicted by the model on a pixel-by-pixel basis.

This resulting model image is compared with the smoothed \higal\ data 
through a linear correlation analysis, the intercept of which provides the 
offset level to be added to the {\em Herschel} data to best match the IRIS 
and {\em Planck} data. This analysis also provides gain corrections (i.e. 
a slope of unity between the data and the model); however, these are well below 
the cumulative relative uncertainties in the datasets used, as well as in the 
dust modelling assumptions, and within 10\%, on average, in all bands. 
The standard {\em Herschel} photometric 
calibration was therefore assumed, and no additional gain corrections were 
applied. Note also that the {\em Planck} data used does not have the same 
absolute calibration as the publicly available version. A forthcoming 
processing of the \higal\ data will use the latest {\em Planck} 
calibration and will allow for a global gain correction. 

\section{Generation of Photometric Catalogues from Hi-GAL maps}
\label{catalogues}

In comparison to the ground-based submillimetre-continuum surveys, the {\it 
Herschel} instruments do not suffer from the need to correct for varying atmospheric emission and absorption, allowing recovery of the rich and highly structured large-scale emission 
from Galactic cirrus and extended clouds.  Such variable and complex 
backgrounds, however, severely hinder the use of traditional methods to detect 
compact sources based on the thresholding of the intensity image.  Such methods are widely 
used by large-scale millimetre and radio surveys from 
ground-based facilities, like the Bolocam GPS \citep{Rosolowsky2010}, 
CORNISH \citep{Purcell2013} or ATLASGAL \citep{Contreras2013}, where 
diffuse emission is filtered out either by atmospheric variation correction or 
the instrumental transfer function. The possibility of processing {\em Herschel} images using high-pass filtering was discarded for various reasons. First of all, it would be difficult to choose a threshold in spatial scale.  Dust cores and clumps are compact but, depending on their distance and physical scale, may not be point-like (i.e., unresolved). A spatial filtering scale threshold too close to the PSF will remove power from compact but resolved sources, while a threshold large enough to make sure that no power is removed from scales corresponding to 2-3 times the PSF will prove ineffective to improve source detection in crowded fields. A second reason is that any high-pass spatial filtering will introduce negative lobes with intensities proportional to the brightness of the extended emission, severely hindering the detection of faint sources that fall within those features.

In a previous work, \cite{moli11a} introduced a new method to detect 
sources and extract their fluxes tailored to the case of the complex and 
structured background present in IR/sub-mm observations.  With respect to 
other popular algorithms, the \cutex \footnote{see 
http://herschel.asdc.asi.it/index.php?page=cutex.html} photometry code, 
standing for Curvature Thresholding Extractor, adopts a different design 
philosophy, looking for the pixels in the map with the highest curvature 
by computing the second 
derivative of the map. All the ``clumps'' of pixels above a defined 
threshold are analyzed and the ones larger than a certain area are kept as 
candidate detections. The pixels of the large ``clumps'' are checked to 
determine enhancement of curvature in the case of multiple sources. For 
each detection, an estimate for the size of the source is determined 
by fitting an ellipse to the positions of the minima of the second 
derivative in each of the 8 principal directions. The output fluxes and sizes are 
determined by simultaneously fitting elliptical Gaussian functions plus a 
2$^{nd}$-order 2D surface for the background. All the sources whose detected 
centres are closer than twice the instrumental PSF are fitted together 
to disentangle their fluxes.

The Gaussian fitting is carried out for each source by considering a 
fitting window centred on each source and with a width of 3 times the 
instrumental PSF to make sure to include sufficient space surrounding the 
source for a reliable estimate of the background. This has the drawback 
that the pixels used to constrain the background are numerically 
predominant with respect to the pixels characterising the source; to 
counterbalance this effect, the pixels located within a distance equal to 
the initial guesstimated source size from the source position are given a 
higher weight in the fit.

\subsection{The characterization of the photometric algorithm}
\label{cutex_char}

\begin{figure}[t]
\includegraphics[width=0.5\textwidth]{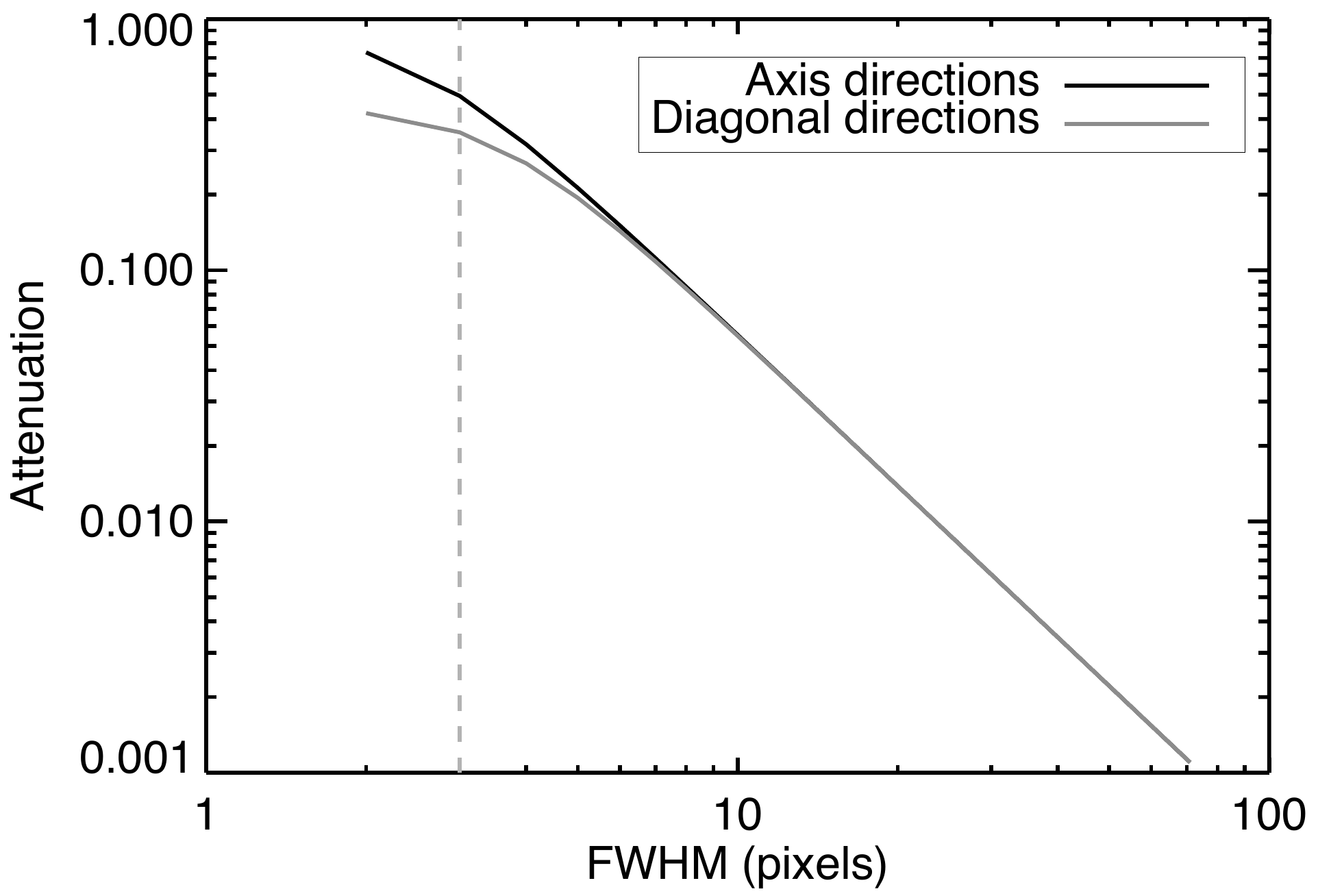}
\caption{Relative attenuation of the peak intensity induced by the 
derivative filtering as a function of the scale of the structure. The 
diagonal directions have been divided by $\sqrt{2}$ to take into account 
the longer distance along the diagonals with respect to the normal axis. 
For scales longer than $\sim$ 6 pixels, the damping increases as a power 
law function with an exponent $-2$. The dark grey dashed line refers to the typical 
size in pixels of the PSF in Hi-GAL maps.}
\label{Attenuation}
\end{figure}

CuTEx, as a derivative-based detection algorithm, acts as a high-pass spatial filter; however, contrary to simple median or boxcar filtering, derivative filtering has inherent multiscale capabilities by selectively filtering out the larger the spatial scales in a continuous way with higher efficiency. Such behaviour is shown in Fig.\,\ref{Attenuation}, where we report, for Gaussians with increasing widths, the ratio between the second derivative image and the original one at the peak position, as a function of the spatial scale expressed in pixels.  The results shown are obtained on a simulated image where the FWHM of the PSF is sampled by three pixels, and therefore is a general result applicable to any map that shares this characteristic, like the {\em Herschel} maps we present here. Fig.\,\ref{Attenuation} shows that the peak intensity of a point-like source, with a FWHM of $\sim$3 pixels (i.e. 1 PSF), is damped in the second derivative image to  $\sim$40\% of its original value, while an extended source with FWHM of $\sim$ 7.5 pixels (i.e. 2.5 $\times$ PSF) and the same peak intensity is damped to $\sim$10\% of the original value. In other words, a point source in the intensity map that is, say, 10 times fainter (contrast 0.1) than the surrounding background, with typical scale of order 15 pixels, i.e.\ 5 $\times$ PSF,  will appear in the derivative map as $\sim$1.7 times brighter than the background (contrast 1.7). Given the trend in Fig.\,\ref{Attenuation}, where attenuation declines following a power-law behaviour with an exponent --2,  it is then possible to detect sources with less favourable contrast the larger is the background typical scale. Clearly, the method has the inherent drawback of being most effective for more compact objects (see below).

\begin{figure}[h]
\includegraphics[width=0.4\textwidth]{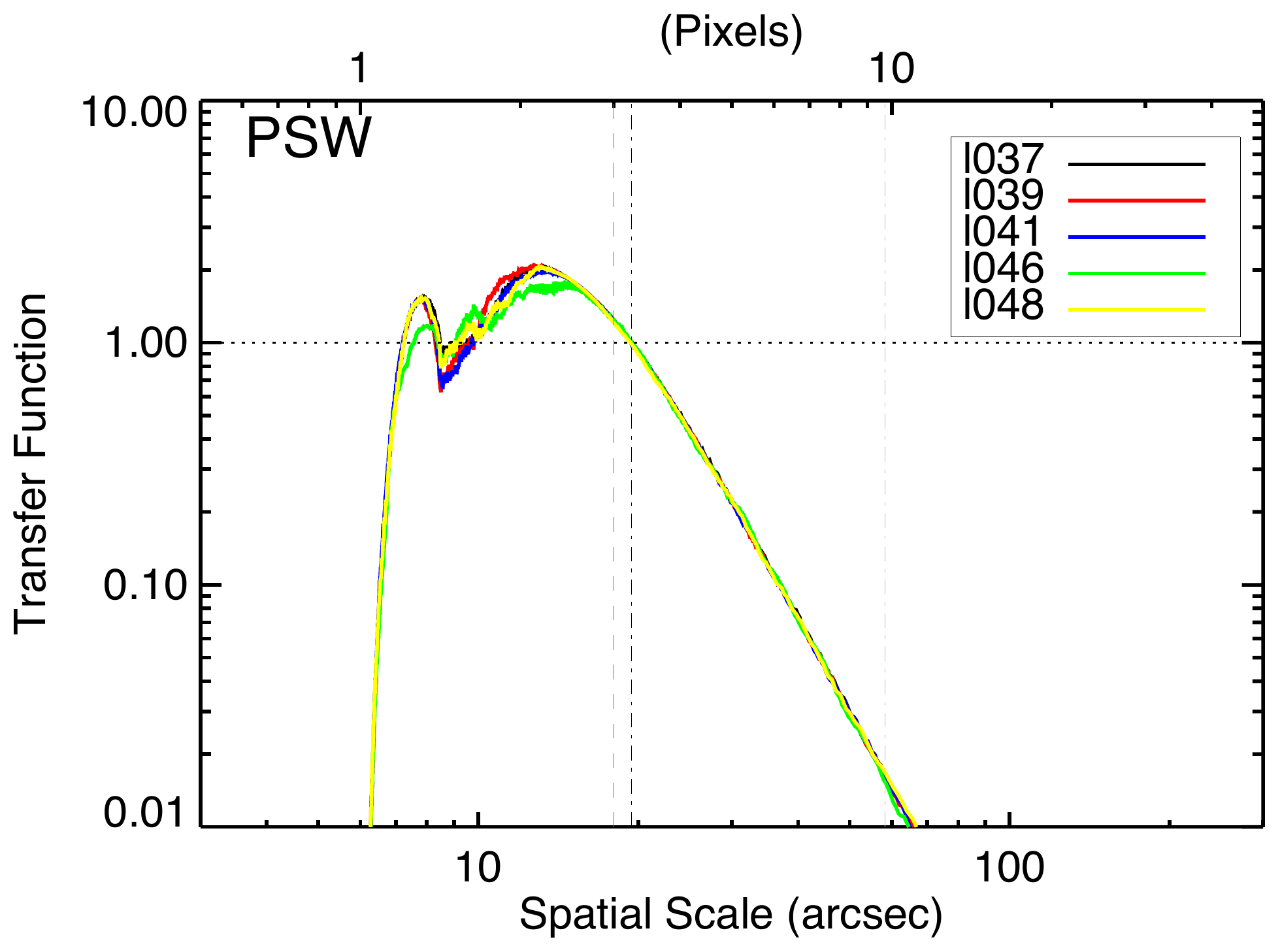}
\caption{Ratio between the power spectrum of the derivative images (averaged over the four directions) computed by CuTEx, and the power spectra of the intensity image for SPIRE 250\um, as a function of spatial scale expressed in arcseconds and in pixels (upper $x$ axis).  Each colour corresponds to a different map; respectively, {\it black l037}, {\it red l039},{\it blue l041}, {\it green l046}, {\it yellow l048} field. All the functions overlap for scales larger than the PSF, indicated as a dark grey dashed line, and decrease following a power-law with an exponent $\sim$ --3.9. The black dot dashed line indicates the scale at which the transfer function is equal to unity. Scales smaller than that result in an overall amplification in the second derivative maps. The light grey dot-dashed line traces the scale above which the extended sources start to get confused with the background in the derivative image. Such value  corresponds approximately to $\sim$ 3 times the PSF.  Similar plots are found for other wavelengths and the functions completely overlap when the spatial scales are expressed in pixels.}
\label{TransferFunction}
\end{figure}

To confirm the performances of CuTEx's derivative operator in the case of real maps, we computed the power spectrum of the second derivative image for each map, averaging the spectra obtained for each derivative direction. We then divided each derivative power spectrum by the power spectrum of the parent intensity image. These ratios are proportional to the module square of the transfer function of the derivative operator used by CuTEx. Fig.\,\ref{TransferFunction} shows these ratios for 5 different maps (indicated with different colours in the figure) in the case of  250-\um\ observations. Similar plots are found  for the other wavelengths, where the only difference is a shift in angular spatial scale due to the different pixel scales.  The scale in the upper $x$ axis is in pixels and insensitive to the specific pixel angular scale. 

Several conclusions can be drawn from the analysis of these functions. First, the transfer function is the same, regardless of the mapped region, for scales larger than the PSF. Second, the damping introduced by  the derivative operator found in Fig.\,\ref{Attenuation} is confirmed also for real maps. From an investigation of a sample of very extended sources in the Hi-GAL maps, we estimated that CuTEx is not able to recover most of the sources with sizes larger than 3 times the PSF (see also Fig. \ref{sourcesize}), being completely insensitive to any source larger than $\sim$5 times the PSF. The third conclusion resulting from Fig.\,\ref{TransferFunction} is  that the derivative filtering introduces an amplification for scales smaller than the PSF. This means that any pixel-to-pixel noise present in the intensity map is increased in the second derivative maps. Slight differences between the different tested fields are only visible at scales below the PSF (the dashed line in the figure) but are not relevant for the detection of real sources. To quantify such an increase, we tested the effect of the derivative operator on pure Gaussian noise maps and
found  that the noise in the second derivative follows the same distribution with a standard deviation 1.13 times the initial one. Such behaviour is not unexpected, due to the linearity properties of the derivative filtering.  

\subsection{Choice of the extraction threshold}
\label{choice_th}

In similar way to source extraction performed on images of surface brightness 
distribution, it is useful to set an extraction threshold as a function 
of the local curvature r.m.s. instead of adopting a constant absolute 
value. In this way, the depth of the extraction is adapted to the 
complexity of the morphological properties and to the intensity of the 
background that constitutes the dominant flux contribution in the far 
infrared toward the GP.

Although the adoption of a detection threshold in the second derivative
image is certainly less intuitive than adopting a threshold on the flux 
brightness map, we have shown above that the noise statistical properties 
do not change when going from flux maps to flux curvature maps (except for 
a small increase in the width of the noise distribution), so that the 
notion of a threshold that adapts to the local noise properties can be 
applied also to detection on the curvature images.

The choice of an optimal source extraction threshold always results from a compromise between the need to extract the faintest real sources, and the need to minimize the number of false detections. Pushing the detection threshold to lower and lower values to extract fainter and fainter sources is of course of minimal use if the majority of such faint extracted sources have a high probability of being false positives, therefore considerably limiting the catalogue completeness and reliability. Unfortunately there is no exact way to control the number of false positives extracted from real images, as there is no control list for real sources present, so that a number of \textit{a posteriori} checks are needed to determine this optimal threshold value.

The procedure we adopted to estimate the optimal extraction threshold is to make extensive synthetic source experiments 
to characterise the flux completeness levels obtained for different 
\cutex\ extraction thresholds $\sigma _c$ in all five Hi-GAL photometric 
bands, where $\sigma _c$ is in units of the r.m.s. of the local values of the second derivatives of the image brightness averaged over 4 directions (see \citealt{moli11a}). 

As it is clearly impractical to make these studies over the entire 
set of Hi-GAL tiles, we chose three tiles at Galactic longitudes of 
19, 30 and 59 degrees that are representative of the widely variable 
fore/background conditions that can be found over the entire survey.
For each of these tiles and for each observed 
band, hundreds of synthetic sources were injected at different flux 
levels. We then ran \cutex\ for a set of extraction thresholds $\sigma _c$ 
from 3 to 0.5, estimating for each threshold the flux for which 90\% of 
the synthetic sources were successfully recovered. We verified that, for 
each of the three tiles, the 90\% completeness fluxes decrease with 
decreasing extraction threshold. In the case of the three SPIRE bands, we 
see that this decrease flattens, starting at $\sigma _c \sim 2$, 
meaning that we do not gain in depth of extraction by going to lower 
thresholds. We emphasize that our artificial source experiments provide 
the same optimal value for the extraction threshold independently of the 
tile used, in spite of the very different properties of the diffuse and 
structured background exhibited by the Hi-GAL images in the longitude 
range covered in DR1. This is a convenient feature of the detection 
method, that is clearly able to deliver similar performances with very 
similar parameters in widely different fields. We then adopt $\sigma _c = 
2$ as the extraction threshold for the SPIRE bands.

\begin{figure}[t]
\includegraphics[width=0.5\textwidth]{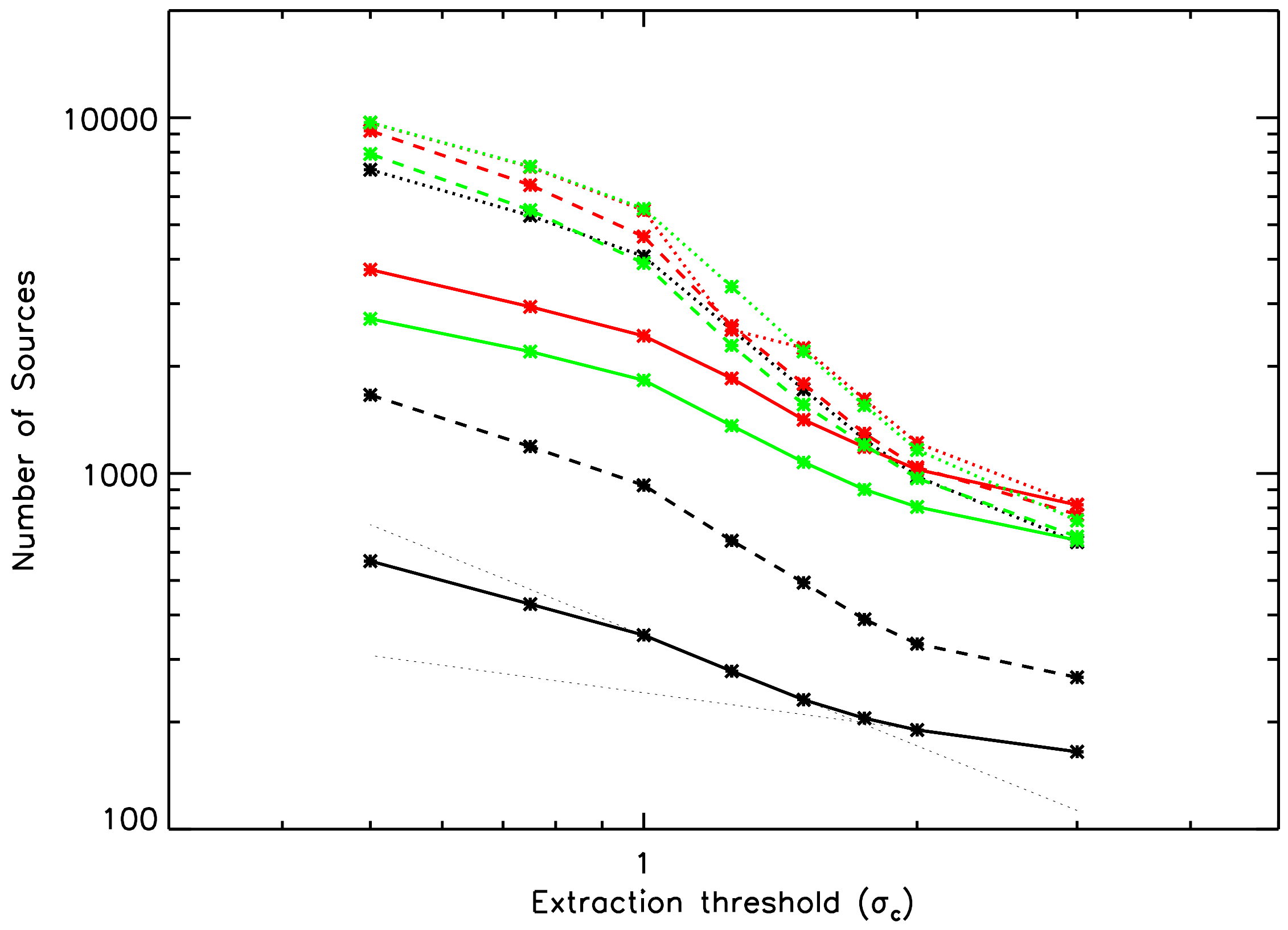}
\includegraphics[width=0.5\textwidth]{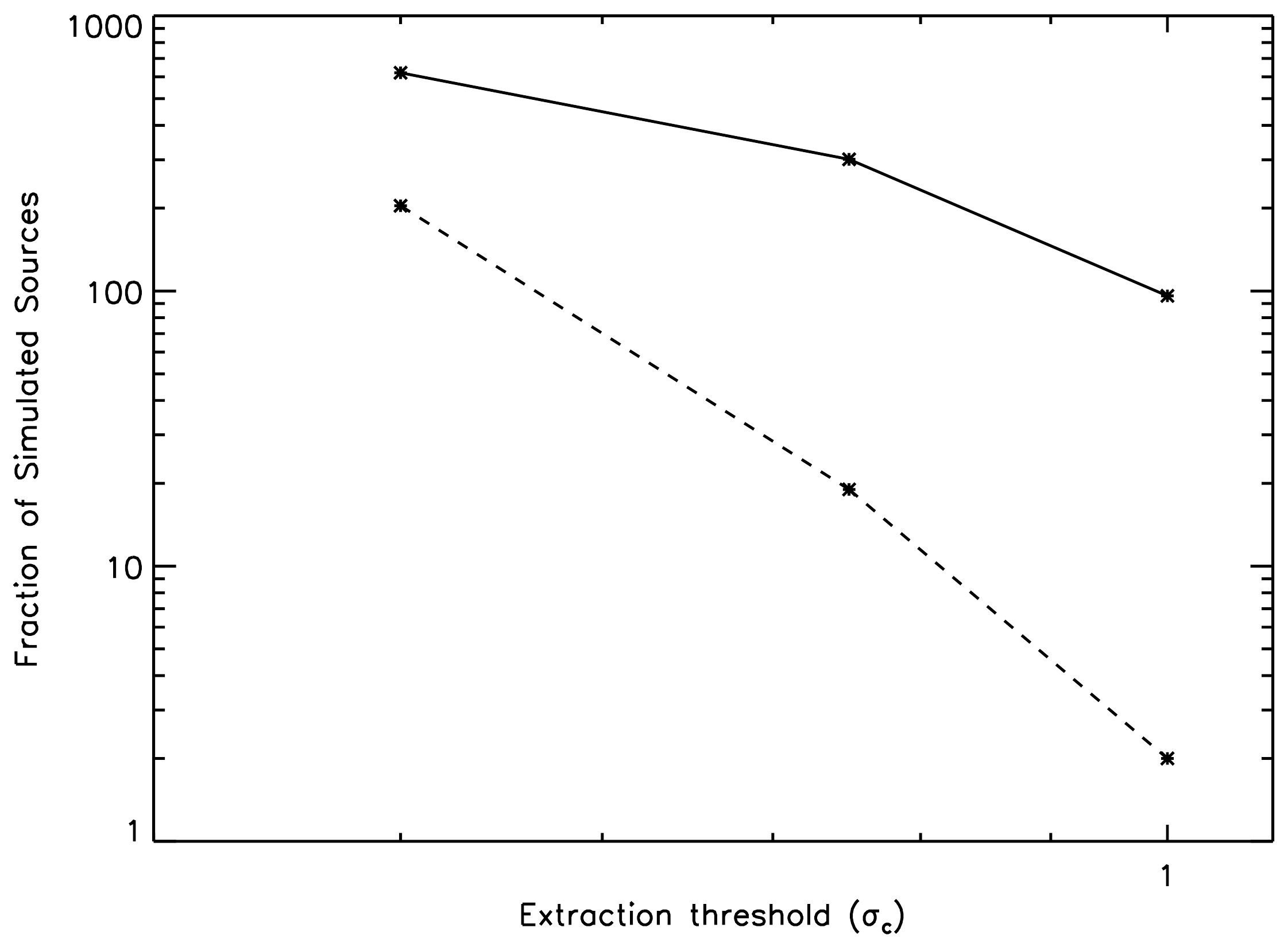}
\caption{\textit{Top Panel:} Number of sources extracted with \cutex\ as a function of extraction threshold for the 70-\um\ (thick solid lines), 160-\um\ (thick dashed lines) and 250-\um\ bands (dotted thick lines), for three Hi-GAL tiles with very different background conditions: $\ell$=19\adeg\ (green lines), $\ell$=30\adeg\ (red lines) and $\ell$=59\adeg\ (black lines). The thin dotted lines are power-law fits to the initial and mid portions of the N-$\sigma _c$ relationship at 70\um\ for the $\ell$=59\adeg\ tile, and are shown to emphasize the change in slope that is visible for all functions for $\sigma _c \lesssim 2$. \textit{Bottom Panel:} Detection statistics for the simulated source experiments reported in Fig. 7 of \cite{moli11a} for a flux of 0.1\,Jy; the total number of simulated sources is 1000. The full line reports the number of true sources recovered, while the dashed line reports the number of false positives as a function of extraction threshold. It is noticeable how the number of false positives increases faster for decreasing thresholds than the number of true sources detected, qualitatively explaining the change of slope in the real fields detections (top panel).}
\label{falsepos}
\end{figure}

In the case of the PACS 70-\um\ and 160-\um\ bands, the decrease of the 90\% 
completeness fluxes continues below $\sigma _c$=2. This apparent gain in 
the number of reliable sources detected at lower and lower thresholds is 
likely to be due to increasing numbers of false-positive detections. 
We characterize the impact of false positives by evaluating the number of extracted sources in the different bands as a function of the extraction threshold. Fig. \ref{falsepos}-top reports the number of 
sources detected in the three tiles (indicated by the different colours) at 
70, 160 and 250\um\ (solid, dashed and dotted lines) as a function of the 
extraction threshold. The figure shows that in all cases the N$-\sigma _c$ 
relationships tend to get steeper below $\sigma _c \sim 2$; we emphasize 
this in Fig. \ref{falsepos} in one case by fitting two power-laws to two 
portions of the N-$\sigma _c$ for the 70-\um\ case of $\ell$=59\adeg\ (the two 
thin dotted lines). A similar behaviour is exhibited for all the other 
cases, and we interpret this increase of rate in detected sources for 
$\sigma _c \leq 2$ as an indication of increased contamination of false 
detections. It is, strictly speaking, impossible to verify this claim on real images, because we do 
not have a truth table for the sources that are effectively present. We then make use of a subset of the extensive simulations that we did in \cite{moli11a} where we presented and characterized the CuTEx package; the bottom panel of Fig. \ref{falsepos} reports the number of true detected sources (full line) and the number of false positives (dashed line) as a function of the extraction threshold for a simulation of 1000 synthetic sources (that were reported in the top-left panel of Fig. 7 in \citealt{moli11a}). It can indeed be seen that for decreasing extraction thresholds, the number of false-positive detections increases faster than the number of real sources. It is irrelevant here to compare the absolute values of the slopes between the real and simulated cases in Fig. \ref{falsepos}, nor the thresholds where the false positives may become dominant, because the two cases refer to very different situations (see \cite{moli11a} for more informations on the simulations carried out). What is important here is that the faster increase of false positives with respect to real sources as a function of decreasing threshold may qualitatively explain the change of slopes in the detection rates with thresholds that we see in the real fields in the top panel of Fig. \ref{falsepos}.

In order to be conservative for this first catalogue release, we choose to 
adopt an extraction threshold of $\sigma _c$=2 also for the 70\um\ and 
160-\um\ PACS bands. We believe that the detection threshold could be 
pushed to lower values especially in the PACS bands and toward low 
absolute Galactic longitudes; this requires more extensive studies of the 
completeness level analysis and characterizations of the real impact of 
false positives contamination and will be deferred to the 
release of subsequent photometric catalogues.

\subsection{Generation of the source catalogues}
\label{generation}

Sources were extracted independently for each Hi-GAL tile and for each band 
using \cutex\ with extraction threshold $\sigma _c =2$. As 
each map tile results from the combination of two observations of the same area 
scanned in nearly orthogonal directions, and since the area scanned in 
the two different directions is never exactly the same, the 
marginal areas of the combined maps will generally be covered only in one direction, 
resulting in very poor quality compared to the majority of the map area. 
For this reason, we exclude such areas
from the source extraction.  The selection of the optimal map regions 
is performed manually for each tile and separately for the PACS and SPIRE images. 
These regions will always be at the margins of the tiles but this does not result in gaps in longitude coverage, since the contiguous border region of any tile will be optimally covered by the adjacent tile.

The full source extraction was carried out on an IBM BladeH cluster with 7 
blades, each equipped with Intel Xeon Dual QuadCores, for a total of 56 
processors. Each independent tile and band extraction job was dynamically 
queued to each processor, allowing us to complete the extraction from 63 
2\adeg $\times$ 2\adeg\ tiles in five bands in one day. The different photometry 
lists for each band were then merged together to create complete 
single-band source catalogues. As there is always a small overlap between 
adjacent Hi-GAL tiles, some sources may be detected in two tiles. 
In this case, where source positions match within one half of the instrumental beam, 
the detection with the higher signal-to-noise (SNR) ratio was accepted into the source catalogue. The number of compact sources extracted over 
the longitude range considered in this release are reported in Table 
\ref{catnum}.

\begin{table}[h]
\caption{Source numbers in the Hi-GAL photometric catalogues}
\begin{center}
\begin{tabular}{lr} \\ \hline\hline
Band & N$_{sources}$ \\ \hline
PACS-70\um & 123,210 \\
PACS-160\um & 308,509 \\
SPIRE-250\um & 280,685 \\ 
SPIRE-350\um & 160,972 \\
SPIRE-500\um & 85,460 \\ \hline\hline
\end{tabular}
\end{center}
\label{catnum}
\end{table}

The CuTEx algorithm detects sources by thresholding on the values of the curvature of the image brightness spatial distribution, and as such is optimised to detect compact objects that may be more extended than the instrumental beam. The analysis reported in section \ref{cutex_char} shows that the 2$^{nd}$-order derivative processing ensures differential enhancement of smaller spatial scales with respect to larger scales also above the instrumental PSF. In section \ref{par_size} below, we verify that the majority of extracted sources have sizes that span the range between 1 and 3 times the instrumental PSF, with most of the objects below 2-2.5 times the beam (see Fig. \ref{sourcesize}) and axis ratio below 2 (see Fig. \ref{sourceellipse}).  In the rest of the paper we will refer to the Compact Source Catalogues, to signify that the catalogues include relatively round objects with sizes generally below 2-2.5 times the beam.

The catalogues contain basic information about the detection and the flux estimation for all sources, including source position, peak and integrated fluxes, estimated source size and uncertainty computed as the brightness residuals after subtraction of the fitted source+background model. The calibration accuracy of the PACS photometer is of the order of 5\% in all bands \citep{Balog+2014}, due to the uncertainties in the theoretical models of the SED of the stars used as calibrators. For SPIRE the main calibrator is Neptune and, as for PACS, the main uncertainty comes from the theoretical model of the planet emission and it is estimated at 4\% in all the bands \citep{Bendo+2013}.

Hi-GAL photometric catalogues are ASCII files in ``IPAC Table'' format, and contain information on source position, peak and integrated fluxes, source sizes, locally estimated noise and background levels,  and a number of flags to signal specific conditions found during the extraction. The full list of the 60 table columns, with explanation of the column contents, can be found in Appendix \ref{cat_expl}; given the number of columns, it is not possible to show a preview of the catalogue tables in a printed form. The DR1 single-band photometric catalogues are delivered to ESA for release through the \textit{Herschel} Science Archive, and are available via a dedicated image cutout and catalogue retrieval service accessible from the VIALACTEA project portal \url{http://vialactea.iaps.inaf.it}.

\subsection{Catalogue Flux Completeness}
\label{completeness}

To quantify the degree of completeness of the extracted source lists we 
carried out an extensive set of artificial source experiments by injecting 
simulated sources into real {\it Hi-GAL} maps. Given the very 
time-consuming nature of these experiments, we chose to carry them out for 
each band but only for a subset of the entire range of longitudes
that is the subject of the present release.  We visually selected one from every 2-3 
tiles, depending on the variation of the emission seen in the maps 
as a function of Galactic longitude. We used a similar methodology as in 
\S\ref{choice_th} for the determination of the optimal extraction 
threshold, but this time we use only one detection threshold and an 
adaptive grid of trial fluxes for the synthetic sources.

For each band of this subsample, we injected 1000 sources modelled as 
elliptical Gaussians of constant integrated flux, with sizes and axis 
ratios equal to the majority of the compact sources determined from the 
initially extracted list (see Figs. \ref{sourcesize} and 
\ref{sourceellipse}). In this way, we are able to test the ability to
recover a statistically comparable population of sources from the same 
map. The sources are randomly spread on the map, with the only constraint 
being to avoid overlap with the positions of the real sources.

\begin{figure}[t]
\includegraphics[width=0.5\textwidth]{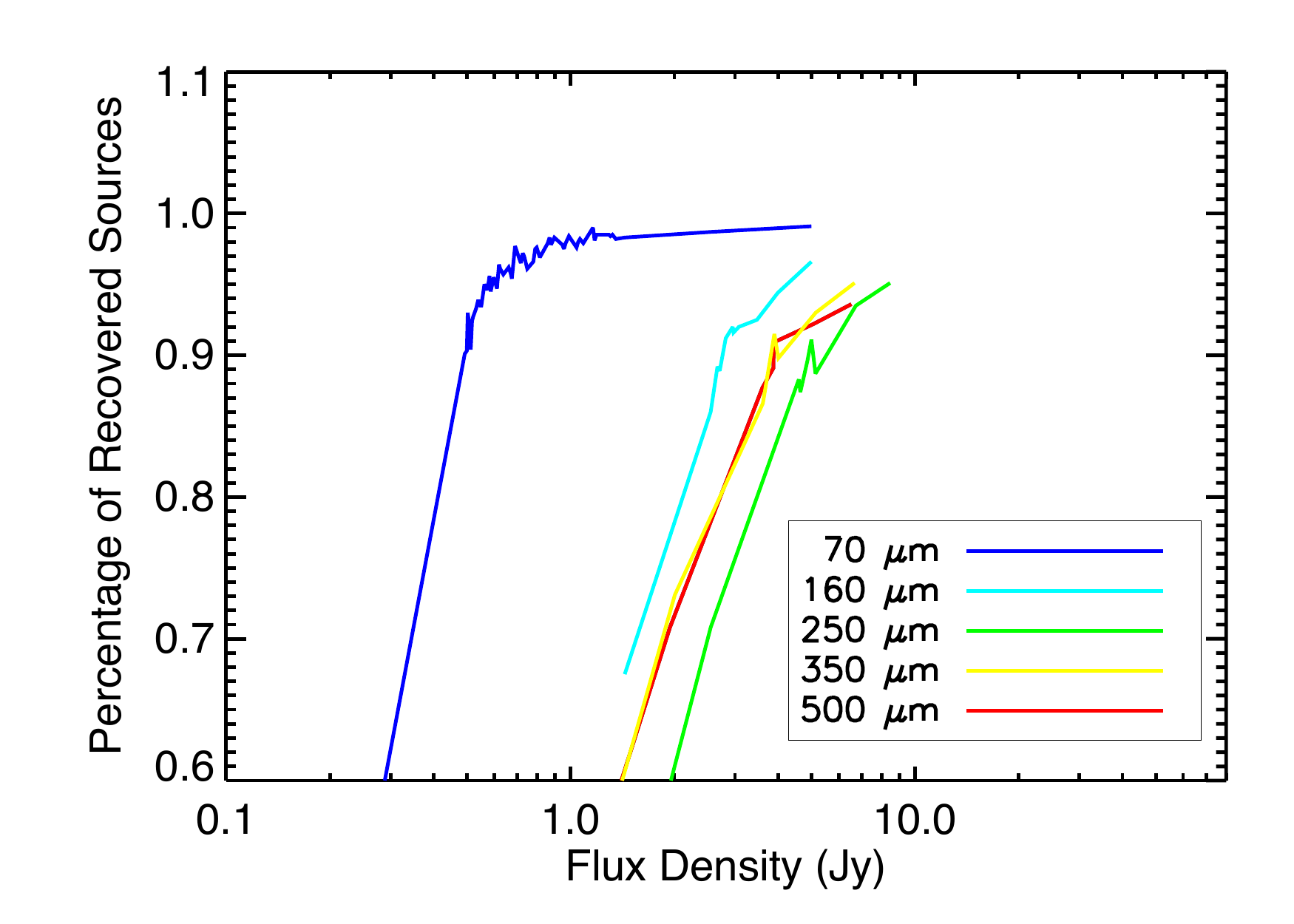}
\caption{Completeness fractions as a function of flux density for the map 
centred at $(\ell,b)$ = (19,0), a field with a very intense and complex 
background at the boundary of the Central Molecular Zone, in the different 
{\em Herschel} bands for sources 
with statistically the same sizes as those in the extracted catalogue.}
\label{CompleteOne}
\end{figure}

The simulated data are processed with \cutex\ , adopting the same setup of 
parameters used for the initial list and the outputs are compared with the 
truth table of the injected sources. To have an estimate on the errors we 
iterated the experiment 10 times and determined how the fraction of 
recovered sources varies. The same process is iterated for different 
values of integrated flux until the fraction of recovered sources is 90\%\ 
(with a tolerance of 1\%). An example of the recovery fraction as a 
function of the integrated flux density of the injected sources is given 
in Fig.\,\ref{CompleteOne}.

\begin{figure}[t]
\includegraphics[width=0.5\textwidth]{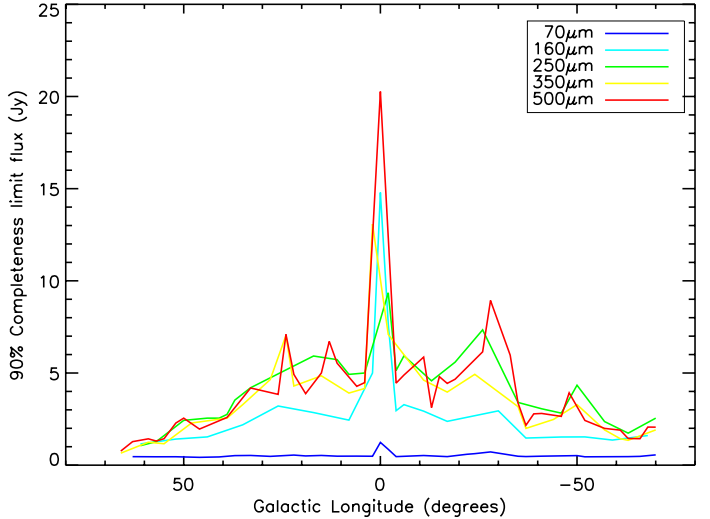}
\caption{90\% completeness limits in flux density for a population of 
sources with the same distribution of sizes as the one extracted by 
\cutex\, as a function of Galactic longitude. The significant increase in 
the completeness limit in the inner Galaxy and especially close to the Galactic Centre is 
due to the brighter background emission in such regions.}
\label{CompleteGal}
\end{figure}

In Fig. \ref{CompleteGal}, we show the estimated completeness limit as a
function of Galactic longitude. The 
limits for the PACS 70-\um\ and 160-\um\ bands are quite regular along the whole 
range of longitude. However, while the completeness in the 70-\um\ band is 
almost constant, at 160\um\ it is higher for $\abs{\ell} \le$ 40. Such a 
behaviour is more significant in the SPIRE wavebands and increases while 
moving toward the Galactic Centre. It is explained by the overall 
brighter emission at lower longitudes, making the detection 
of fainter objects a harder task, even with the strong damping induced by 
\cutex.

\begin{figure}[t]
\centering
\includegraphics[width=0.5\textwidth]{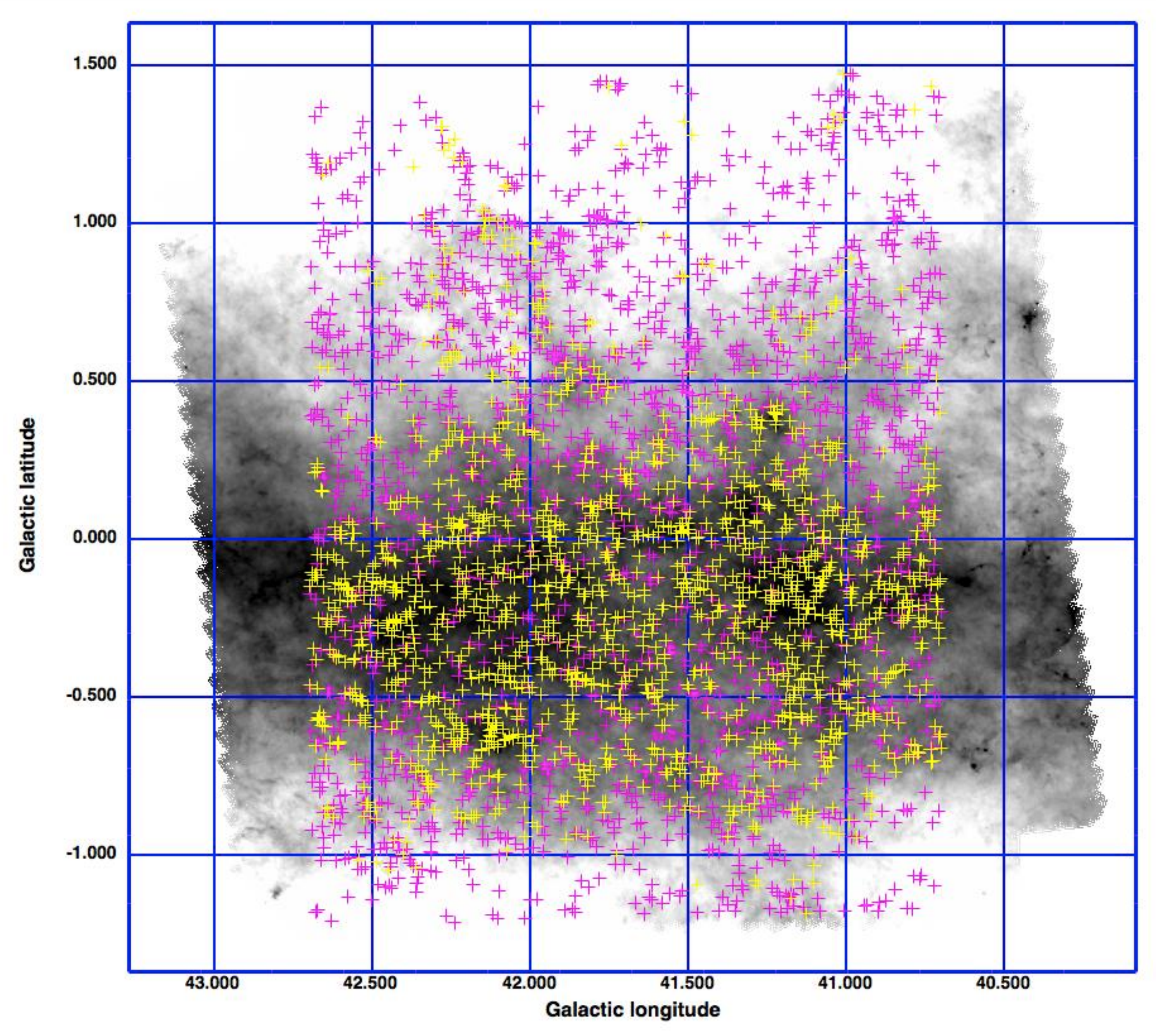}  
\includegraphics[width=0.5\textwidth]{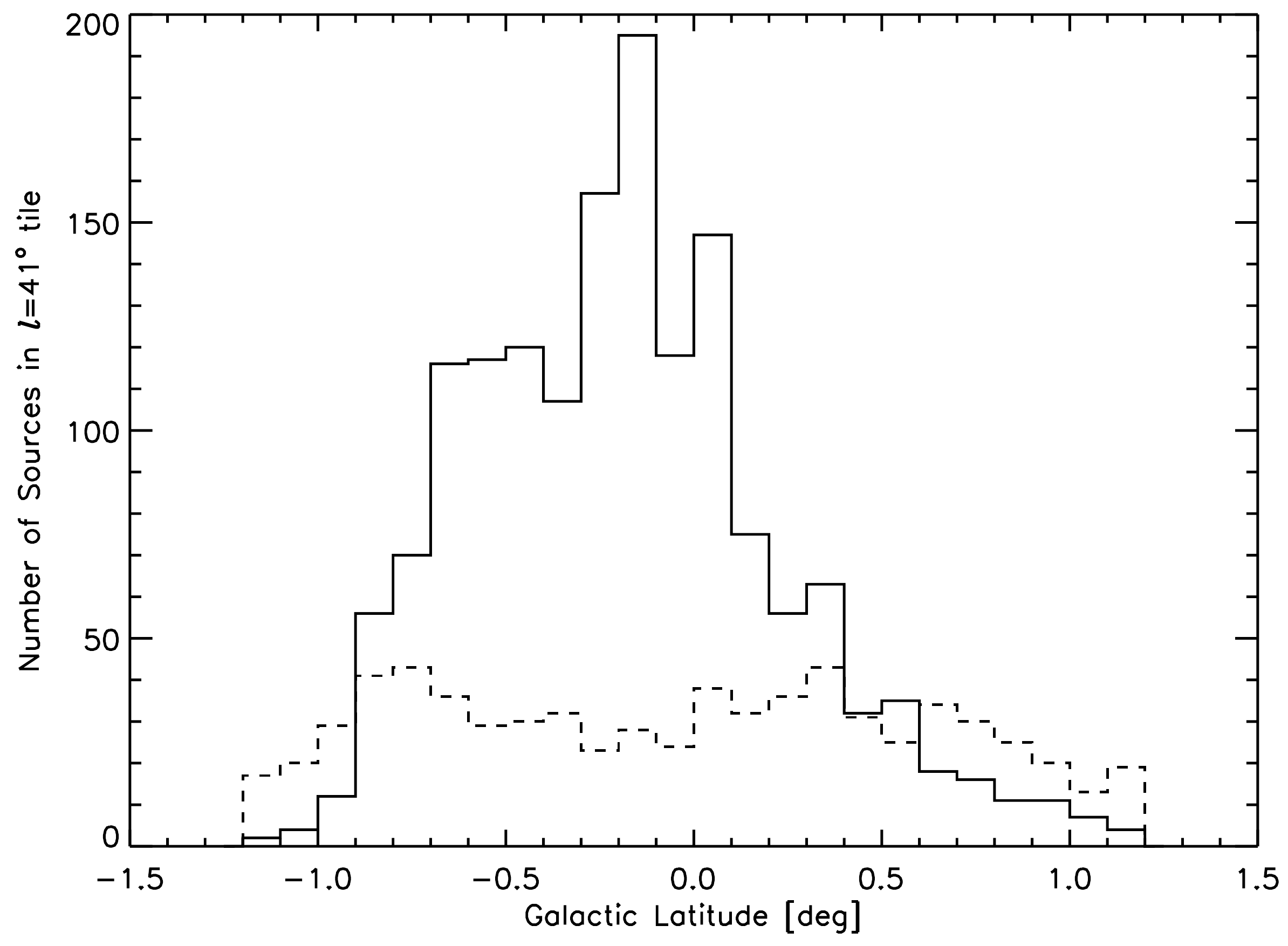}
\caption{\textit{Upper panel:} 250-\um\ image of the Hi-GAL tile at 
$\ell$=41\adeg\ with superimposed the sources detected with \cutex. The 
yellow crosses indicate the sources with fluxes above the completeness 
limit, while the magenta crosses indicate the sources with fluxes below 
the completeness limit. \textit{Lower panel:} histograms of latitude 
distributions for 250-\um\ sources, above (full line) and below (dashed 
line) the completeness limit.}
\label{l041psw_test}
\end{figure}

The completeness limits reported in Fig. \ref{CompleteGal} should be seen 
as conservative because they are determined by spreading the 
synthetic sources randomly over each entire tile. However the diffuse 
background is highly non-uniform in each tile, but it is dominated by the 
strong GP emission with a maximum in the central horizontal 
section of each map, and then decreasing toward the north and south Galactic 
directions. A typical example is offered in Fig. \ref{l041psw_test} where, 
in the upper panel, we show the 250-\um\ image of the tile centred at 
$\ell$=41\adeg.  Superimposed are the extracted 250-\um\ compact sources with 
integrated fluxes above (yellow crosses) and below (magenta crosses) the 
flux completeness limit appropriate for the Galactic longitude at that 
band (3\,Jy, from Fig. \ref{CompleteGal}). This is also shown in the lower 
panel of Fig. \ref{l041psw_test}, where the latitude distribution of the 
two groups of sources is also reported with full/dashed lines for sources 
above/below the confusion limit. 

The two groups of sources have a very different spatial distribution, with 
sources brighter than the completeness limit mostly concentrated at --0\adeg\!.6 $\leq b \leq$ 
0\adeg\!.2 while the fainter sources are uniformly distributed and mostly 
found toward the map areas where the diffuse emission is relatively less 
intense. The dashed line in the lower-panel histogram is flat because fainter sources are better detected in lower surface-brightness regions (above and below the Plane) than in the central band of the Plane.

In subsequent releases of the Hi-GAL photometric catalogues we 
will provide more precise estimates of the catalogue completeness limits 
specific to different background conditions.

\subsection{Deblending}

CuTEx is designed to fit a gaussian function to each position where there 
is an enhancement of the 2$^{nd}$ derivative with respect to its nearby 
environment. While the flux estimate relies on the performance of the 
fitting engine as well as on the fidelity of the gaussian model fit to the 
real source profiles, it is clearly important to quantify the ability of 
the photometric algorithm to separate individual sources in the case 
where they are very close to each other. To quantify the deblending performance
of the algorithm we generated simulations with 2000 sources randomly 
distributed on a region whose size represents the typical footprint of the Hi-GAL 
maps. For every set of positions we produced two different sets of 
simulated populations. In the first case, we injected sources with sizes 
of the order of the beam size. In the second case we simulated a population of 
extended sources modeled as elliptical gaussians with the FWHM of one of 
the two axes drawn from a uniform distribution between 1 and 2.5 times 
the beam size. The other axis is determined by assuming an axis ratio 
randomly drawn from a uniform distribution between 0.5 and 1.5 times the beam size. The 
input sources are randomly oriented. We computed several simulations with 
different positions and increasing source densities in order 
to estimate the deblending performance for cases of both lesser and greater clustering.

\begin{figure}[h]
\centering
\includegraphics[width=0.5\textwidth]{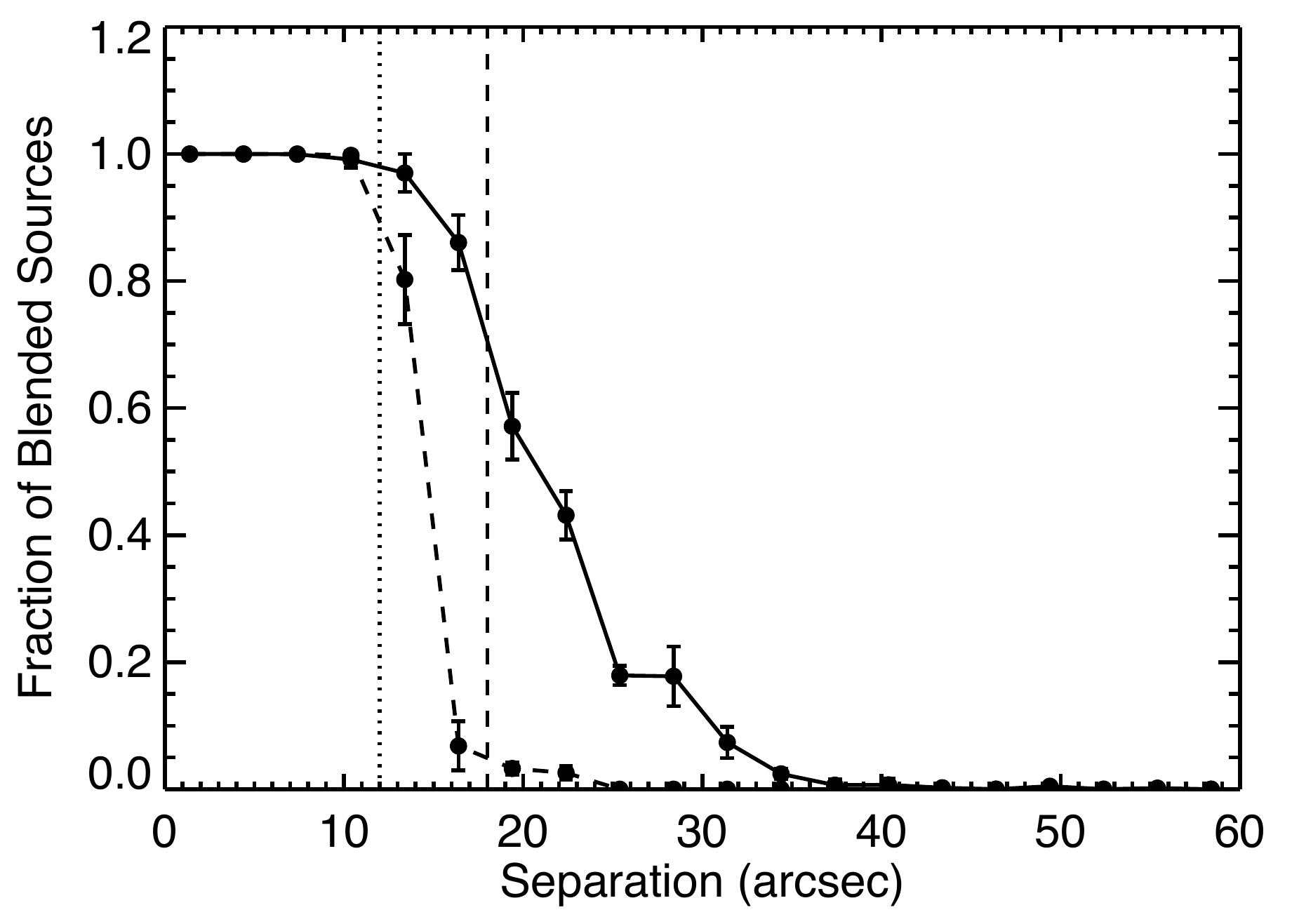}  
\caption{The curves represent the fraction of blended sources that \cutex\ 
is not able to deblend as a function of source separation for a set of 
synthetic sources described in the text; simulations in this 
case are made for the 250\um\ images. The full 
and dashed lines are the results for simulations with extended and 
point-like sources, respectively. Vertical lines represent the size of the 
beam (dashed), and 75\% the size of the beam (dotted).}
\label{deblending_fig}
\end{figure}

We processed the simulations with CuTEx and determined its ability to 
correctly identify individual sources as a function of the source pair 
separation. Due to the large number of sources and their relatively high 
densities,  in each simulation, there are several thousand source pairs 
that can be tested for the effectiveness of our deblending algorithm. We 
plot in Fig.~\ref{deblending_fig} the fraction of source pairs that are 
not resolved into their separated components as a function of their 
relative separation for simulations of the 250\um\ 
data (where the maps have a pixel size of 6\asec). Similar curves are 
found for the other wavelengths. The error bars represent the amplitude of 
such a fraction found in the whole set of simulations. The full line 
refers to the case of the population of extended sources, while the dashed 
line indicates the results for the sample of point sources. The vertical 
dashed line traces the size of the beam, while the dotted line traces 0.75 
times the beam.

Fig.~\ref{deblending_fig} shows that CuTEx is able to deblend sources quite effectively. Point-like sources are 
resolved perfectly up to distances that are $\sim$ 0.8 times the beam, 
while extended sources are properly deblended and identified for distances larger than $\sim$ 1.25 
times the beam. For the extended source case, half of the source pairs that are 
separated by a single beam size are deblended. Clearly, the gaussian fit for a 
blended source pair will result in a larger size estimate than the case 
where the two components are resolved by the detection algorithm.

\begin{figure}[!t]
\centering
\includegraphics[width=0.45\textwidth]{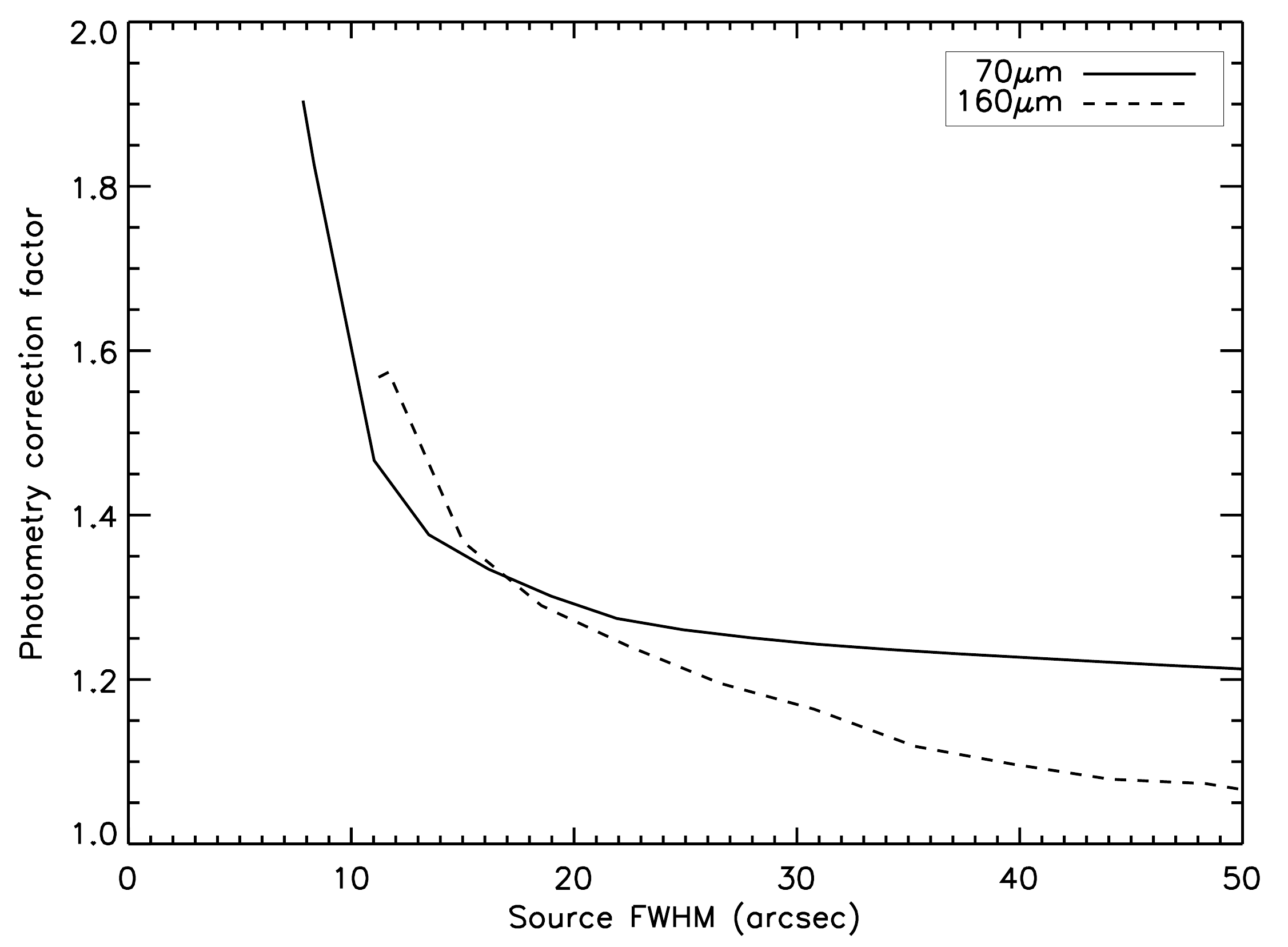}  
\caption{Correction factors to be applied to \cutex\ photometry as a 
function of the source FWHM. The values are only applicable for images 
obtained similarly to \higal.}
\label{phot_corr}
\end{figure}

\begin{figure*}[!h]
\centering
\includegraphics[width=0.33\textwidth]{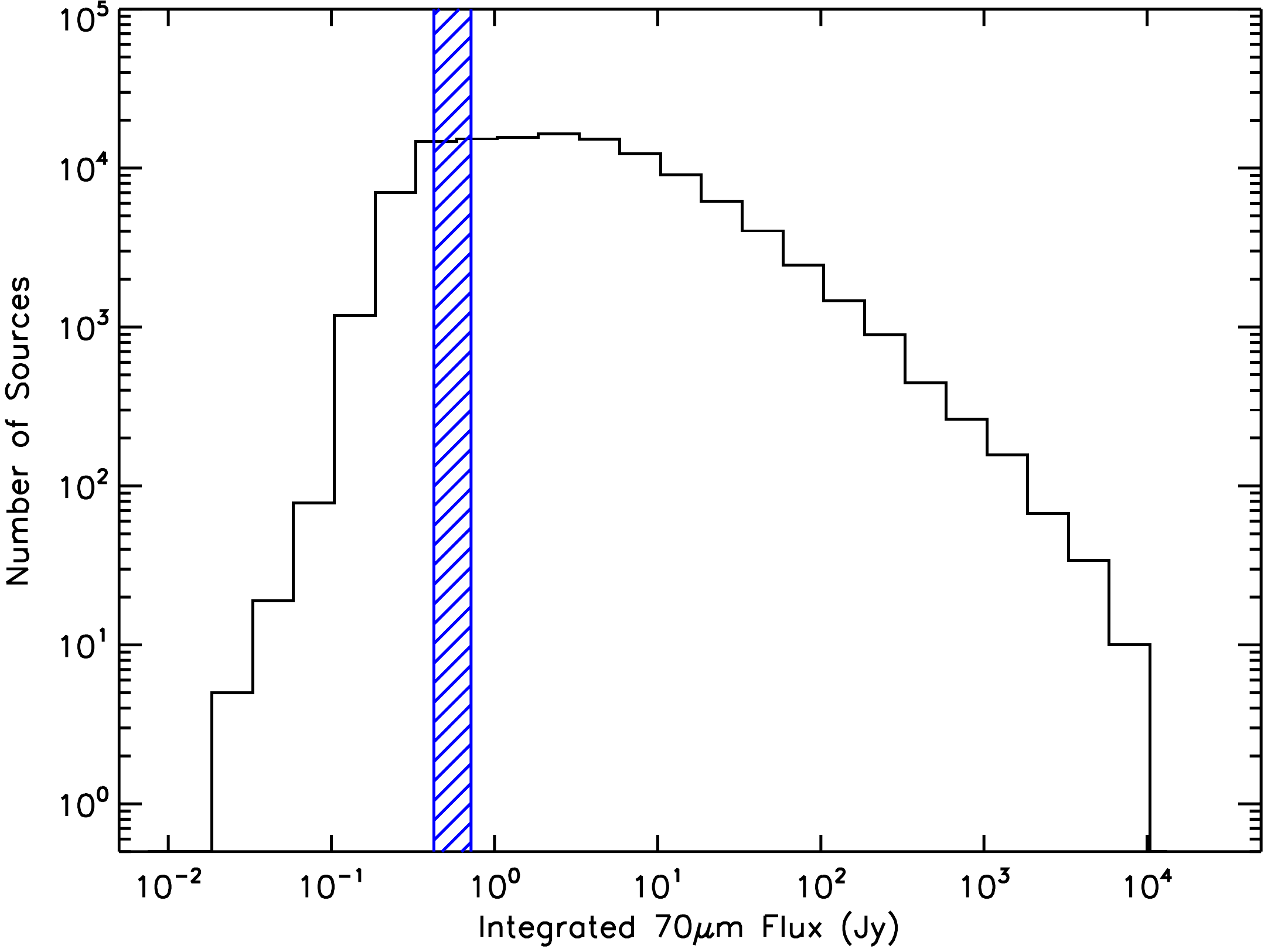} 
\includegraphics[width=0.33\textwidth]{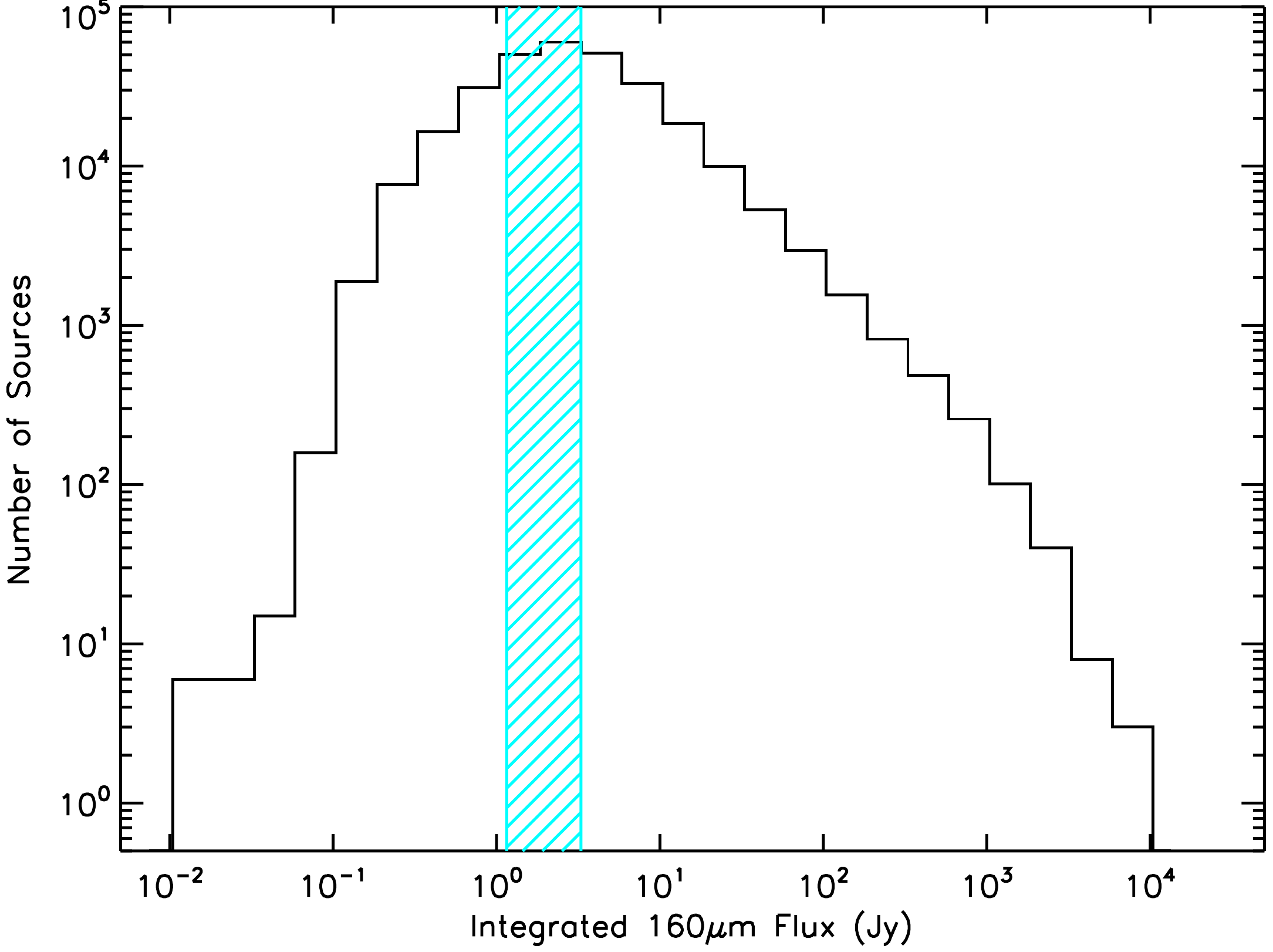} 
\includegraphics[width=0.33\textwidth]{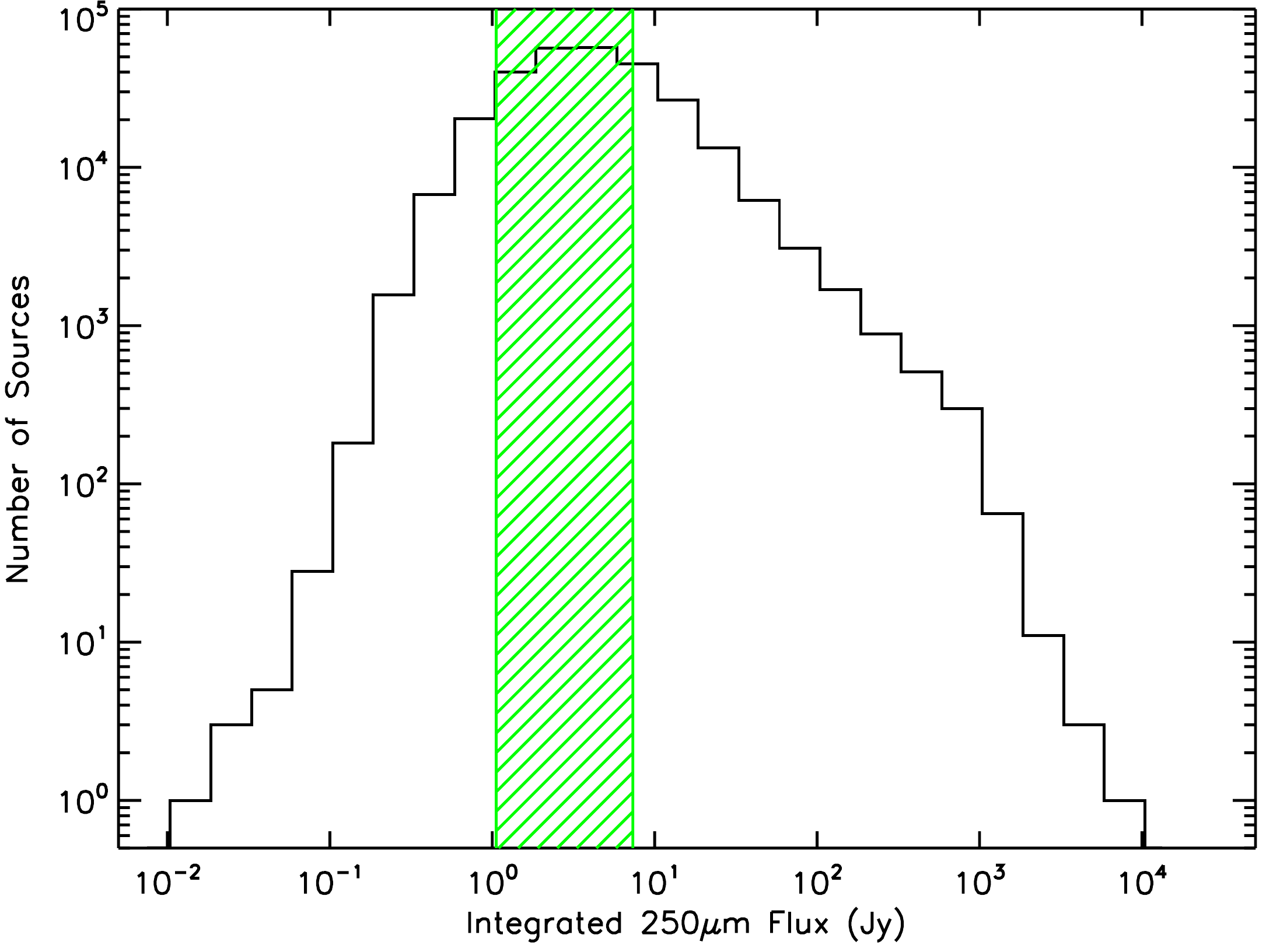} 
\includegraphics[width=0.33\textwidth]{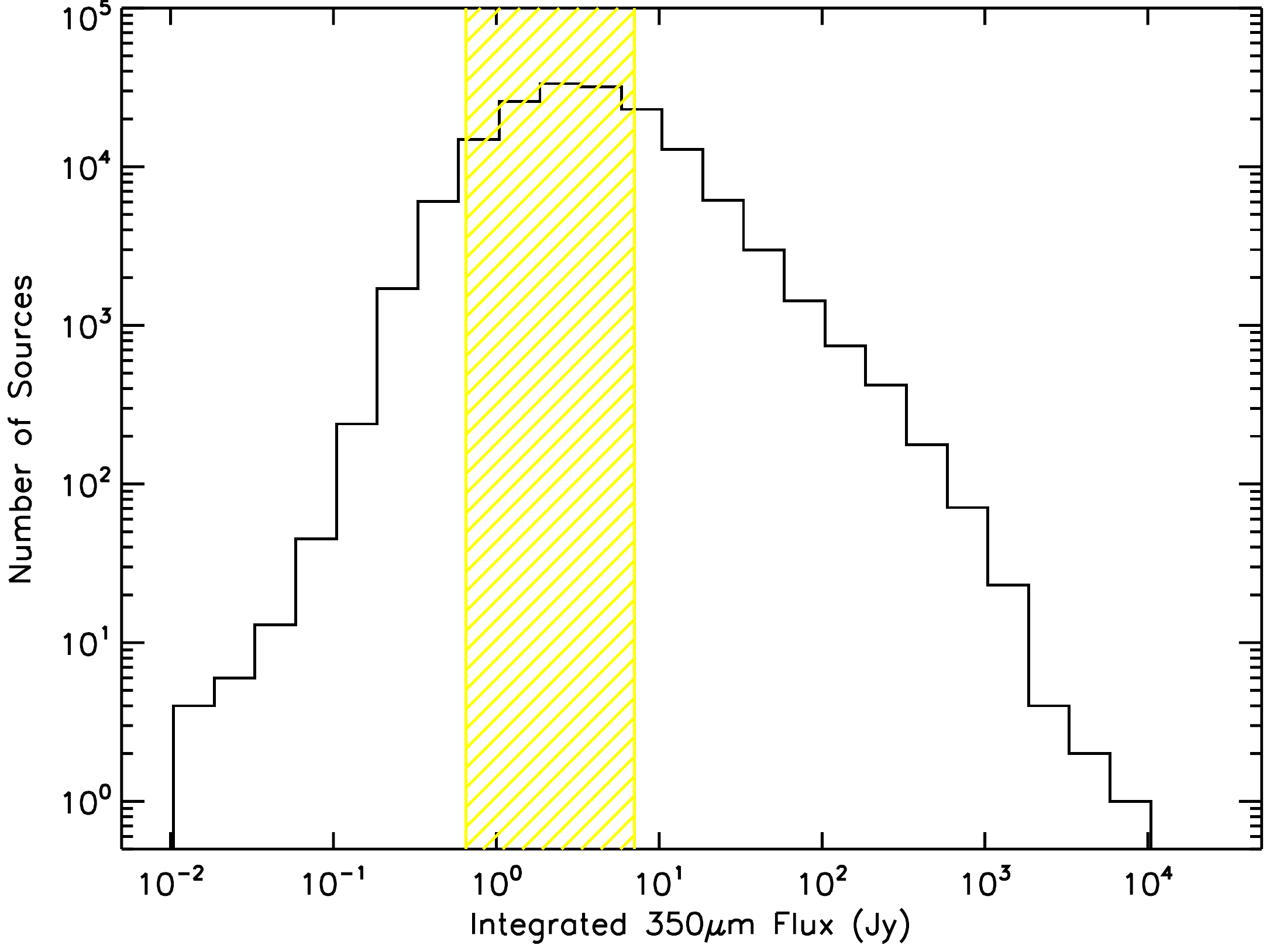} 
\includegraphics[width=0.33\textwidth]{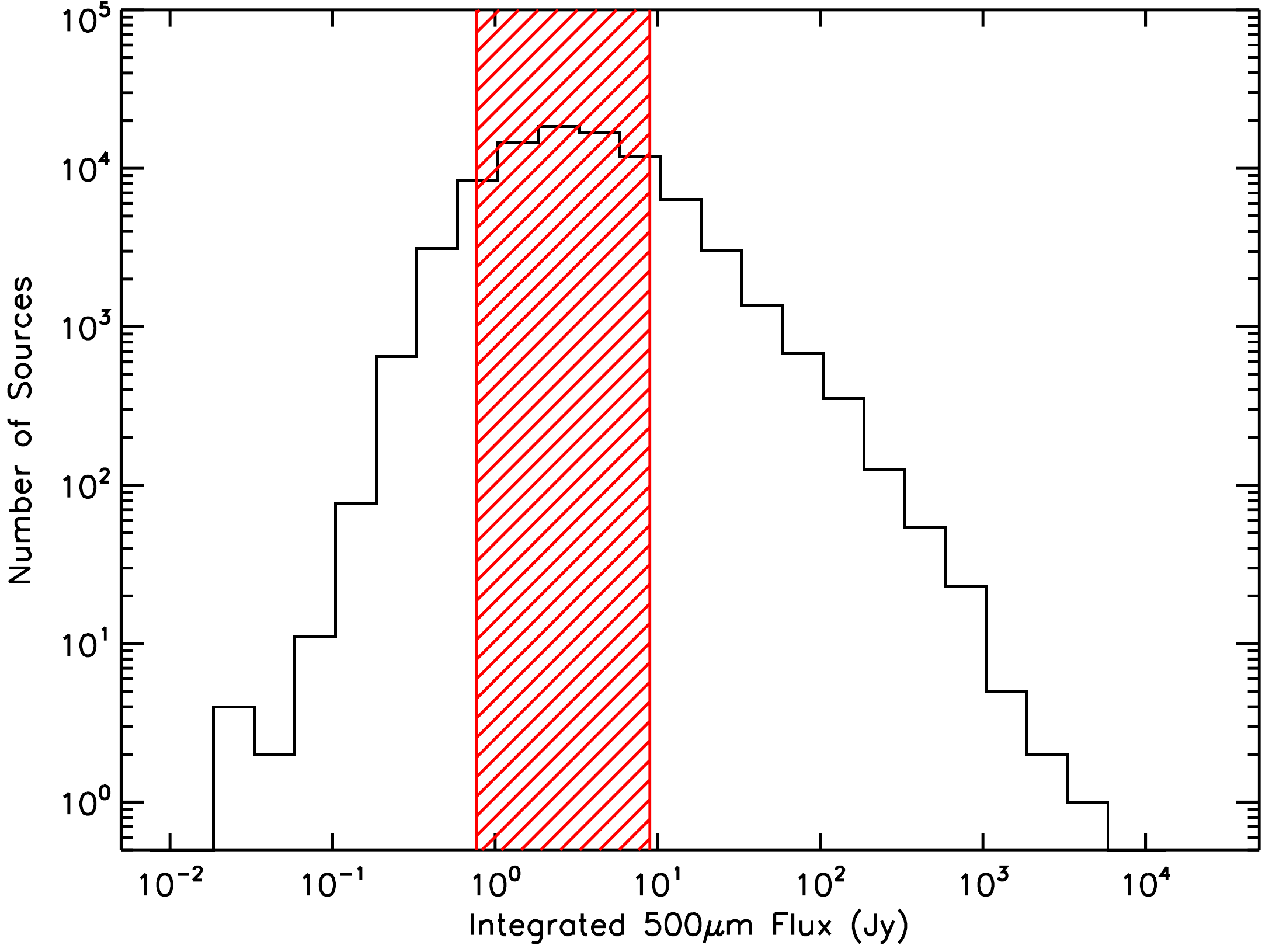} 
\caption{Histograms of the Integrated Flux F$_{\rm Int}$ for all \higal\ 
compact sources in the 5 bands for the entire DR1 Survey area. Flux completeness limits vary with Galactic Longitude (see fig. \ref{CompleteGal}); the spanned longitude range  (with the exception of the central |$l$|$\leq 2$\adeg) is reported by the the colour-coded and shaded areas in fig. \ref{CompleteGal}. } 
\label{hist_fint} 
\end{figure*}

\subsection{Photometric Corrections to Integrated Fluxes}
\label{photchecks}

The flux of the source candidates is derived from the parameters of the 
2D-Gaussian fit found with \cutex. While a 2D Gaussian is a good and 
acceptable approximation for the PSF of SPIRE \citep{spire_obs_man}, the 
same is not true for PACS due to the observing setup adopted for the 
\higal\ survey. The on-board coaddition (in groups of 8 frames at 70\um\ 
and 4 frames at 160\um) while scanning the satellite, results in 
substantially elogated beams (see \S\ref{observations} above) that 
show significant departures from a circularly symmetric morphology. Part 
of this asymmetry is mitigated by the coaddition of scans in orthogonal 
directions, but significant departures from an ideal Gaussian symmetry 
persist. It is then necessary to estimate correction factors to be 
applied to the extracted \cutex\ photometry to account for the (incorrect) 
assumption of Gaussian source brightness profiles assumed by \cutex.

We adopted an empirical approach to estimate the corrections to the 
\cutex\ photometry of PACS images.  This was done by performing \cutex\ photometry, using 
the same settings as used for the \higal\ catalogues, on an image of a primary 
{\em Herschel} photometric calibrator - $\alpha~Bootis$.  $\alpha~Bootis$ was observed during 
OD~269 in the same conditions as the Hi-GAL observations (i.e. with two 
mutually orthogonal scan maps in parallel mode with a scanning speed of 
60\arcsec/s). The $\alpha~Bootis$ images present a nice and clean point-like 
object with no detectable diffuse emission background (ideal photometry 
conditions compared to \higal). To extend the photometric correction factors to the more 
general case of compact but resolved sources, we convolved the images of 
$\alpha~Bootis$  with a 2D-circular Gaussian kernel of increasing size while 
normalizing  integrated flux (i.e. flux conserving). The convolving kernels 
span the interval [0.0,5.0]$\times\theta_0$ in steps of 0.5$\theta_0$, 
where $\theta_0$ is the FWHM derived from the unconvolved $\alpha~Bootis$ 
profile. \cutex\ integrated fluxes for the entire set of simulations were 
then compared with the expected values in the PACS bands as derived from 
theoretical models \citep{Muller+2014}.  After applying a colour correction 
estimated via \cite{Pezzuto+2012}, the fluxes of $\alpha~Bootis$ used for 
the comparison are 15.434 and 2.891 Jy at 70 and 160\um, respectively. 
Figure \ref{phot_corr} reports the correction factors as estimated from 
the above analysis as a function of the FWHM of the compact source 
considered. The correction factors decrease rapidlyfrom point-like 
to minimally resolved sources.  With larger sources, the decrease in the correction factor is a weaker function of source size.
 Beam asymmetries, however, are clearly persistent and detectable 
even for relatively extended sources. 

The integrated fluxes for each source in the 70 and 160\um\ catalogues 
were corrected using the curves in fig. \ref{phot_corr} and the sources' 
circularised size (see\S\ref{par_size}). Both the uncorrected and the 
corrected integrated fluxes are reported in the columns FINT and 
FINT\_UNCORR of the source catalogues (see Appendix \ref{cat_expl}). We 
emphasise that these correction factors are only valid for images obtained 
from two scan maps taken in orthognal directions in pMode with a 60\asec 
/s scanning speed, and for sources extracted using a 2D Gaussian source 
model (i.e., they are not valid if PSF-fitting or aperture photometry is 
performed). The same analysis was carried out for SPIRE, but the correction factors 
estimated were largely within 10\% for the unconvolved $\alpha~Bootis$ image, 
confirming the reliability of the Gaussian approximation for the SPIRE 
beams.  Larger sources could not be simulated due to the high spatial 
density of background compact objects of extragalactic origin but, as 
suggested by fig. \ref{phot_corr}, the effect should be even lower.

\section{Properties of the Compact Source Catalogues}
\label{cat_properties}

\subsection{Source fluxes and reliability}
\label{reliability}
 
In fig. \ref{hist_fint} we report the distribution of the integrated 
fluxes of all extracted compact sources in the 5 photometric bands. The 
histograms report the sources detected within the entire DR1 survey area.  The large spread in detected fluxes, while representative of the entire survey, does not necessarily reflect the flux distribution in any individual tile. For example, the sources in the faint tail of the 
distributions originate mainly from the tiles at larger longitudes 
and are not detected in tiles like the one at $[l, b]$=(19\adeg , 0\adeg ) 
for which we report the completeness limits in fig. \ref{CompleteOne}, or from regions that are removed from the central latitude band around $b=0$\adeg. 
In addition, the objects at the far left side of each histogram (low flux) are those that are potentially most affected by false positives, as dicussed in \S\ref{choice_th}.  We note, however, that even if we combine the sources in the 4 left-most bins of each histogram in fig. \ref{hist_fint}, these souces, combined, only account for 0.8\% of the total number of sources in the 70\um\ band and less than 0.1\% for the other bands.

\begin{figure*}[t]
\centering
\includegraphics[width=0.4\textwidth]{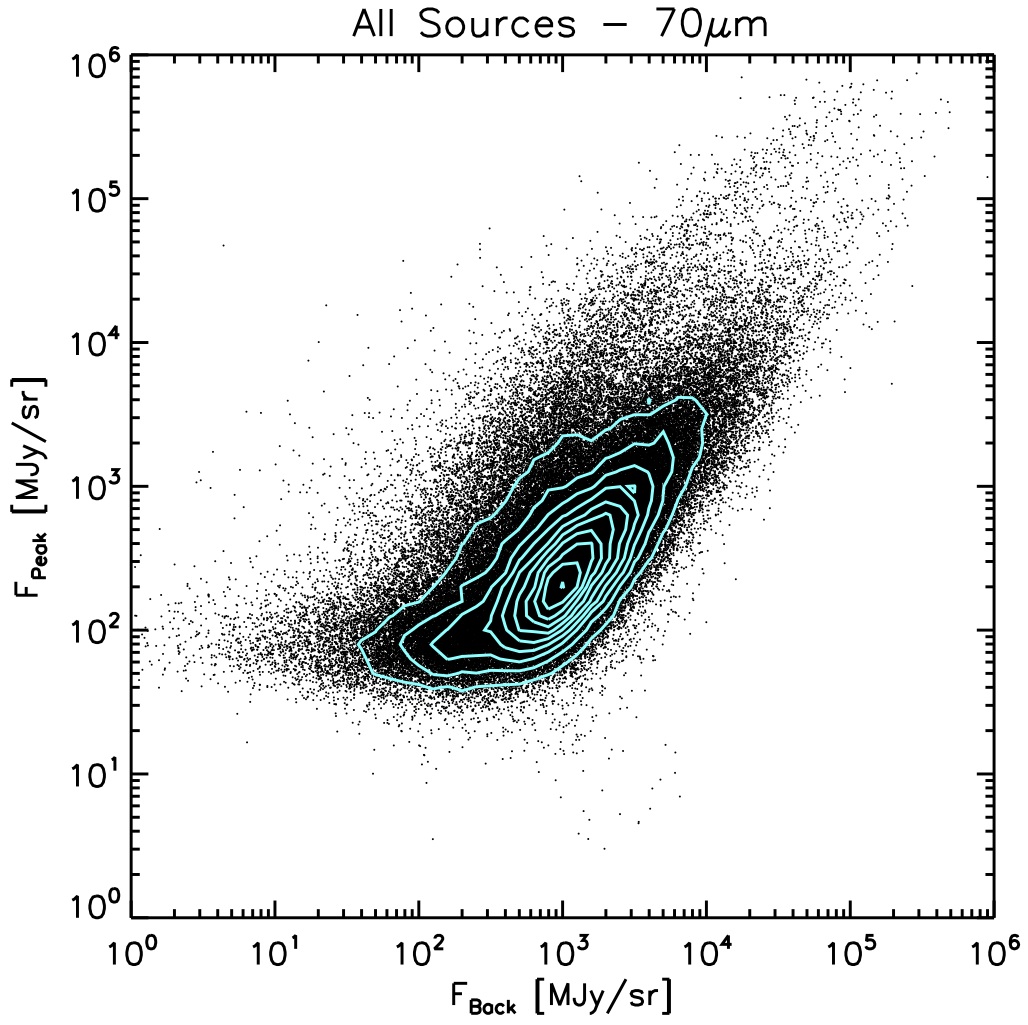}
\includegraphics[width=0.4\textwidth]{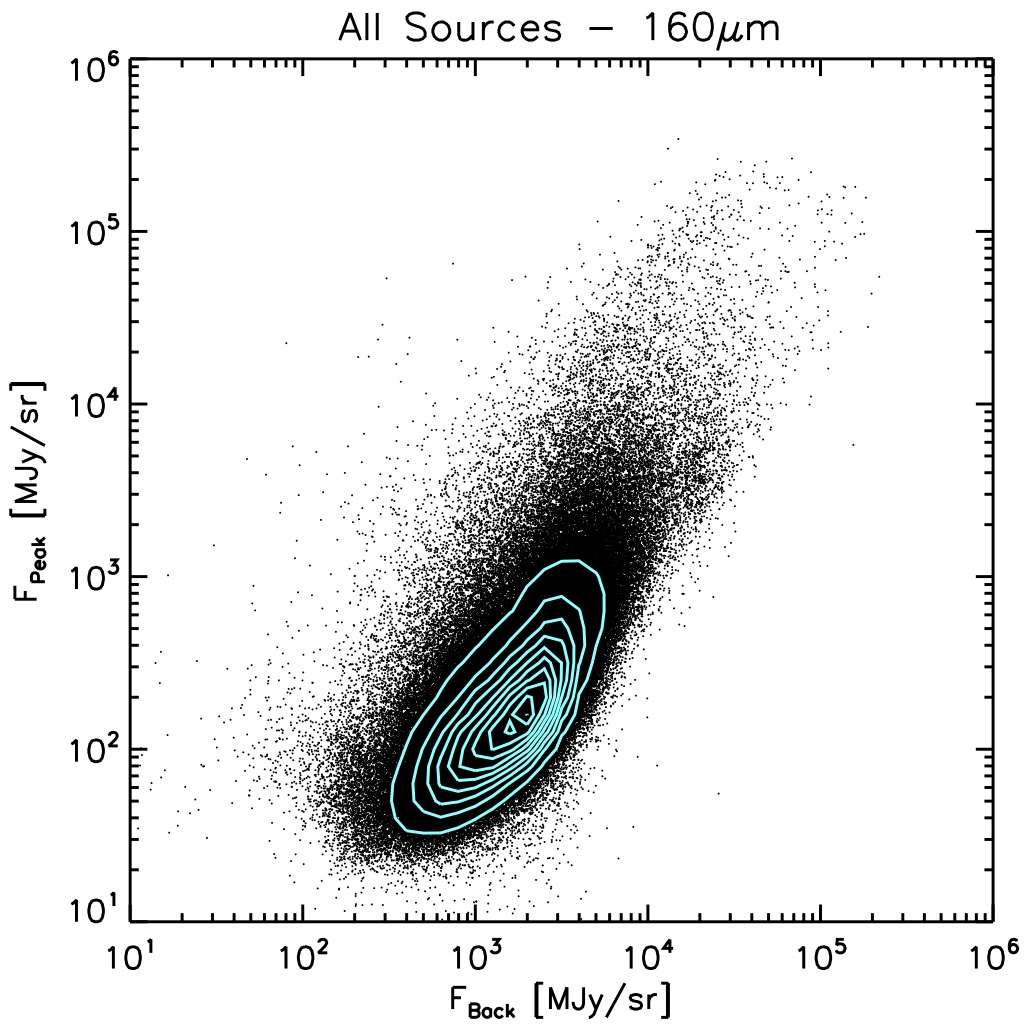}
\includegraphics[width=0.33\textwidth]{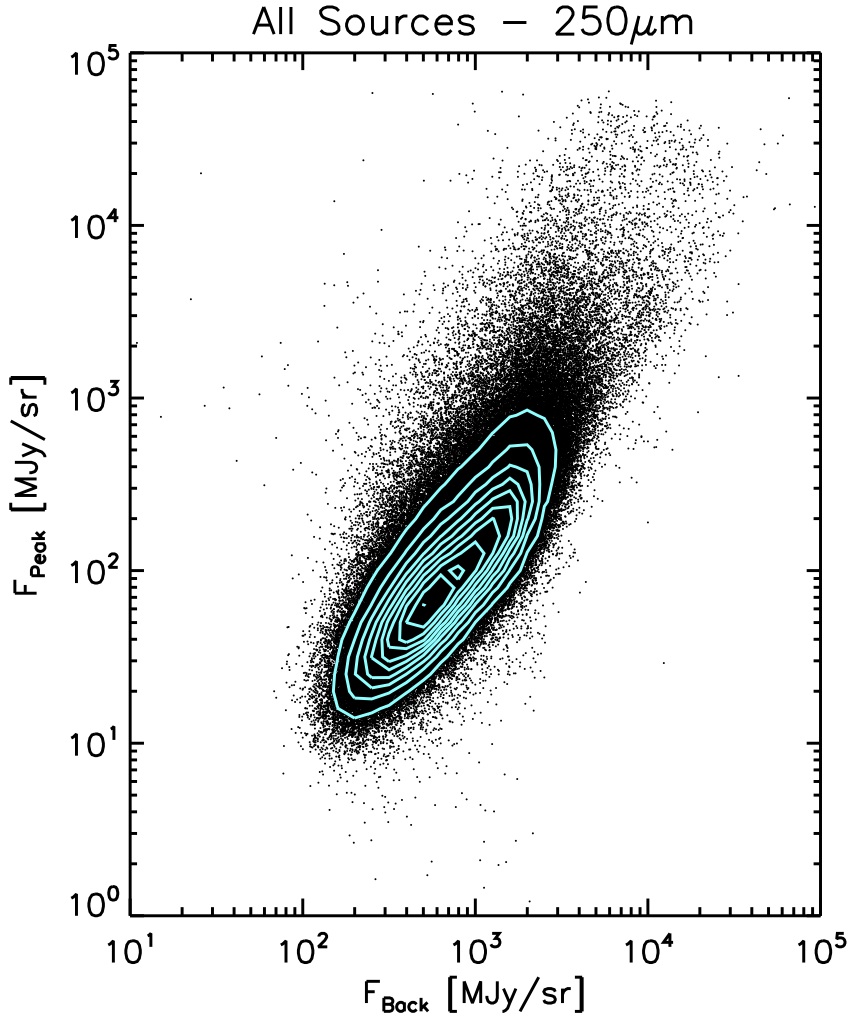}
\includegraphics[width=0.33\textwidth]{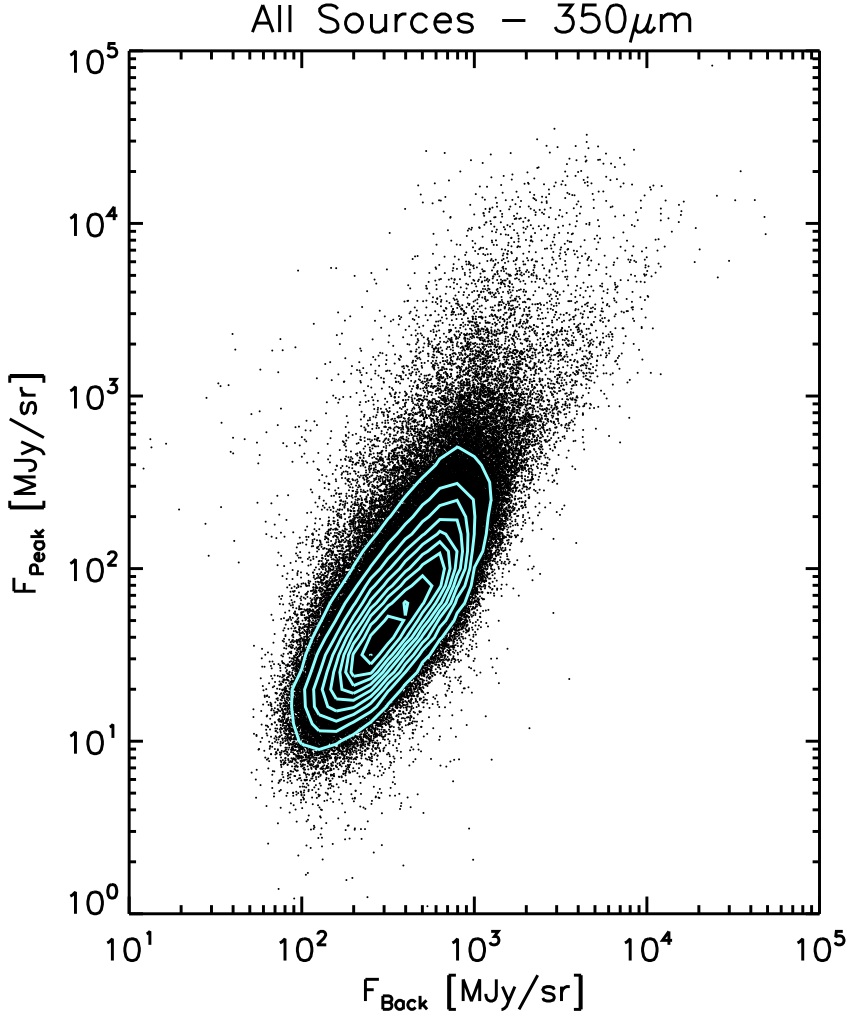}
\includegraphics[width=0.33\textwidth]{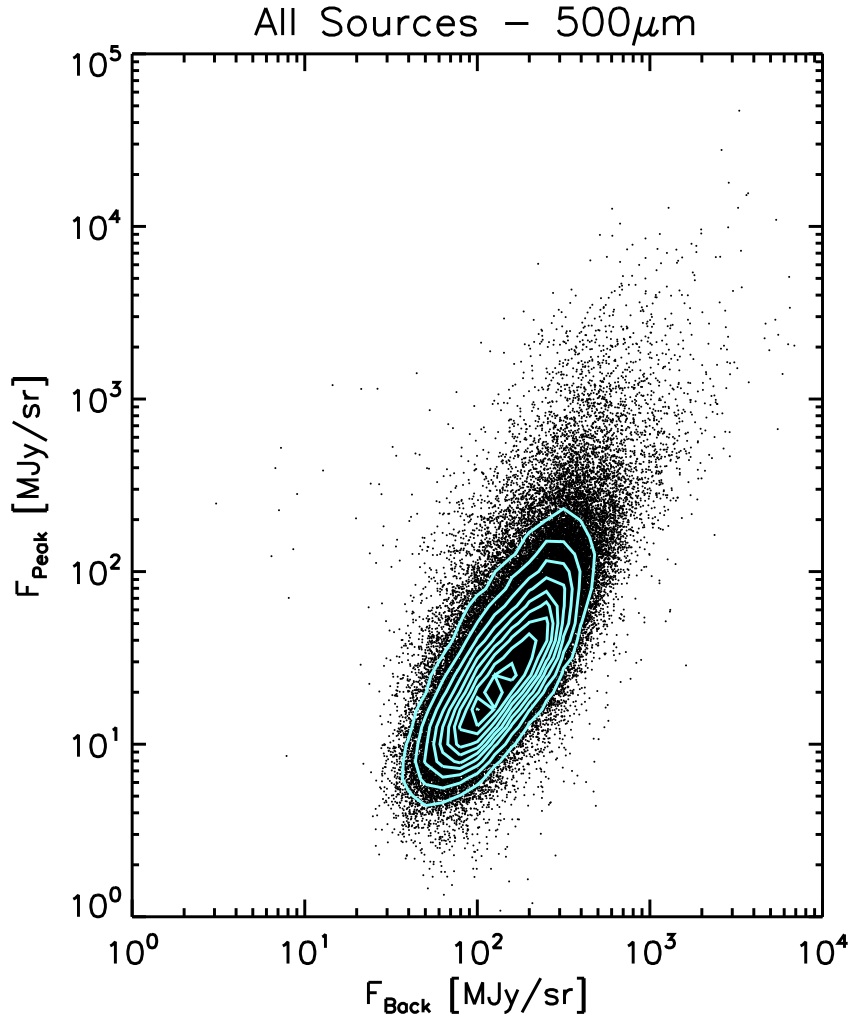}
\caption{Plots of the background-subtracted source peak flux density F$_{\rm Peak}$ as a function of 
the flux density of the underlying background F$_{\rm Back}$, as estimated 
by the source fitting for all Hi-GAL bands as indicated. The 10 cyan contours (equally spaced in source density) indicate the source density in the most crowded area. Note that axis scale is not the same in all panels.}
\label{fpeak-back}
\end{figure*}

\begin{figure*}[t]
\centering
\includegraphics[width=0.4\textwidth]{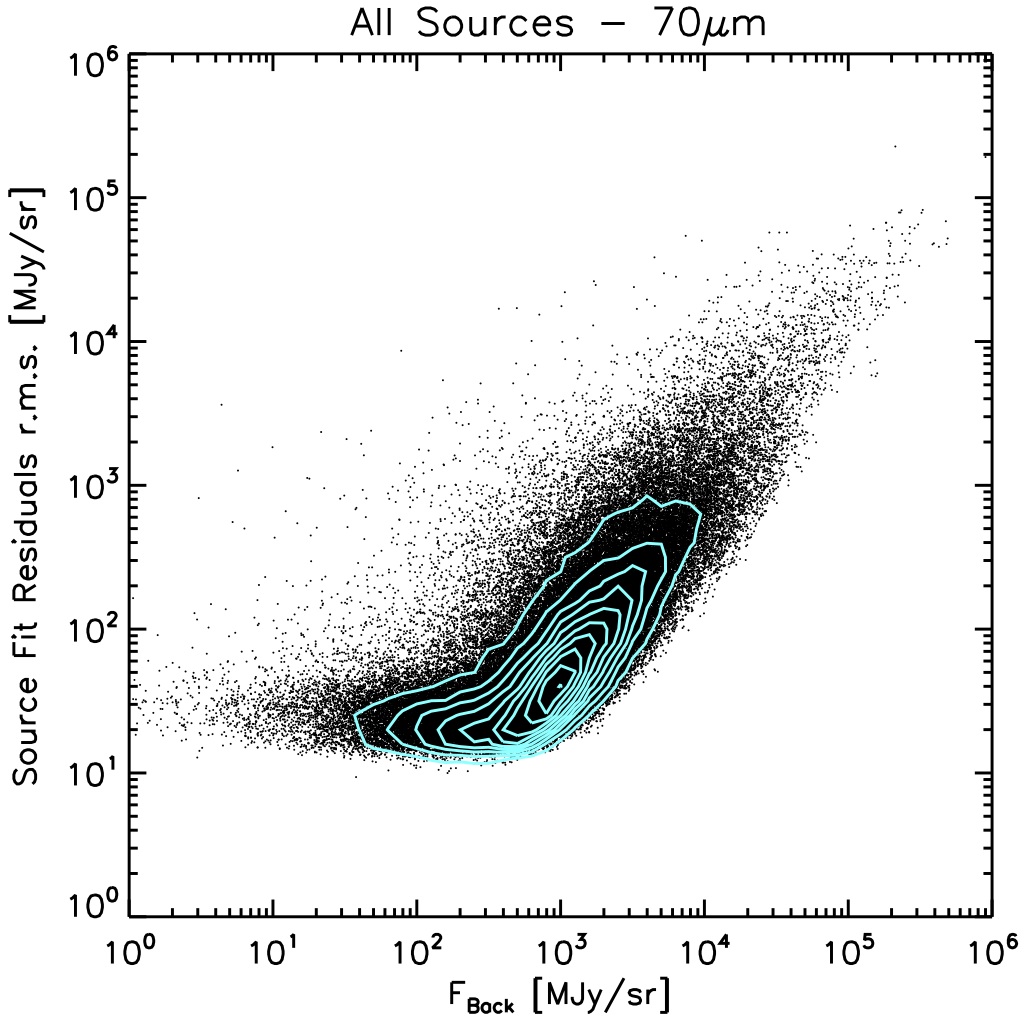}
\includegraphics[width=0.4\textwidth]{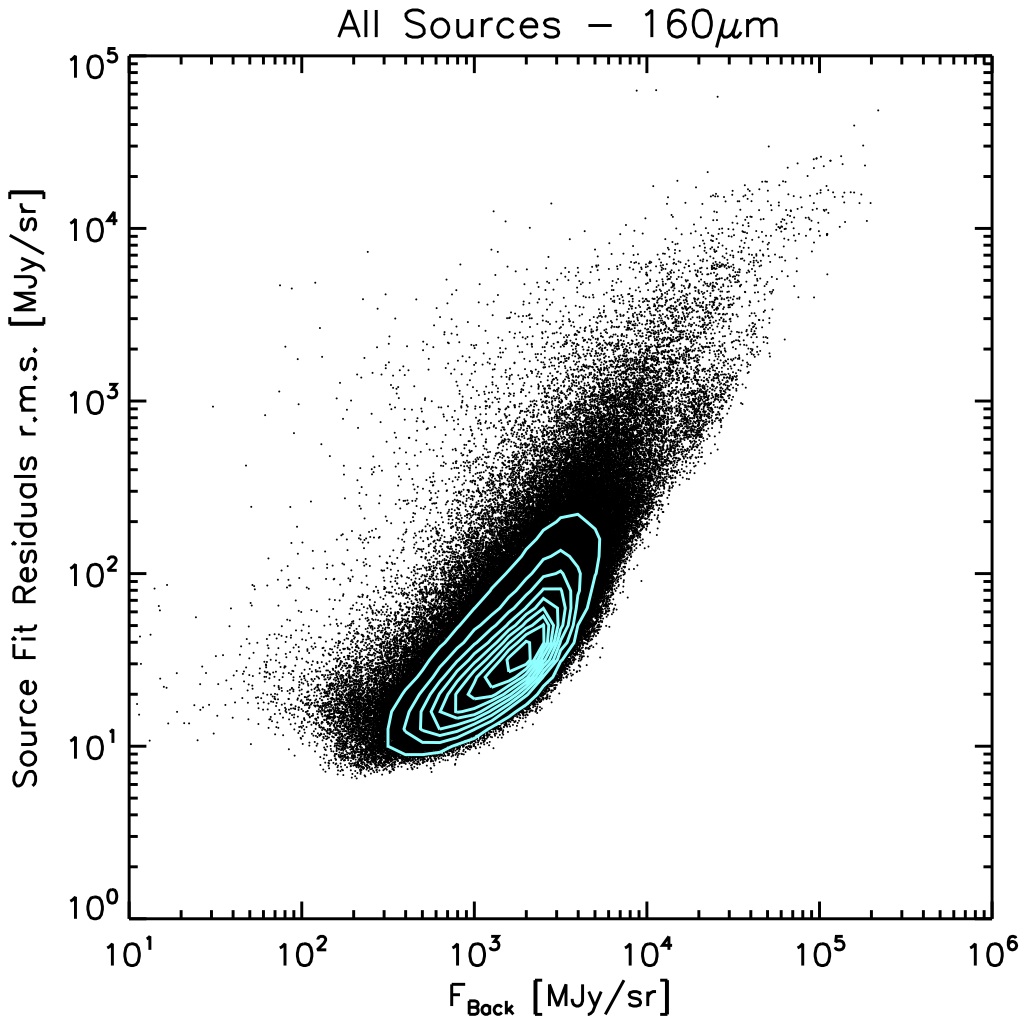}
\includegraphics[width=0.33\textwidth]{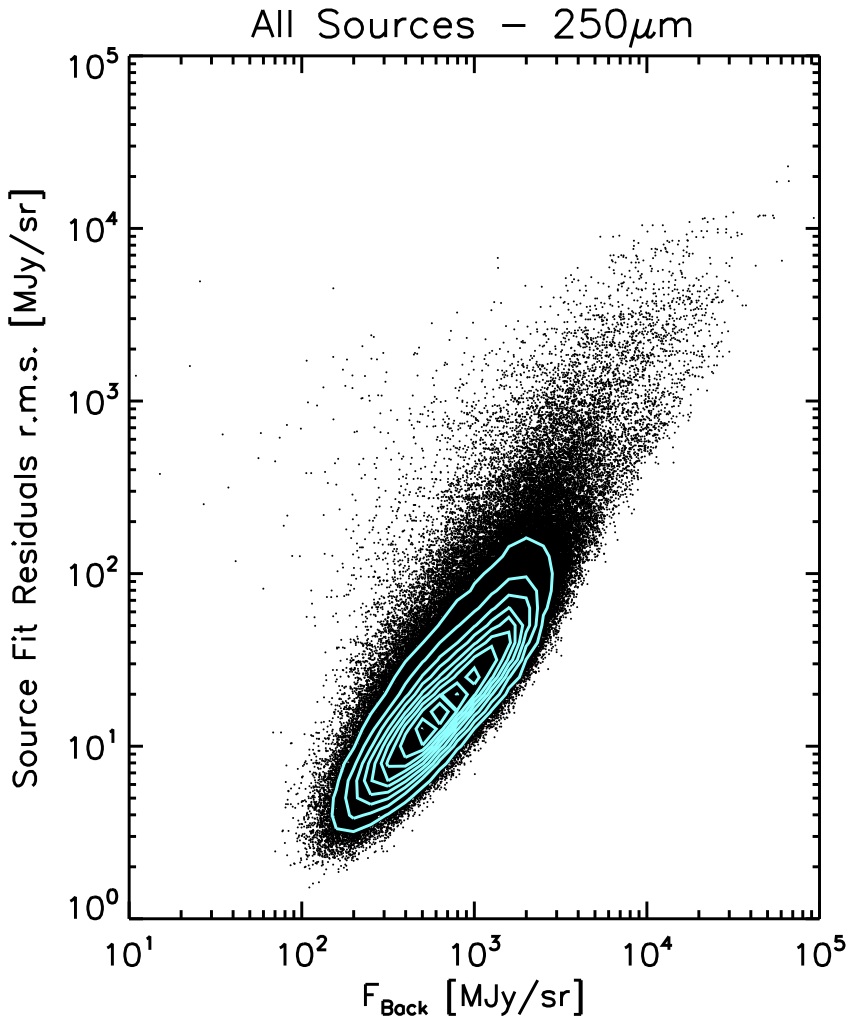}
\includegraphics[width=0.33\textwidth]{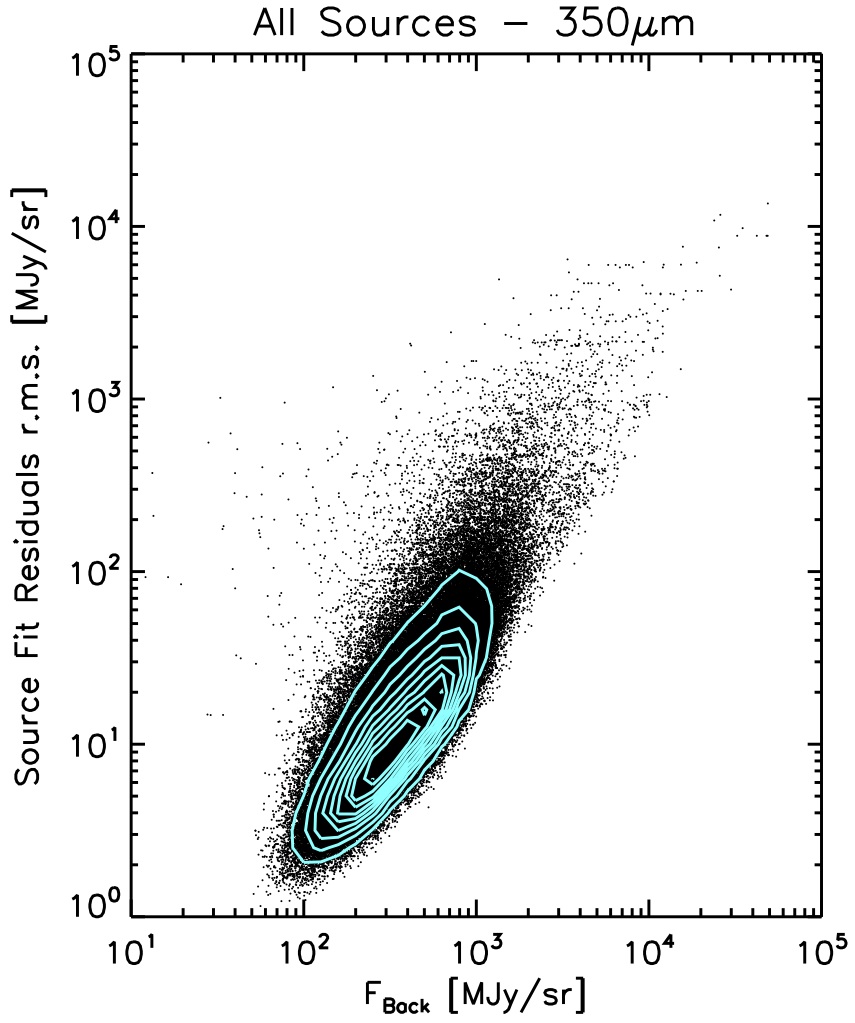}
\includegraphics[width=0.33\textwidth]{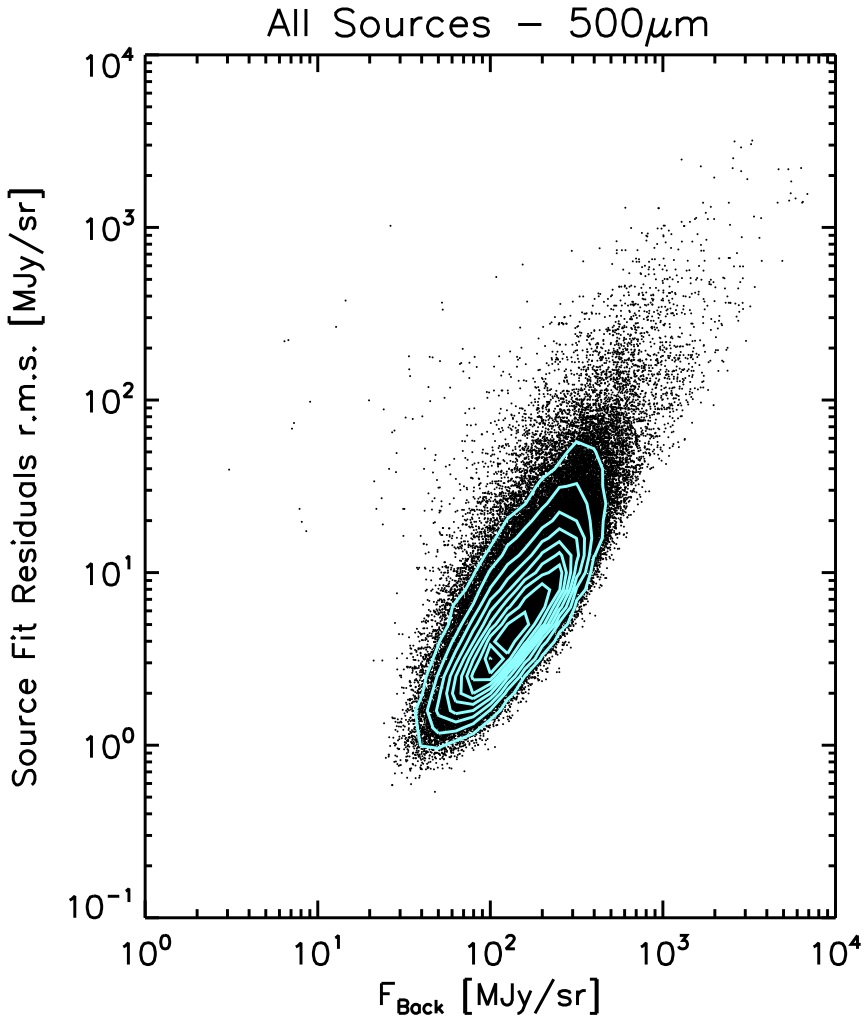}
\caption{Plots of the r.m.s.of the flux density residuals after 
subtraction of the fitted source+background, as a function of the flux 
density of the underlying background F$_{\rm Back}$, as estimated by the 
source fitting for all Hi-GAL bands as indicated. The 10 cyan contours (equally spaced in source density) indicate the source density in the most crowded area. Note that axis scale is not the same in all panels.}
\label{resid-back}
\end{figure*}

It is difficult to identify a parameter that can be uniquely taken as a 
measure of the reliability of a source detection. It is important to 
remember that the background conditions found at {\em Herschel} 
wavelengths in the Galactic Plane are totally unprecedented.  Therefore,   
criteria based on e.g. the S/N of the detected sources (that are 
reliable criteria in conditions of absent or low background) are not 
straightforward to apply, because compact sources have a 
variety of sizes (see \S\ref{par_size}) and sit on a Galactic ISM 
background that shows spatial variations at all scales. Fig. 
\ref{fpeak-back} illustrates the relationship between the background-subtracted peak flux 
densities of the sources and the intensity of the underlying background 
emission as estimated during the 2D Gaussian fitting in \cutex. A direct 
relationship between the two quantities is apparent in all bands and Fig. 
\ref{fpeak-back} further shows 
that the peak flux of the sources is always a factor of a few fainter than 
the value of the background. An additional problem is that, not only does the 
background dominante over the source peak fluxes, but its 
fluctuations increase with the absolute level of the background.  Therefore, since the 
uncertainties in the extracted source fluxes are computed starting 
from the residuals obtained after subtracting the fitted source+background 
(the latter modelled with a 2$^{nd}$-order surface) from the original 
maps, the magnitude of the residuals will be higher the higher the 
absolute level of the background. This is shown in fig. \ref{resid-back} 
where the r.m.s. of the  fitted residuals is reported for the various bands as a 
function of the absolute level of the fitted background.

\begin{figure*}[!t]
\centering
\includegraphics[width=0.33\textwidth]{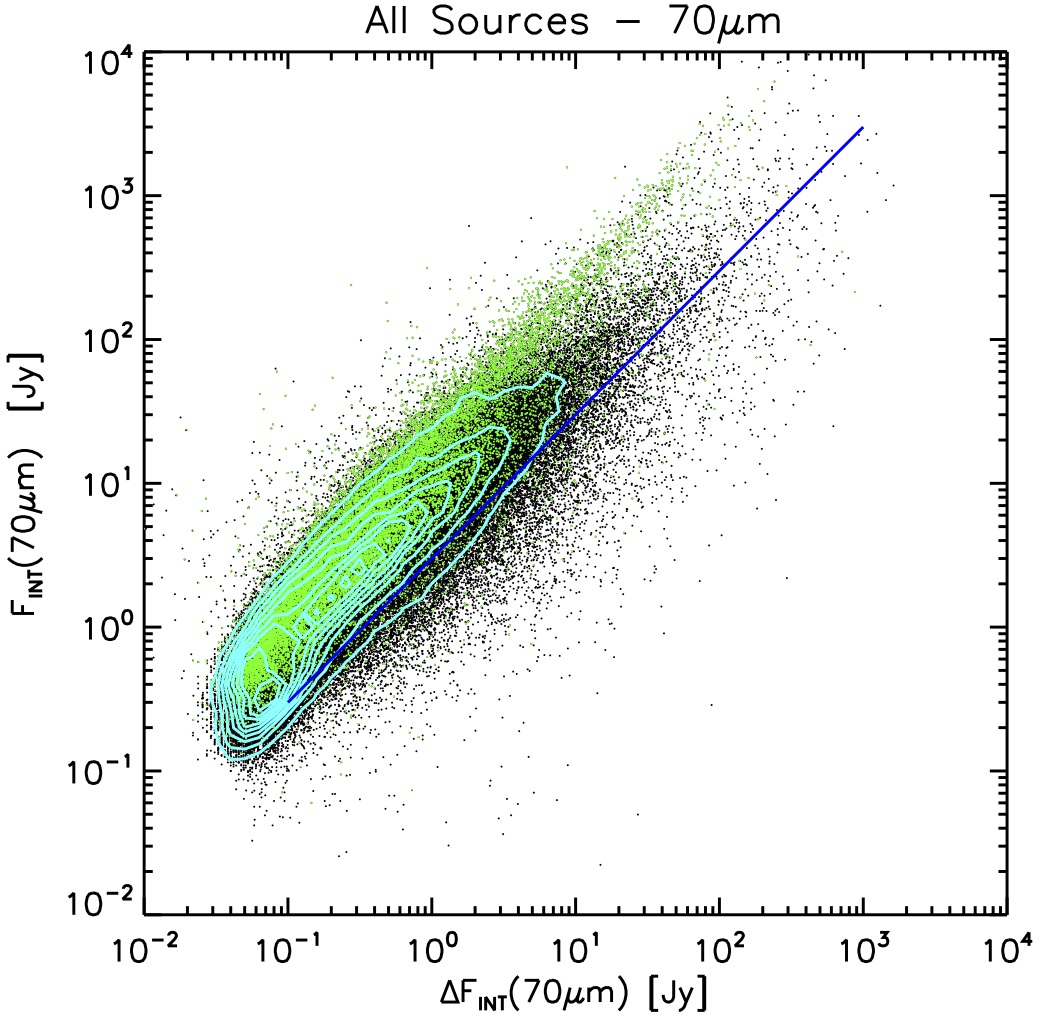}
\includegraphics[width=0.33\textwidth]{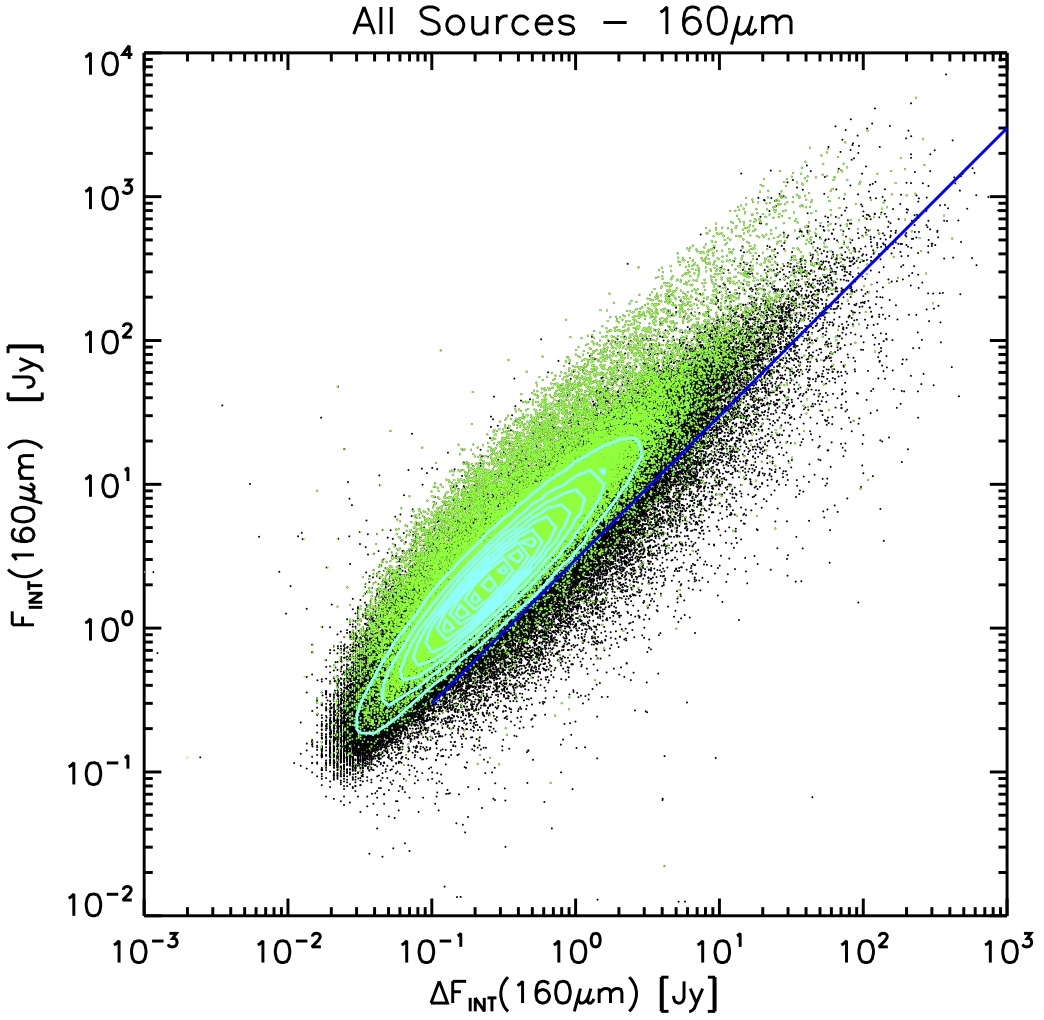}
\includegraphics[width=0.33\textwidth]{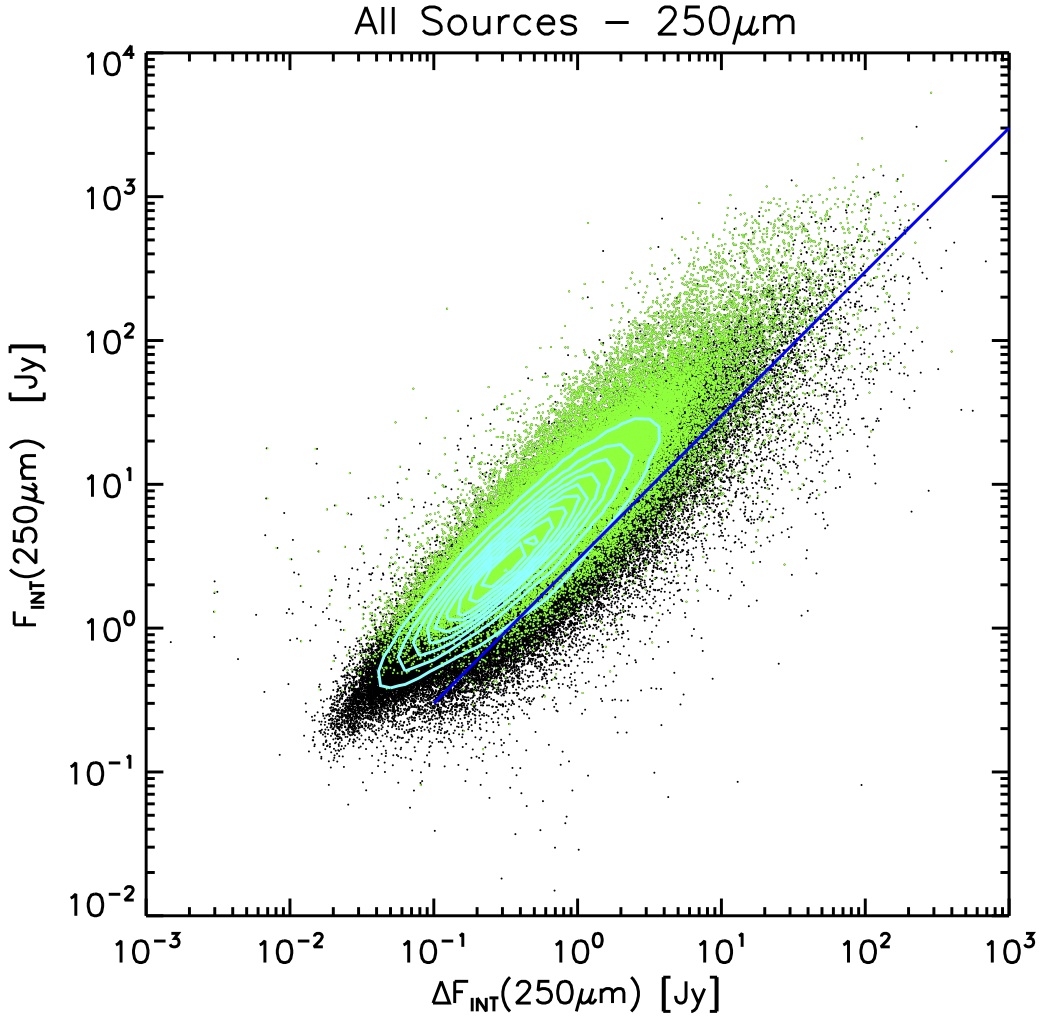}
\includegraphics[width=0.33\textwidth]{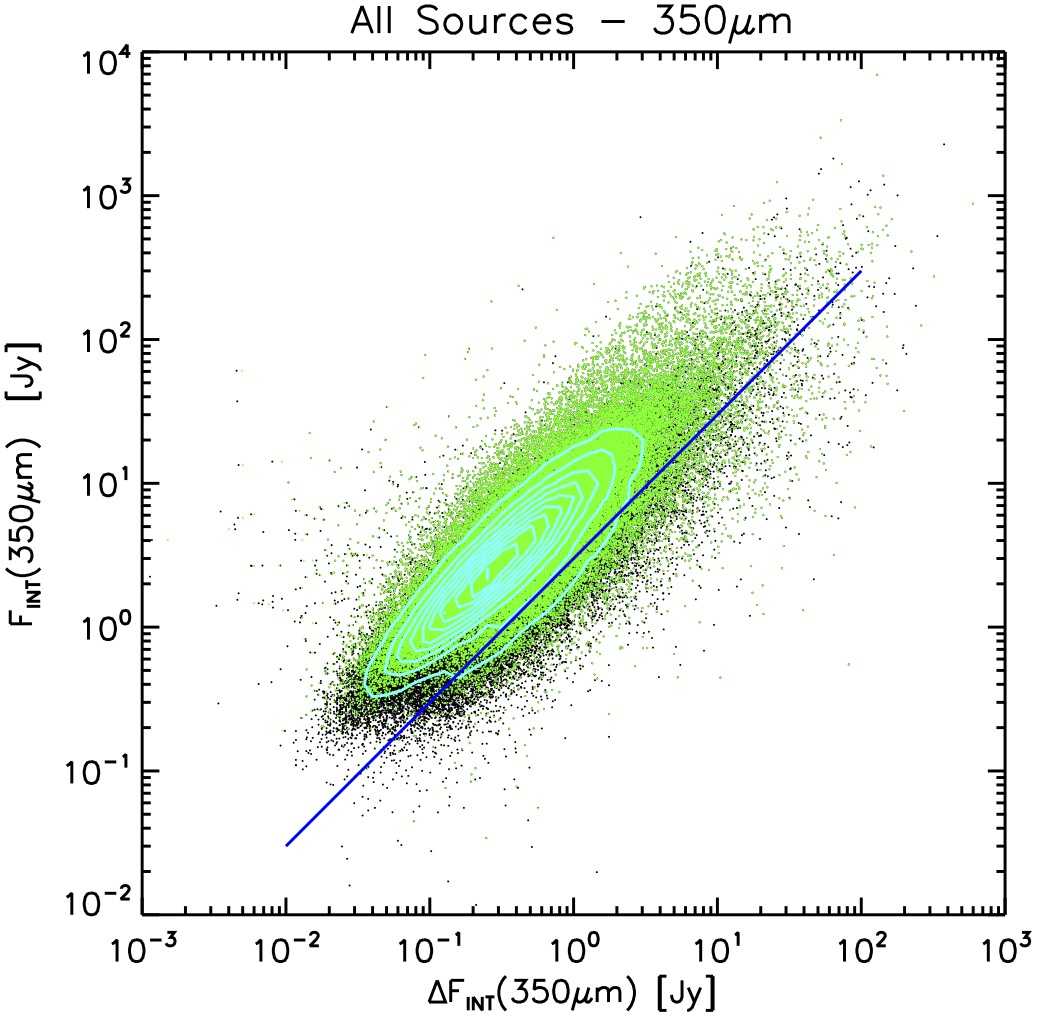}
\includegraphics[width=0.33\textwidth]{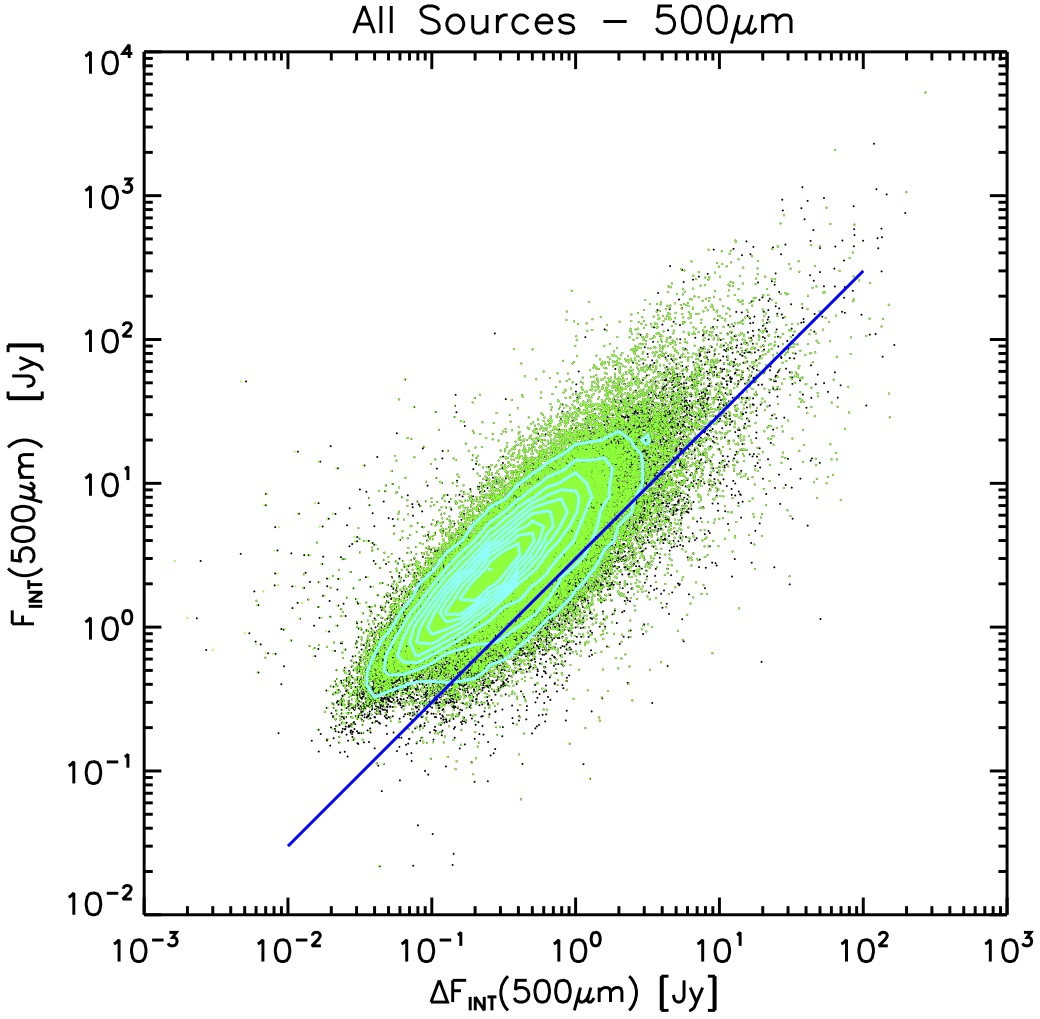}
\caption{Plots of the Integrated Flux F$_{\rm Int}$ as a function of its 
uncertainty $\Delta$F$_{\rm Int}$ for all Hi-GAL bands as indicated. The 
black points are all the sources in each band catalogue; the 10 cyan contours (equally spaced in source density) indicate the source density in the most crowded area. The green 
points are the subset of sources that possess a counterpart in at least 
two adjacent bands (so as to form an SED with at least three  
photometric points, see \citealt{Elia+2016}). The blue line represents 
SNR$_{\rm Int}$=3.}
\label{f-df}
\end{figure*}

\begin{figure*}[!t]
\centering
\includegraphics[width=0.4\textwidth]{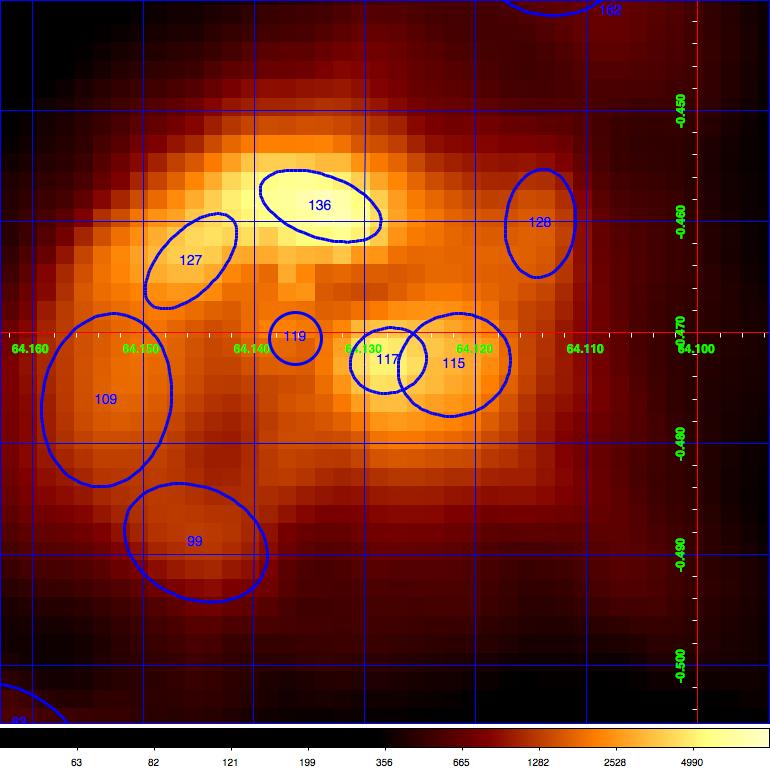} 
\includegraphics[width=0.4\textwidth]{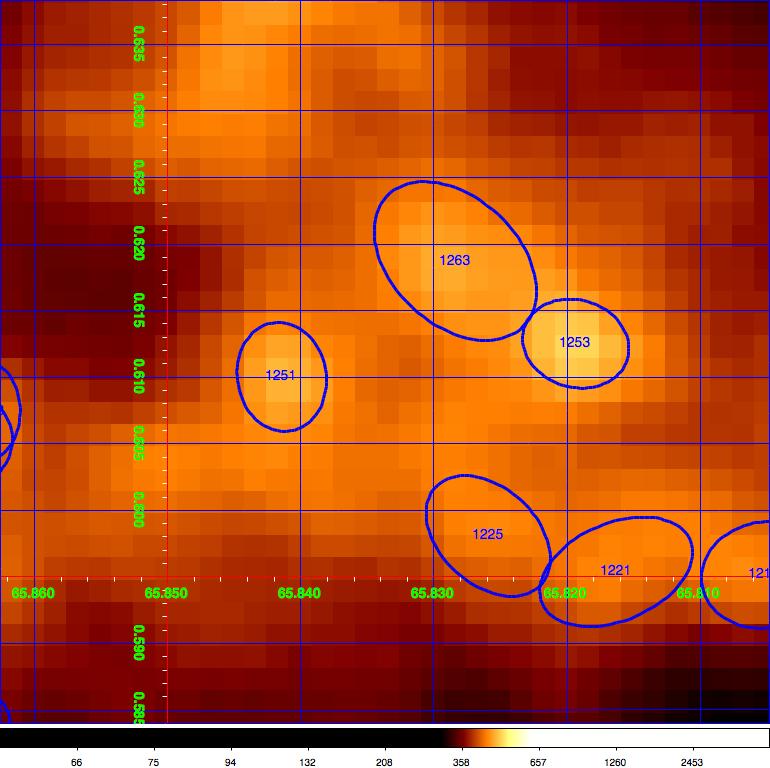}
\caption{\textbf{a} \textit{(left panel) )} Case of a source at 250\um , labeled as \# 117, in which the peak/background contrast is $\sim 1$ (hence relatively high, see fig.\ref{fpeak-back}) but the SNR is only 2.1. \textbf{b} \textit{(right panel) )} Case of a source at 250\um , labeled as \#1251, with a very good SNR of 13.5, but with peak/contrast ratio $\sim 0.15$.}
\label{strangecases}
\end{figure*}

The result is that even relatively very bright objects will have a limited 
S/N. We plot in Fig. \ref{f-df} the relationship between the integrated 
fluxes and their uncertainties.  These uncertainties are the estimated r.m.s. of the 
image residuals computed by subtracting the source as fitted, and integrating the residual
over the source's fitted area. We see that a large majority of the 
extracted sources have SNR$\geq 3$ (the blue line in the figure), but 
rarely does the SNR go above $\sim10$. This is the effect of the complex 
background that makes it difficult to estimate source sizes or even to 
effectively represent, in analytical form, the underlying background during 
the source fitting process.  Therefore, it is possible that even relatively good contrast sources may have low SNRs. For example, Fig. 
\ref{strangecases}a shows source \#117 in the 250\um\ band 
which has a peak/background contrast of $\sim 1$ (which  is relatively high 
compared to the average conditions represented in fig. \ref{fpeak-back}) 
but whose SNR is only $\sim 2$.  Yet, upon visual inspection, the source detection appears entirely reliable. This reinforces the notion that the quoted 
uncertainties should not be taken as a direct indication of the 
reliability of a source detection, but solely of the reliability of the 
estimated integrated flux. In other words, it may be difficult to estimate 
a high-fidelity flux even in the case of a bright source, given the 
intensity and complexity of the background found in the far-IR in the 
Galactic Plane.

One could be tempted then to adopt contrast value as a 
simple-to-use quality indicator for the reliability of a source.  
Unfortunately there are also several cases where relatively low contrast 
sources have high SNRs. This is demonstrated in fig. \ref{strangecases}b, 
where source  \#1251 has a contrast of $\sim$0.15 but a
SNR$\sim$13.5. Therefore,  for the present release, we find ourselves in the very difficult situation where it 
is not possible to define any combination of 
parameters that may offer a reliable "quality flag" for all detected sources. 
We therefore issue this first release of the Hi-GAL catalogues with a 
strong caveat;  for the moment there is no easy shortcut to identify 
the most reliable sources other than attempting combinations of various 
parameters (that likely may give good results in certain background 
conditions but bad results in others) followed by visual inspection of the 
maps. A blind selection of sources with high SNR will definitely result in 
reliable samples, but will certainly miss many reliable objects.

A helping hand in this respect may come from cross-matching sources 
in different bands. The green points in Fig. \ref{f-df} represent the 
subset of all sources for which a counterpart can be positionally matched 
(see \citealt{Elia+2016, Mart+2015}) in at least two adjacent wavelength bands. The 
fact that virtually all the green points are above the SNR=3 line is an 
indication that a positive match with counterparts in other bands is, at present, likely the best criterion to ensure the reliability 
of both the detection and the flux estimate for a source. Several sources 
that appear with high S/N at 70 and 160\um\ in fig. \ref{f-df} do not show 
counterparts in at least three Hi-GAL bands (i.e. the black dots above 
S/N=3).  For the greater part, these sources have relatively strong 
counterparts at shorter wavelengths and exhibit SEDs that decrease 
longward of 100\um\ and are not detected at SPIRE wavelengths. More 
complete statistics in this respect will be presented by \cite{Elia+2016} 
and \cite{Mart+2015} who will discuss the Hi-GAL photometric catalogues in 
the context of ancillary photometric Galactic Plane surveys like ATLASGAL 
\citep{Schuller2009}, MIPSGAL \citep{Carey+2009} and others. We emphasize 
once more that some of the sources with SNR$ \geq 3$ and with counterparts 
in three adjacent bands (the green points) have integrated fluxes below 
the completeness limit pertinent to the specific Galactic longitude if the 
source is located more than 0.3-0.4 degrees latitude on average off the 
midplane.

As experience accumulates in the use of these catalogues we plan to 
improve the quality assessment for the catalogue sources in subsequent data 
releases. Ultimately, since there may be  no  better instrument to judge the reliability of a source than an
astronomer's trained eye, a possible 
strategy could be to deploy machine-learning capabilities.  In such techniques,  
input from a trained user would teach the algorithm  to look for specific patterns in the combination of catalogue parameters, thereby 
allowing it to automatically identify sources that should be discarded.

\subsection{Contamination from extra-galactic sources}
\label{xgal_contamination}

Although the Galactic Plane is inarguably the most unfavourable 
environment in which to detect galaxies, there is no doubt that background galaxies 
could, in principle, contaminate the detection of Galactic sources in 
Galactic Plane surveys \citep{Marleau+2008, Amores+2012}. To evaluate the 
degree of possible contamination from galaxies in our photometric 
catalogues, we take advantage of the shallow cosmological surveys 
carried out by {\em Herschel} using the same observing mode that we used 
for \higal. \citet{Rigby+2011} report the photometric catalogues for the 
Science Demonstration Phase fields of the H-ATLAS survey with SPIRE 
\citep{Eales+2010}, showing that at 250\um\ the density of extragalactic 
sources with integrated flux larger than 0.1 Jy is of the order of 10 
deg$^{-2}$. The distribution of the integrated fluxes in the 250\um\ 
\higal\ catalogues reported in fig. \ref{hist_fint} shows that basically 
all (99.98\%) of the $\sim 280\,000$ sources detected at 250\um\ have 
fluxes above 0.1 Jy; as the present catalogue release encompasses a 
surveyed area of $\sim 270$ square degree, the average density of the 
250\um\ \higal\ sources is therefore $\sim$1000 deg$^{-2}$. The average 
contamination from extragalactic sources is, therefore,  
$\leq$1\%\ at 250\um. Using the same method, the contamination fractions 
at the other SPIRE wavelengths are $\leq$ 0.7\%\ at 350\um\ and 
$\leq$ 0.3\%\ at 500\um. Given the shape of extragalactic source counts 
\citet{Rigby+2011}, these estimated contaminations are concentrated toward the 
faint end of the \higal\ source catalogues.  The extragalactic source 
density decreases by 1 order of magnitude going from 0.1 to 0.4~Jy, while 
the number of \higal\ 250\um\ sources above 0.4~Jy is still 99.3\%\ of the 
total. Therefore, contamination effect from extragalactic background sources is 
 negligible and limited to integrated fluxes below 0.4Jy at 
250\um. The situation is even more favourable at the other SPIRE 
wavelengths. This is marginally visible in the histograms of Fig. 
\ref{hist_fint}.

Concerning the PACS bands, \cite{Lutz+2011} provide photometric 
catalogs from the PEP program which surveyed well-known consmological fields. 
About 125 sources with fluxes above 0.1Jy at 160\um\ are detected in the  2.78 sq. deg. 
PEP fields, corresponding to about 45 
deg$^{-2}$. Using the same approach as for the SPIRE bands, this 
corresponds to a contamination fraction from extragalactic sources of the 
order of 4\%. However, the PEP PACS maps were taken in Prime Mode with a 
scan speed of 20\asec /s, achieving much higher sensitivities than in 
Parallel Mode observations, especially in the virtually background-free
conditions typical of cosmological fields. \cite{Lutz+2011} quote 
3$\sigma$ noise levels of 8~mJy, which agrees very well with the expected 
noise levels predicted by the HSpot tool for the PEP observing mode at the 
centre of the maps where the coverage is higher. On the other hand, the 
HSpot tool provides a 1$\sigma$ sensitivity of $\sim$46~mJy at 160\um\ for 
observations in Parallel Mode, or a factor $\sim$16 worse than for 
observations in Prime Mode. If we artificially degrade  the 
flux uncertainties reported by \cite{Lutz+2011} by this factor, the number of sources 
that would have been detected at a 3$\sigma$ level would decrease from 125 
to 29, bringing the contamination level down to $\leq$1\%. PEP 70\um\ 
source catalogues have been made available for the GOODS-S field only 
\citep{Lutz+2013}, and only 1 source has been detected with a flux greater 
than 0.1~Jy, which is not a large enough sample with which to assess possible contamination.  
Given this single detection, however, we deem the contamination to be negligible in this band.

We conclude that contamination from distant background galaxies is 
extremely low and concentrated toward the faint end of the flux 
distribution of the \higal\ sources.  This effect is, perhaps, visible in 
Fig. \ref{hist_fint} as a tentative flattening of the flux distributions 
at F$_{int} \leq $0.1~Jy for the 160, 250 and 350\um\ bands. Local 
universe galaxies have larger fluxes, but are also far from compact and 
have a very low spatial density (see \cite{Ciesla+2012}  and \citealt{Boselli+2010})
Therefore,  it is unlikely that they have been included in the present catalog.

\subsection{Source sizes}
\label{par_size}

\begin{figure}[h]
\centering
\includegraphics[width=0.5\textwidth]{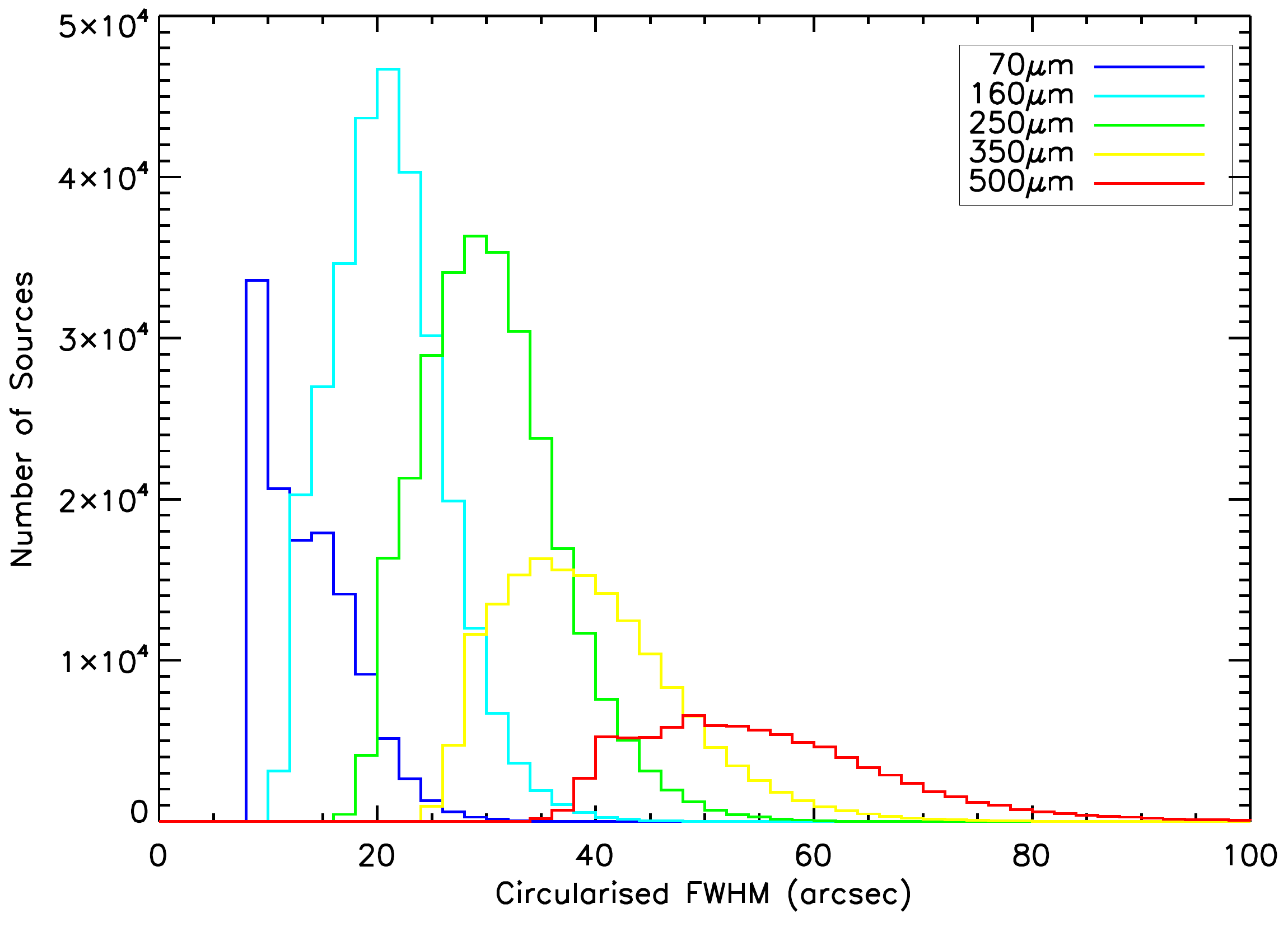}
\includegraphics[width=0.5\textwidth]{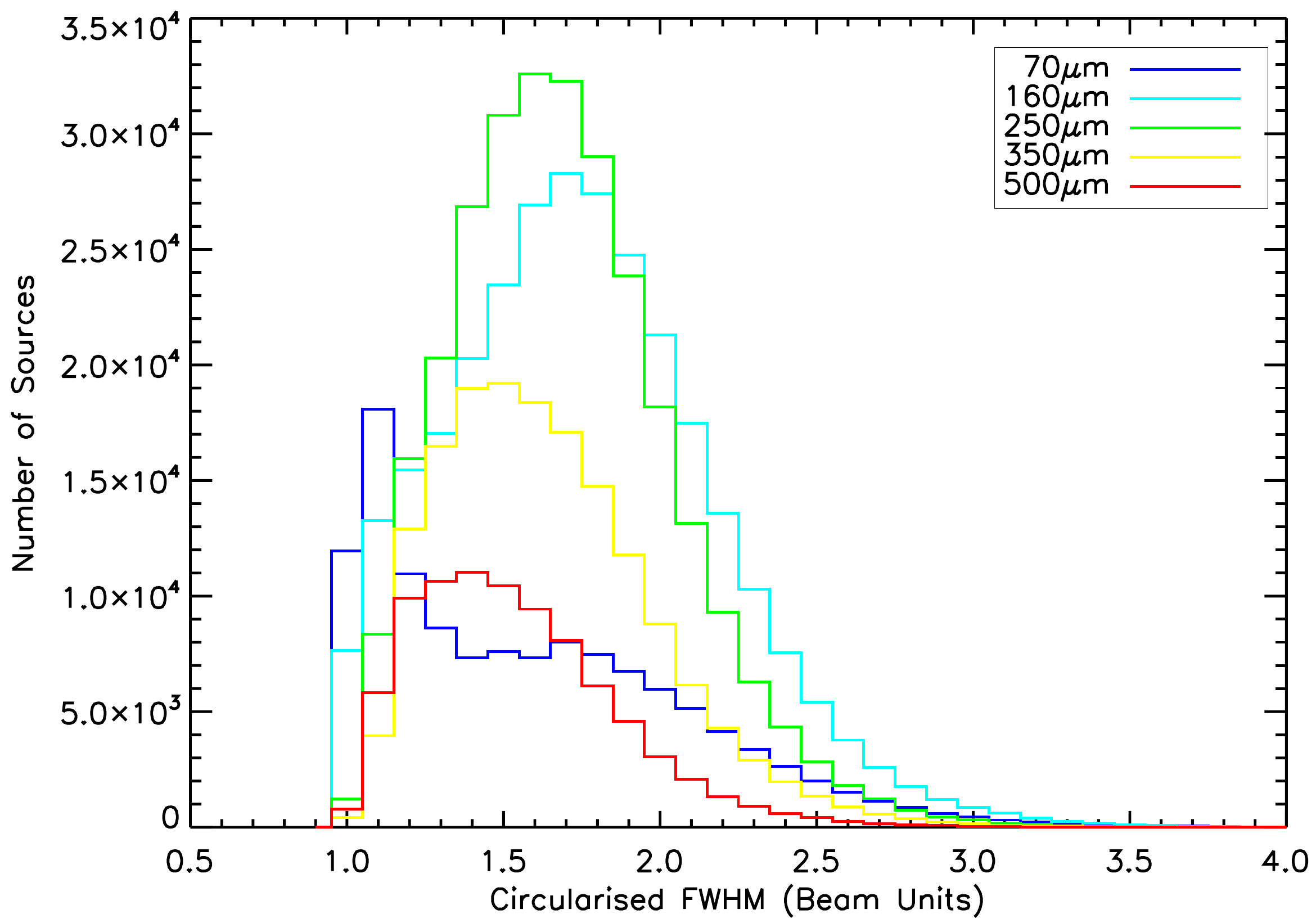}
\caption{\textit{Top-panel}) Distribution of the circularized FWHM of the catalogue sources in the five bands. Sizes are computed as the geometric mean of the FWHMs estimated by  2D Gaussian fitting in two orthogonal directions. \textbf{\textit{Bottom-panel}) Same as above, but in units of the beam-size}.}
\label{sourcesize}
\end{figure}

\begin{figure}[h]
\centering
\includegraphics[width=0.5\textwidth]{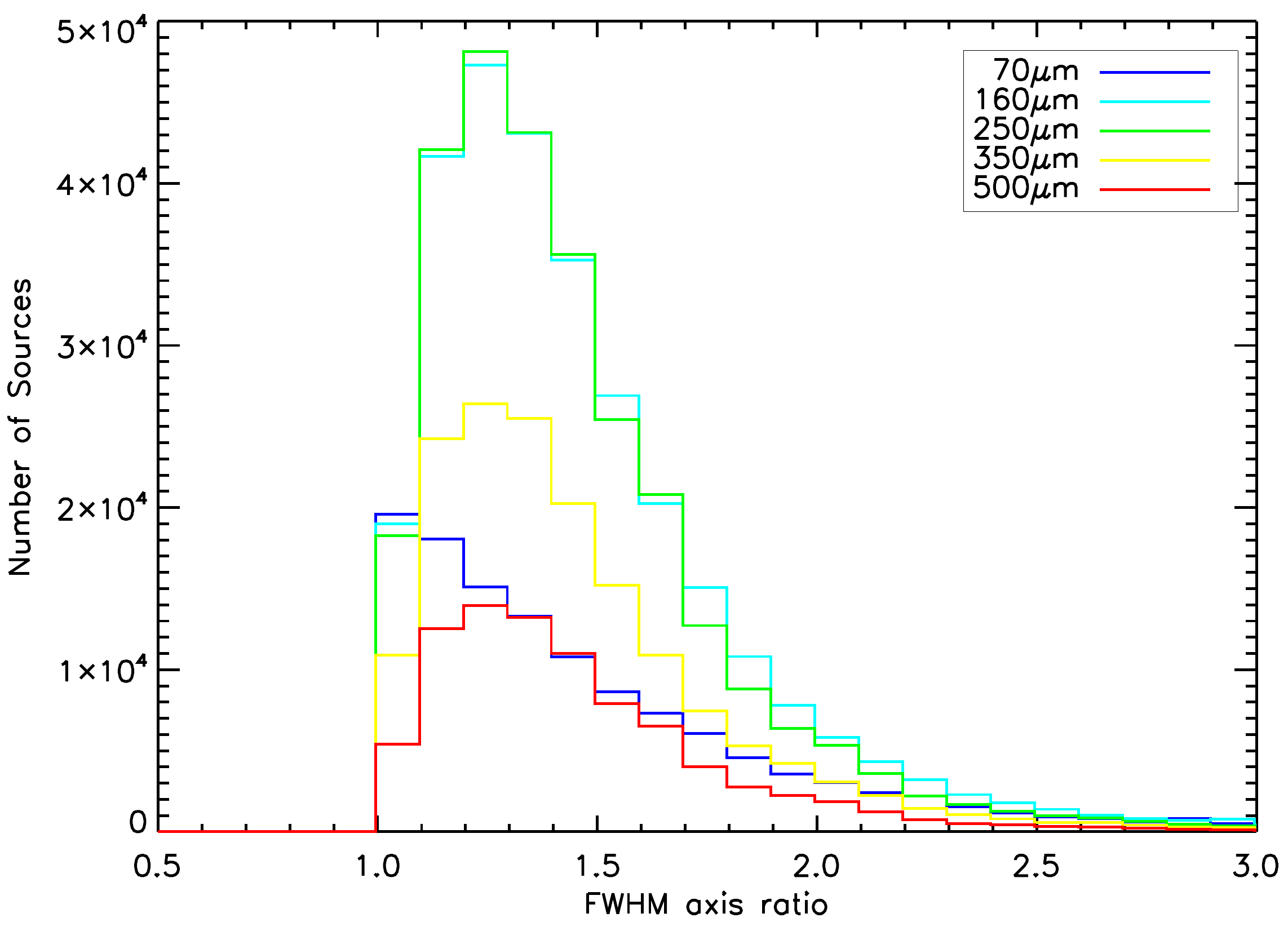}
\caption{Distribution of the axis ratio, computed as the ratio of 
FWHM$_{Maj}$/FWHM$_{min}$ of the catalogued sources in the five bands.}
\label{sourceellipse}
\end{figure}

Fig. \ref{sourcesize} reports the distribution of the circularized source sizes 
 in the different bands, calculated by taking the square root of the 
product of the major and minor axis as estimated by \cutex. Source sizes 
span a range of values, from that of the PSF to about twice the PSF for 
most of the sources. The broad distributions in Fig. \ref{sourcesize} show 
that the sources are generally mildly resolved, with the exception of the 
70\um\ band where a peak at the PSF value is clearly visible. We stress 
that Fig. \ref{sourcesize} reports the circularised sizes; sources may be 
unresolved in one direction and resolved in the other, therefore resulting in being moderately
resolved on average. This is confirmed by Fig. \ref{sourceellipse} showing 
that extracted sources are mildly elliptical, with axis ratios peaking 
between 1.2 and 1.3, and with the large majority of the sources showing 
values below 1.5.

\begin{figure*}[!t]
\centering
\includegraphics[width=0.47\textwidth]{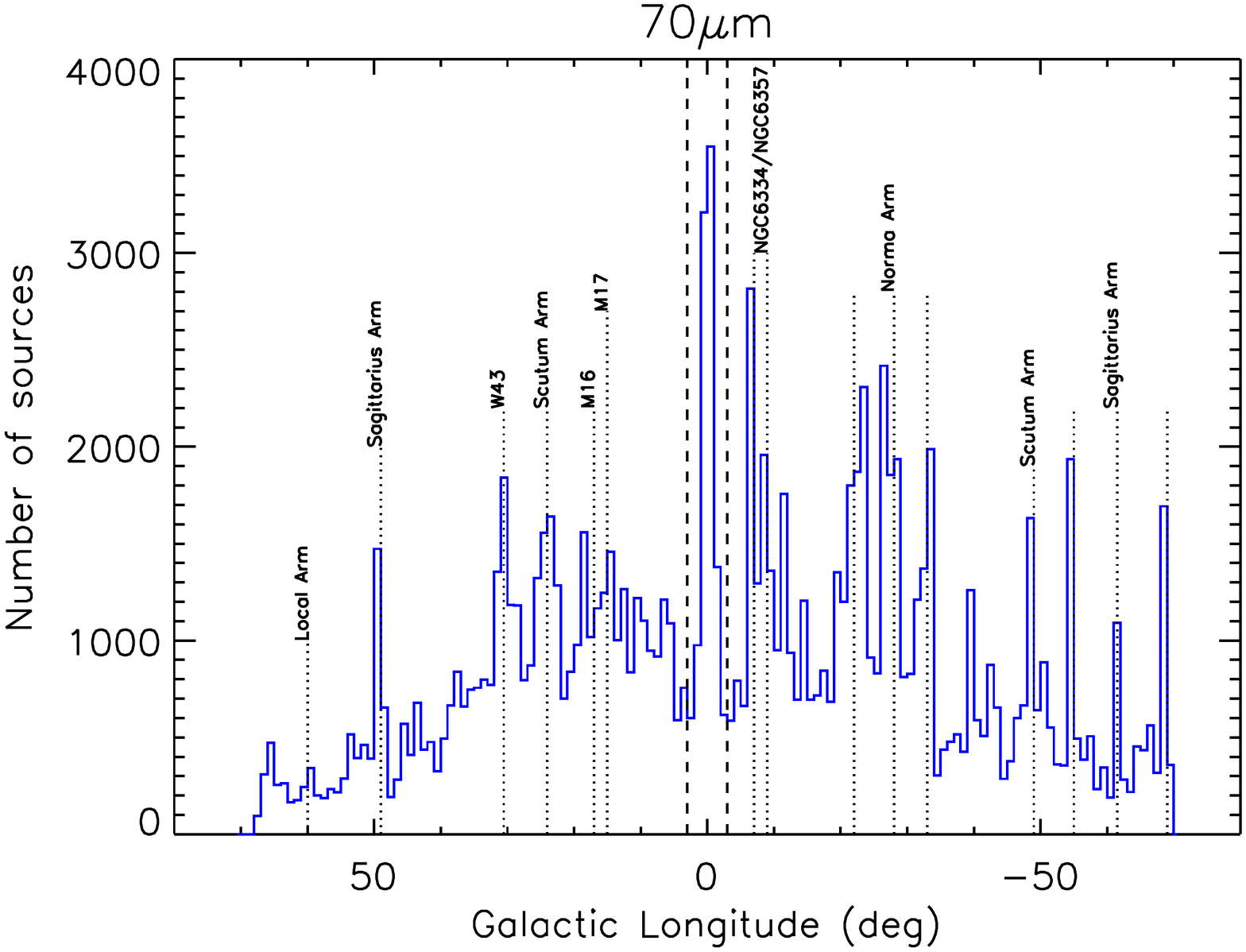} 
\includegraphics[width=0.47\textwidth]{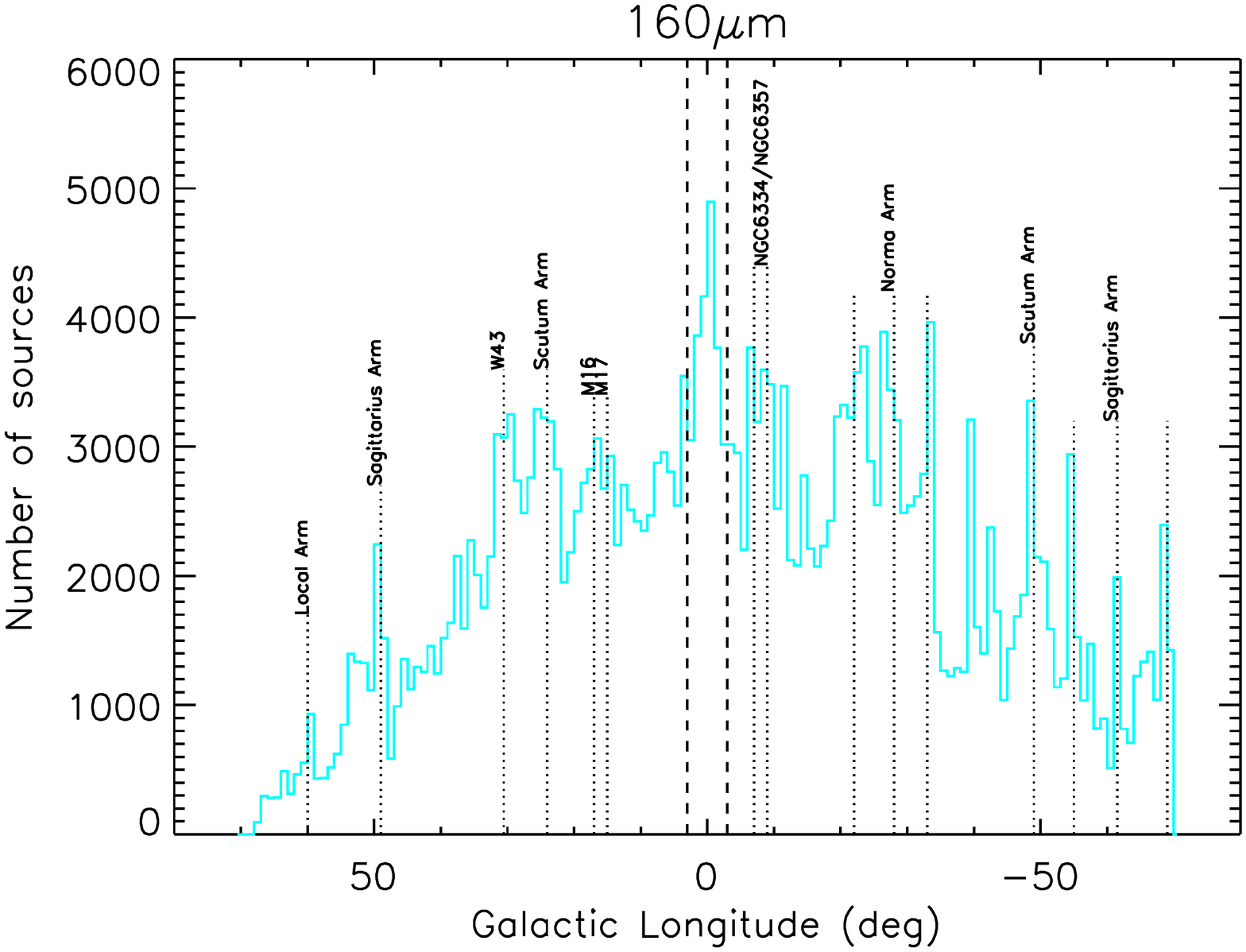} 
\includegraphics[width=0.47\textwidth]{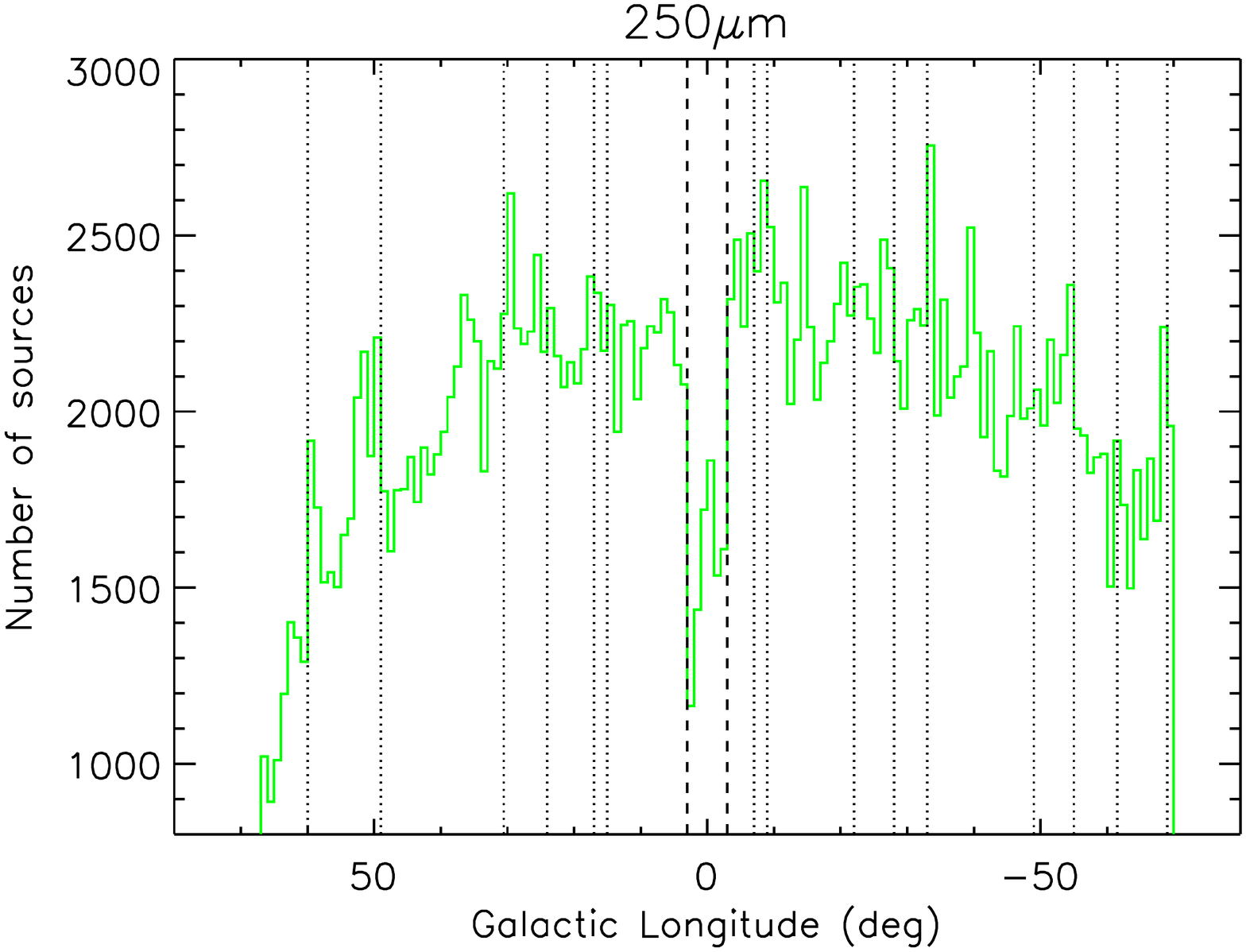} 
\includegraphics[width=0.47\textwidth]{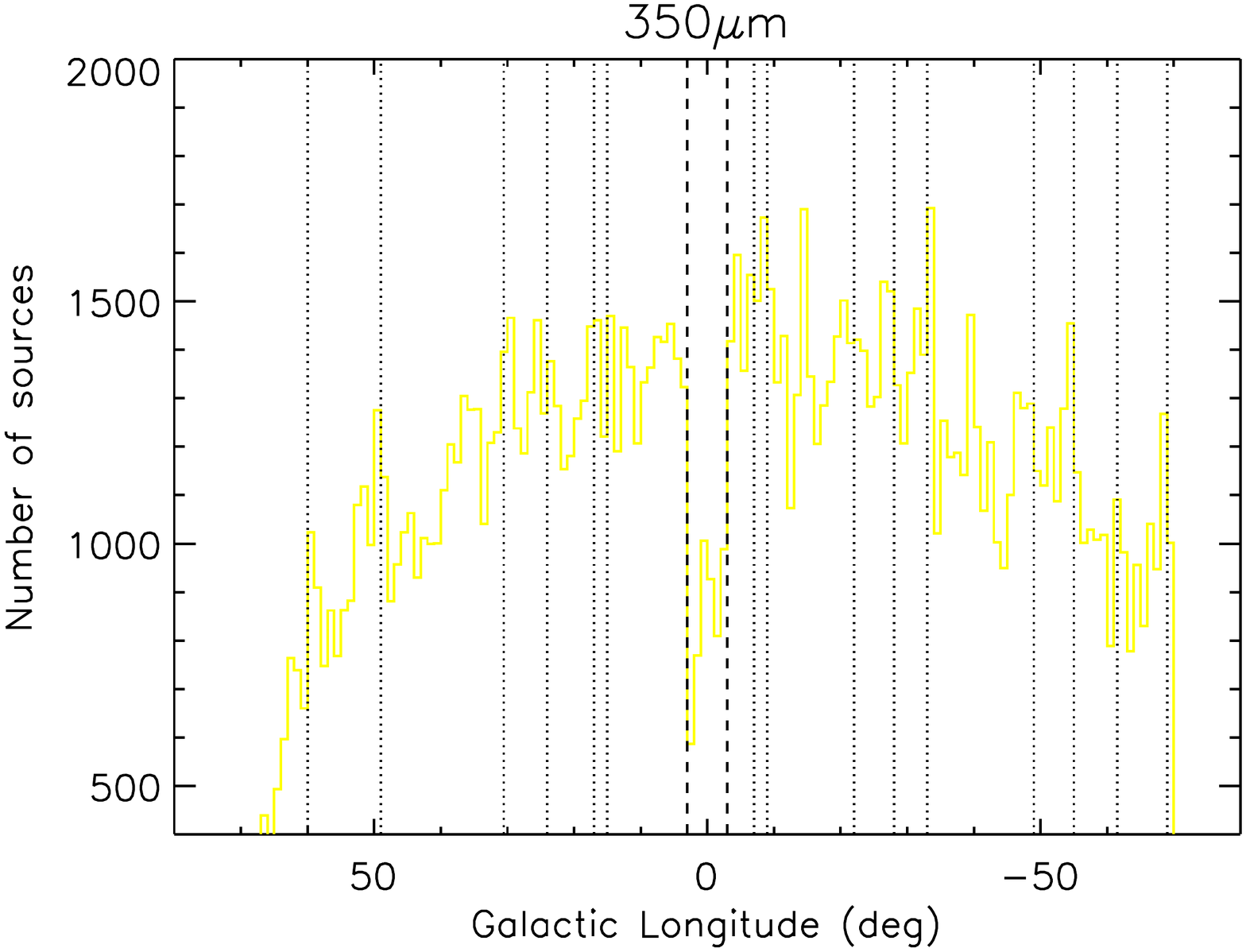} 
\includegraphics[width=0.47\textwidth]{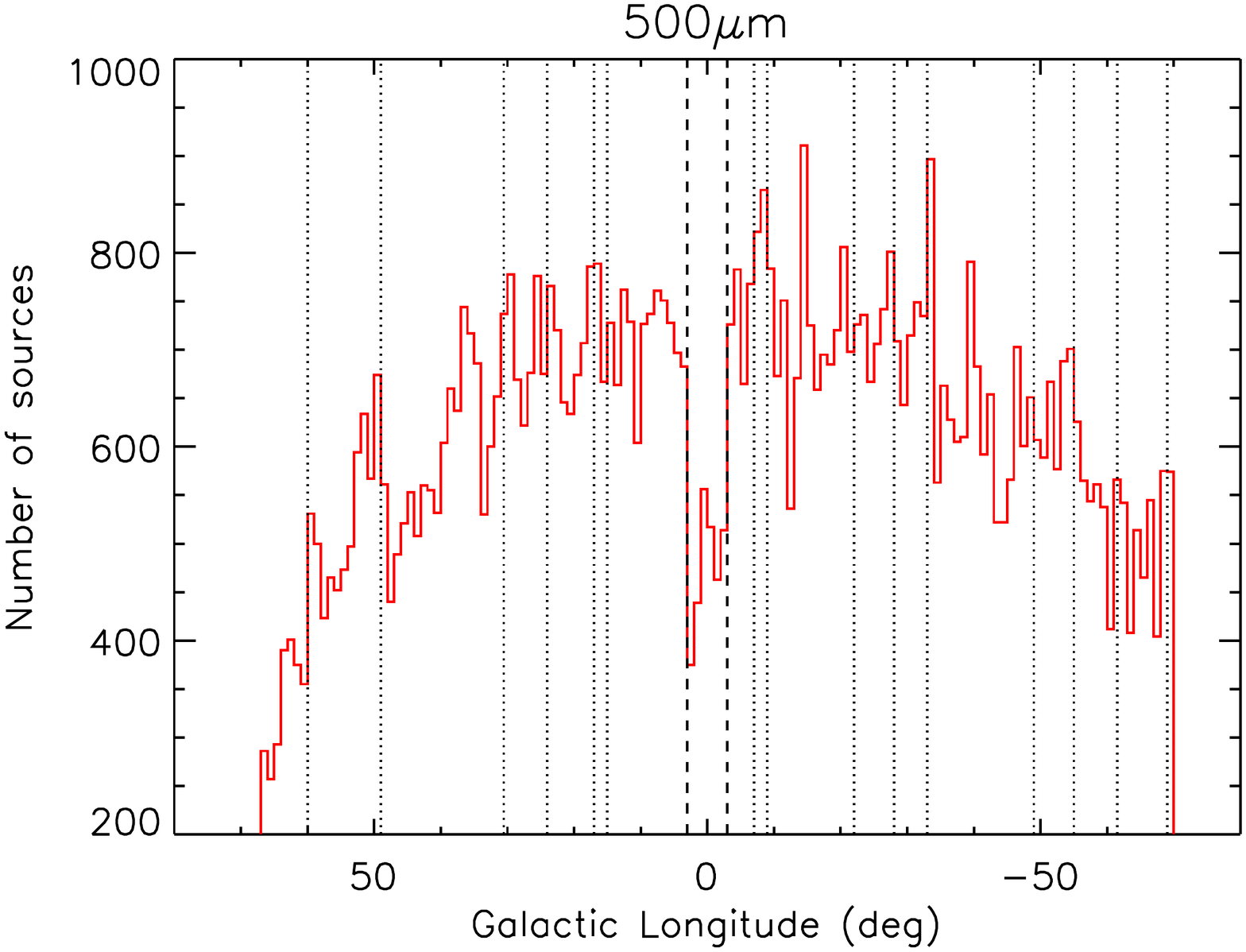}  
\caption{Longitude distribution of source counts (in 1\adeg\ bins) at 70, 
160, 250, 350 and 500\um\ colour-coded as in fig. \ref{CompleteGal}. The 
vertical dashed lines denote the longitude range close to the Galactic 
centre where SPIRE was used in ``bright source" mode to avoid saturation 
and non-linearities in detector response. Labeled dotted lines indicate 
major spiral arms, tangent point/intersections or star forming complexes.}
\label{glondistr_all}
\end{figure*}

It is not surprising to find such a low number of PSF-like sources in 
the \higal\ catalogues. Compact dust clumps around young star forming objects 
do not show abrupt transitions in density when they merge into the ISM 
filaments or clouds in which they are embedded, so there is no reason 
\textit{a priori} to expect these objects to be unresolved. The physical 
size of dense clumps hosting protoclusters, on average between 0.1 and 1 
parsecs, should indeed be resolvable for a large span of heliocentric 
distances with the angular resolutions accessible to the {\em Herschel} 
cameras. This aspect will be discussed in more detail in \cite{Elia+2016}. On the 
other hand, it has been noted several times that the intensity of extended 
emission background in the DR1 \higal\ maps is generally higher (fig. 
\ref{fpeak-back}) than the peak flux of the detected compact sources. The 
ideal flat and faint background conditions of the $\alpha$~Boo image that 
was used in \S\ref{photchecks} to calibrate the departures of the 
brightness profile for point-like and mildly extended compact sources, are 
never found on the Galactic Plane at {\em Herschel} wavelengths, 
with the exception of some spots in the most peripheral tiles at 70\um\ . 
Under these conditions, it is certainly difficult for any adaptive 
brightness profile-fitting algorithm to converge to PSF-like source sizes.

\section{Global properties of the Galactic structure.}


Figure \ref{glondistr_all} shows the distribution  of Hi-GAL sources in 
Galactic longitude for the five wavelength bands. All 
histograms show decreasing source counts as a function of distance from the 
Galactic Centre, comparing very well with similar plots from other 
infrared and submillimetre surveys. A variety of peaks  can be 
seen throughout the longitude range, with greater dynamic range for the 70 
and 160\um\ bands. The abrupt dips in source count over the 6\adeg -wide region centred on 
the Galactic centre which are clearly visible in the 
SPIRE bands  is due to the fact that SPIRE was used in ``bright 
source'' mode for the three tiles of the survey close to the Galactic 
Centre (see \S\ref{observations}).

Similar to \cite{Beuther2012} we identify on fig. \ref{glondistr_all} 
features that can be associated with major star formation complexes or to 
source accumulations along the line of sight correspondening to 
tangent points or major intersections of the line of sight with known 
spiral arms. 

A comparison with the ATLASGAL survey \citep{Schuller2009}, covering the Galactic Plane 
between roughly +60\adeg and $-60$\adeg in the 870\um\ continuum, shows substantial similarities in the source count
distributions, confirming that both surveys are mostly tracing dense, 
star-forming (or potentially star-forming) regions. A similar distribution was also found by \cite{Rygl+2010} using high-extinction clouds identified with Spitzer colours excess
Therefore, it is reasonable to posit that ATLASGAL is typically tracing the higher-flux 
fraction of the Hi-GAL sources, although this depends on the 
intrinsic SEDs of the various objects. We defer a detailed analysison on this topic to a subsequent paper \citep{Elia+2016}. .

\begin{figure}[h]
\centering
\includegraphics[width=0.5\textwidth]{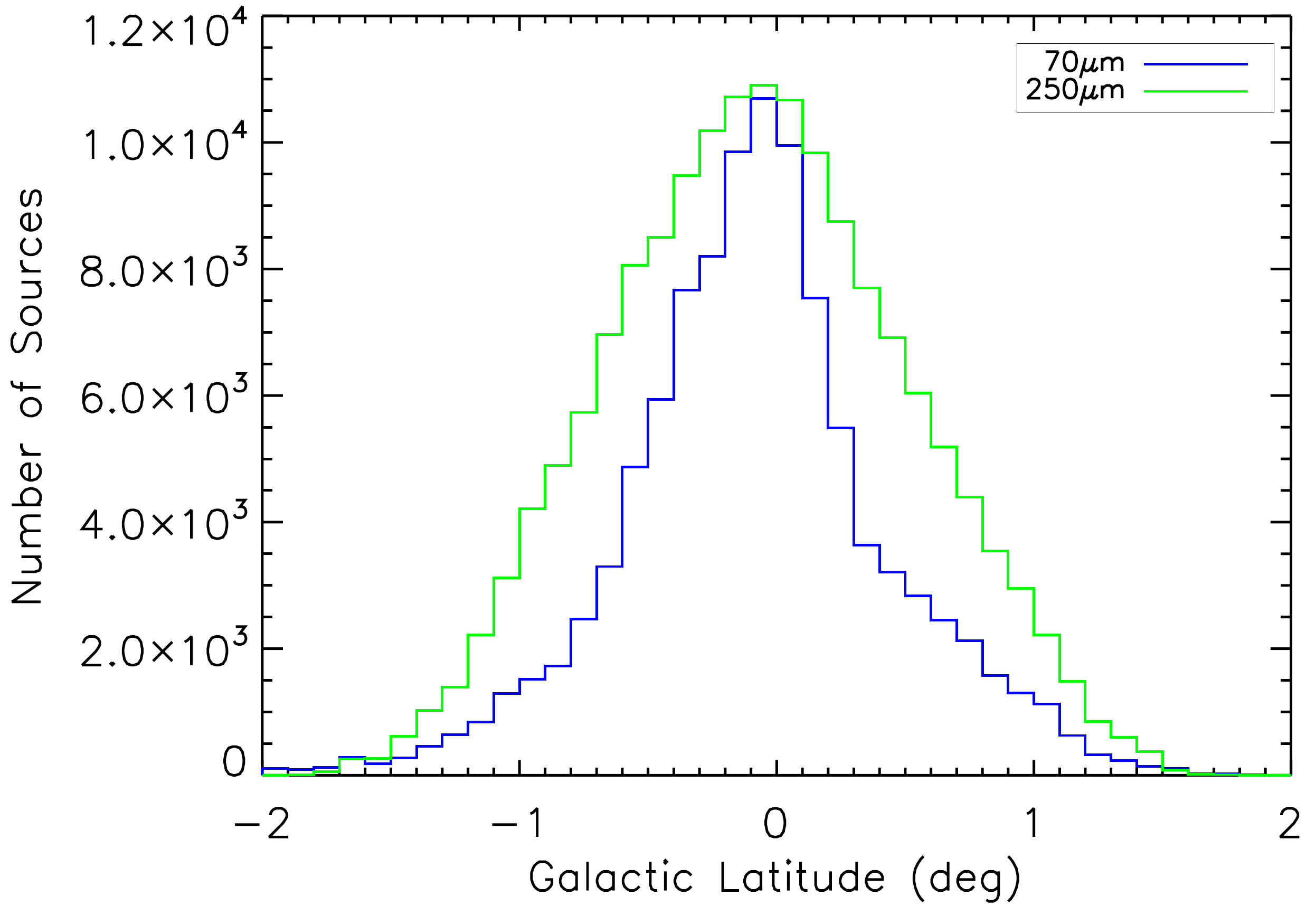}
\caption{Distribution of Galactic latitude values for the Hi-GAL sources with F$_{Int} \geq$0.5Jy at 70\um\ (blue line), and with F$_{Int} \geq$3.0Jy at 250\um\ (green line).}
\label{hist_glat}
\end{figure}

The latitude distribution of the Hi-GAL compact sources is reported in fig. \ref{hist_glat} for the 70\um\ and the 250\um\ catalogue sources. Both histograms peak at slightly negative values, similar to what was recently reported for the ATLASGAL submillimetre sources distribution as well as for other infrared and molecular line data Galactic plane surveys (see \citealt{Beuther2012} and references therein).  The median values for Hi-GAL source latitude is $\sim -$0\adeg .06 below the nominal midplane, in excellent agreement with the value reported for ATLASGAL by \cite{Beuther2012}. We therefore confirm that the current definition of the Galactic midplane may need to be revisited to account for a latitude shift that most likely reflects an overall bias; possibly due to an incorrect assumption of the Sun's vertical position in the Milky Way. A more in-depth and statistically significant analysis of the latitude distribution of the \higal\ sources as a function of longitude  is deferred to a companion paper \citep{Molinari+2015}.

\section{Conclusions}
\label{conclusions}

This is the first public Data Release of high-quality products from the 
{\em Herschel} \higal\ survey. The release comes 2 years after the end of the 
{\em Herschel} observing, and is the result of extensive testing of the data reduction and extraction procedures created by members of the Hi-GAL Consortium. The complexity and the large variation 
of the background conditions in all Herschel wavelength bands makes source extraction on the Galactic Plane a non-trivial task. With Hi-GAL DR1, we provide access, via a cutout service, to high-quality images and Compact Source Catalogues for the Galactic Plane at 70, 160,250, 350 and 500\um\ in the region 68\adeg $\gtrsim l \gtrsim -70$\adeg\ and $|b| \leq1$\adeg. the service is accessible from the VIALACTEA Project portal at \url{http://vialactea.iaps.inaf.it}.

The catalogues were generated using the CuTEx software package that was specifically designed to operate in the  intense and highly spatially variable background conditions found in the Galactic Plane at Far-infrared wavelengths. Source detection is carried out on the  2$^{nd}$-derivative of the brightness images which is particularly sensitive to curvature in the continuum brightness spatial distribution.  The detection is optimised for compact objects with FWHM typically ranging from 1 to three times the instrumental PSF(but mostly within 2 times the PSF). The impact of false positives is estimated, and a careful analysis of the flux completeness limits is presented, independently for each photometric band. The source catalogues contain 123210, 308509, 280685, 160972 and 85460 sources in the five bands, respectively.

After considerable time, effort and experience gathered through use of the photometric 
calatogues by \higal\ Consortium astronomers, we are not yet at a stage 
where we feel we can confidently define a figure of merit that can 
uniquely and definitively be used to assess the degree of reliability of a 
source detection. Thresholding on the S/N ratio that we assign to each source 
appears to be the best way to select \textit{bona fide} compact objects.  
Although the user should be warned that, due to the complex background 
conditions, there may be sources with a formal S/N$<$3 that have relatively 
good contrast ratios over the background and appear reliable on visual 
inspection. An additional crriterion to assess the reliability of sources 
is their persistence in other adjacent photometric bands.

Subsequent releases are planned that will cover the entire Galactic Plane, 
with even higher quality catalogues based on improved handling of the 
problems present in source extraction for variable-size objects in extreme 
background conditions. Additional products will include carefully 
intercalibrated large map mosaics and dust column density maps.

\begin{acknowledgement}
We thank an anonymous referee for valuable comments that led to an improvement of the original manuscript.
This work is part of the VIALACTEA Project, a Collaborative Project under Framework Programme 7 of the European Union, funded under Contract \# 607380 that is hereby acknowledged.  {\it Herschel} Hi-GAL data processing, maps production and source catalogue generation is the result of a multi-year effort that was initially funded thanks to Contracts I/038/080/0 and I/029/12/0 from ASI, Agenzia Spaziale Italiana. 

{\it Herschel} is an ESA space observatory with science instruments provided by European-led Principal Investigator consortia and with important participation from NASA. 

PACS has been developed by a consortium of institutes led by MPE (Germany) and including UVIE (Austria); KUL, CSL, IMEC (Belgium); CEA, OAMP (France); MPIA (Germany); IAPS, OAP/OAT, OAA/CAISMI, LENS, SISSA (Italy); IAC (Spain). This development has been supported by the funding agencies BMVIT (Austria), ESA-PRODEX (Belgium), CEA/CNES (France), DLR (Germany), ASI (Italy), and CICYT/MCYT (Spain). 

SPIRE has been developed by a consortium of institutes led by Cardiff Univ. (UK) and including Univ. Lethbridge (Canada); NAOC (China); CEA, LAM (France); IAPS, Univ. Padua (Italy); IAC (Spain); Stockholm Observatory (Sweden); Imperial College London, RAL, UCL-MSSL, UKATC, Univ. Sussex (UK); Caltech, JPL, NHSC, Univ. colourado (USA). This development has been supported by national funding agencies: CSA (Canada); NAOC (China); CEA, CNES, CNRS (France); ASI (Italy); MCINN (Spain); Stockholm Observatory (Sweden); STFC (UK); and NASA (USA).
\end{acknowledgement}


\appendix

\newcommand{\sizecola}{2.5cm}
\newcommand{\sizecolb}{0.5cm}
\newcommand{\sizecolc}{1.5cm}

\section{Explanatory note for the Hi-GAL photometric catalogs}
\label{cat_expl}

\begin{table*}[h]
   \centering
    \caption{Field description for the single-band photometric catalogs.}
   \begin{tabular}{llll} \hline\hline
      Field name    & Format & Units & Description \\ \hline
      \multicolumn{4}{c}{\textbf{Source identification and position information}}\\ \hline
\parbox{\sizecola}{DESIGNATION \\ \\ \\ \\}& \parbox{\sizecolb}{A25 \\ \\ \\ \\ } & \parbox{\sizecolc}{$-$ \\ \\ \\ \\ }&  \parbox{12cm}{Designation of the source based on its Galactic position in the form LLL.llll$\pm$b.bbbb. The naming convention for the Hi-GAL cataloguehas the form HIGALPXLLL.llll$\pm$b.bbbb, where HIGALP stands for the preliminary catalog, X stands for the band where the source has been identified among the possible choices: B - blue band; R - red band; S - PSW band; M - PMW band; L - PLW band.} \\
\parbox{\sizecola}{GLON} & F12.6 & degrees & Galactic longitude of the source \\
\parbox{\sizecola}{GLAT} & F12.6 & degrees & Galactic latitude of the source \\
\parbox{\sizecola}{DGLON \\ } & \parbox{\sizecolb}{F5.2 \\ } & \parbox{\sizecolc}{arcsec \\ } & \parbox{12cm}{Uncertainty in the Galactic longitude coordinate derived from the fitting procedure. 0 indicates that the fitting process hit the boundary limits imposed to the fit.} \\
\parbox{\sizecola}{DGLAT \\ } & \parbox{\sizecolb}{F5.2 \\ } & \parbox{\sizecolc}{arcsec \\ } & \parbox{12cm}{Uncertainty in the Galactic latitude coordinate derived from the fitting procedure. 0 indicates that the fitting process hit the boundary limits imposed to the fit.} \\
\parbox{\sizecola}{RA} &  \parbox{\sizecolb}{F12.6} & \parbox{\sizecolc}{degrees} & \parbox{12cm}{J2000 Right Ascension for the source} \\
\parbox{\sizecola}{DEC} &  \parbox{\sizecolb}{F12.6} & \parbox{\sizecolc}{degrees} & \parbox{12cm}{J2000 Declination for the source} \\
\parbox{\sizecola}{ATLAS\_IMAGE} &  \parbox{\sizecolb}{A40} & \parbox{\sizecolc}{$-$} & \parbox{12cm}{Atlas Image file identifier from which source was extracted.}\\
\parbox{\sizecola}{X} &  \parbox{\sizecolb}{F9.3} & \parbox{\sizecolc}{pixel} & \parbox{12cm}{x-pixel coordinate of this source on the original image.} \\
\parbox{\sizecola}{Y} &  \parbox{\sizecolb}{F9.3} & \parbox{\sizecolc}{pixel} & \parbox{12cm}{y-pixel coordinate of this source on the original image.} \\
\parbox{\sizecola}{DX} &  \parbox{\sizecolb}{F9.3} & \parbox{\sizecolc}{pixel} & \parbox{12cm}{Uncertainty in the x coordinate of this source derived from the fitting procedure.}\\
\parbox{\sizecola}{DY} &  \parbox{\sizecolb}{F9.3} & \parbox{\sizecolc}{pixel} & \parbox{12cm}{Uncertainty in the y coordinate of this source derived from the fitting procedure.}\\
\parbox{\sizecola}{SOURCE\_ID \\ \\ }  & \parbox{\sizecolb}{A10 \\ \\ }  & \parbox{\sizecolc}{$-$ \\ \\ } & \parbox{12cm}{Unique source identification in the form lLLL NNNN, where lLLL is a unique identifier of the original image over which source extraction was carried out, and NNNN is a progressive 4 digit, zero-filled, number indicating the sequential order of extraction.}\\ \hline
\multicolumn{4}{c}{\textbf{Primary Photometric information}} \\ \hline
\parbox{\sizecola}{FINT\_UNCORR} &  \parbox{\sizecolb}{F15.3} & \parbox{\sizecolc}{Jy} & \parbox{12cm}{Source Integrated flux measured from the fitting process} \\
\parbox{\sizecola}{FINT \\ \\} & \parbox{\sizecolb}{F15.3 \\ \\} & \parbox{\sizecolc}{Jy \\ \\} & \parbox{12cm}{Source Integrated flux measured from the fitting process after applying photometric corrections as a function of the source size, to account for source non-Gaussianity and for scan speed.}\\
\parbox{\sizecola}{ERR\_FINT \\ \\} & \parbox{\sizecolb}{F15.3 \\ \\} & \parbox{\sizecolc}{Jy \\ \\} & \parbox{12cm}{Uncertainty on the integrated flux computed by multiplicating the fitted source residual r.m.s. (RMS\_TOTAL) by the fitted source area as estimated by FWHMA and FWHMB.} \\
\parbox{\sizecola}{FPEAK} &  \parbox{\sizecolb}{F15.3} & \parbox{\sizecolc}{MJy/sr} & \parbox{12cm}{Source peak flux measured from the fitting process} \\
\parbox{\sizecola}{FWHMA \\} & \parbox{\sizecolb}{F10.2 \\} & \parbox{\sizecolc}{arcsec \\} & \parbox{12cm}{Full Width Half Maximum of the source along axis $a$ of the elliptical Gaussian as determined by fitting engine.} \\
\parbox{\sizecola}{FWHMB \\} & \parbox{\sizecolb}{F10.2 \\} & \parbox{\sizecolc}{arcsec \\} & \parbox{12cm}{Full Width Half Maximum of the source along axis $b$ of the elliptical Gaussian as determined by fitting engine.} \\
\parbox{\sizecola}{PA} & \parbox{\sizecolb}{F6.1} & \parbox{\sizecolc}{degrees}  & \parbox{12cm}{Position angle of the elliptical Gaussian (N$\rightarrow$E).} \\
\parbox{\sizecola}{BACKGROUND} & \parbox{\sizecolb}{F15.3} & \parbox{\sizecolc}{MJy/sr} & \parbox{12cm}{Background value determined at the source peak position.} \\
\parbox{\sizecola}{BACK\_ACOEFF \\} & \parbox{\sizecolb}{F12.5 \\} & \parbox{\sizecolc}{MJy/sr \\} & \parbox{12cm}{Coefficient \textbf{a} of the 0$^{th}$-order term of the background obtained by the fit at the source peak position.}\\
\parbox{\sizecola}{BACK\_BCOEFF \\} & \parbox{\sizecolb}{F12.5 \\} & \parbox{\sizecolc}{MJy/sr/pxl \\} & \parbox{12cm}{Coefficient \textbf{b} of the 1$^{th}$-order term $x$  of the background obtained by the fit at the source peak position.} \\
\parbox{\sizecola}{BACK\_CCOEFF \\} & \parbox{\sizecolb}{F12.5 \\}& \parbox{\sizecolc}{MJy/sr/pxl \\} & \parbox{12cm}{Coefficient \textbf{c} of the 1$^{th}$-order term $y$  of the background obtained by the fit at the source peak position.} \\
\parbox{\sizecola}{BACK\_DCOEFF \\} &  \parbox{\sizecolb}{F12.5 \\} &  \parbox{\sizecolc}{MJy/sr/pxl$^2$ \\} & \parbox{12cm}{Coefficient \textbf{d} of the 2$^{nd}$-order term $x^2$ of the background obtained by the fit at the source peak position.}\\
\parbox{\sizecola}{BACK\_ECOEFF \\} & \parbox{\sizecolb}{F12.5 \\} & \parbox{\sizecolc}{MJy/sr/pxl$^2$ \\} & \parbox{12cm}{Coefficient \textbf{e} of the 2$^{nd}$-order term $y^2$ of the background obtained by the fit at the source peak position.}\\
\parbox{\sizecola}{BACK\_FCOEFF \\} & \parbox{\sizecolb}{F12.5 \\} & \parbox{\sizecolc}{MJy/sr/pxl$^2$ \\} & \parbox{12cm}{Coefficient \textbf{f} of the 2$^{nd}$-order term $xy$ of the background obtained by the fit at the source peak position.}\\
\parbox{\sizecola}{RMS\_TOTAL \\ } & \parbox{\sizecolb}{F12.5 \\ } & \parbox{\sizecolc}{MJy/sr \\ } & \parbox{12cm}{Standard Deviation, $\sigma _{loc}$, of the residuals computed within the source area defined by FWHMA and FWHMB after the subtraction of the best fit.} \\
\parbox{\sizecola}{RMS\_SURROUND \\ \\} & \parbox{\sizecolb}{F12.5 \\ \\} & \parbox{\sizecolc}{MJy/sr \\ \\} & \parbox{12cm}{Standard Deviation, $\sigma _{loc}$, of the residuals computed within the fitting window after the subtraction of the best fit, excluding both the pixels that belong to the source and the pixels belonging to other sources that fall within the fitting window.} \\
\parbox{\sizecola}{SNR \\}& \parbox{\sizecolb}{F12.5 \\} & \parbox{\sizecolc}{$-$ \\} & \parbox{12cm}{Signal-to-noise ratio obtained by dividing FPEAK by the residual r.m.s. over a source area with FWHMA and FWHMB as \textit{semi-axes}}. \\ \hline
\multicolumn{4}{c}{\textbf{Basic detection information}} \\  \hline
\parbox{\sizecola}{DET\_X \\ \\} & \parbox{\sizecolb}{F10.3 \\ \\ } & \parbox{\sizecolc}{$-$ \\ \\} & \parbox{12cm}{Relevance of the source in the 2$^{nd}$ derivative map along $x$-axis defined as the ratio between the measured 2$^{nd}$ derivative at source peak position and the adopted local threshold value} \\   
\parbox{\sizecola}{DET\_Y \\ \\} & \parbox{\sizecolb}{F10.3 \\ \\ } & \parbox{\sizecolc}{$-$ \\ \\ } & \parbox{12cm}{Relevance of the source in the 2$^{nd}$ derivative map along $y$-axis defined as the ratio between the measured 2$^{nd}$ derivative at source peak position and the adopted local threshold value} \\   
\hline
   \end{tabular}
   \label{cat_descr}
\end{table*}

\addtocounter{table}{-1}
\begin{table*}[h]
   \centering
    \caption{Continued}
   \begin{tabular}{llll} \hline\hline
      Field name    & Format & Units & Description \\ \hline
\multicolumn{4}{c}{\textbf{Basic extraction information}} \\ 
\hline
\parbox{\sizecola}{DET\_X45 \\ \\} & \parbox{\sizecolb}{F10.3  \\ \\} & \parbox{\sizecolc}{$-$ \\ \\} & \parbox{12cm}{Relevance of the source in the 2$^{nd}$ derivative map along the bisector of the $xy$-axis defined as the ratio between the measured 2$^{nd}$ derivative at source peak position and the adopted local threshold value} \\   
\parbox{\sizecola}{DET\_Y45 \\ \\} & \parbox{\sizecolb}{F10.3  \\ \\} & \parbox{\sizecolc}{$-$ \\ \\} & \parbox{12cm}{Relevance of the source in the 2$^{nd}$ derivative map along the bisector of the $yx$-axis defined as the ratio between the measured 2$^{nd}$ derivative at source peak position and the adopted local threshold value} \\   
\parbox{\sizecola}{DETLIM\_X \\ } & \parbox{\sizecolb}{F10.3  \\ } & \parbox{\sizecolc}{MJy/sr/pxl$^2$ \\ } & \parbox{12cm}{Absolute value for the local detection limit threshold adopted for the 2$^{nd}$ derivate along the $x$-axis coordinate} \\   
\parbox{\sizecola}{DETLIM\_Y \\ } & \parbox{\sizecolb}{F10.3  \\ } & \parbox{\sizecolc}{MJy/sr/pxl$^2$ \\ } & \parbox{12cm}{Absolute value for the local detection limit threshold adopted for the 2$^{nd}$ derivate along the $y$-axis coordinate} \\   
\parbox{\sizecola}{DETLIM\_X45 \\ } & \parbox{\sizecolb}{F10.3  \\ } & \parbox{\sizecolc}{MJy/sr/pxl$^2$ \\ } & \parbox{12cm}{Absolute value for the local detection limit threshold adopted for the 2$^{nd}$ derivate along the bisector of the 1$^{st}$ and 3$^{rd}$ quadrant.} \\   
\parbox{\sizecola}{DETLIM\_Y45 \\ } & \parbox{\sizecolb}{F10.3  \\ } & \parbox{\sizecolc}{MJy/sr/pxl$^2$ \\ } & \parbox{12cm}{Absolute value for the local detection limit threshold adopted for the 2$^{nd}$ derivate along the bisector of the 2$^{nd}$ and 4$^{th}$ quadrant.} \\   
\parbox{\sizecola}{CLUMP\_FLAG \\ \\ \\} & \parbox{\sizecolb}{I5  \\ \\ \\} & \parbox{\sizecolc}{$-$ \\ \\ \\} & \parbox{12cm}{Flag for confusion at detection level. A value equal to 0 means that the source was identified from an isolated group of pixels above the threshold in all the four derivative directions. Sources belonging to the extraction of the same Atlas Image having the same value of this flag belong to the same group of pixels above the threshold.} \\  
\parbox{\sizecola}{NCOMP \\ \\} & \parbox{\sizecolb}{I2 \\ \\} & \parbox{\sizecolc}{$-$ \\ \\} & \parbox{12cm}{Number of gaussian components used simultaneously in the fitting process. This number includes the source, so the minimum value is 1, such number is greater than 1 if the source is fit with other nearby detections.} \\   
\parbox{\sizecola}{XCENT }& \parbox{\sizecolb}{F9.1} & \parbox{\sizecolc}{pxl} & \parbox{12cm}{The $x$-pixel coordinate of the centre of the source fitting window on the original image.} \\
\parbox{\sizecola}{YCENT} & \parbox{\sizecolb}{F9.1} & \parbox{\sizecolc}{pxl} & \parbox{12cm}{The $y$-pixel coordinate of the centre of the source fitting window on the original image.} \\
\parbox{\sizecola}{XWINDOW \\} & \parbox{\sizecolb}{I2\\} & \parbox{\sizecolc}{pxl \\} & \parbox{12cm}{Half-width size of the source fitting window along $x$ coordinate and centred at XCENT.} \\
\parbox{\sizecola}{YWINDOW \\} & \parbox{\sizecolb}{I2\\} & \parbox{\sizecolc}{pxl \\} & \parbox{12cm}{Half-width size of the source fitting window along $y$ coordinate and centred at YCENT.} \\
\parbox{\sizecola}{NCONTAM \\ \\ } & \parbox{\sizecolb}{I2 \\ \\} & \parbox{\sizecolc}{$-$ \\ \\} & \parbox{12cm}{Number of other sources falling inside the fitting window whose presence is taken into account at fitting stage. Not all those other sources might have been fitted at the same time} \\   
\parbox{\sizecola}{CENT\_TOL \\ \\ } & \parbox{\sizecolb}{F5.2 \\ \\} & \parbox{\sizecolc}{pxl \\ \\} & \parbox{12cm}{Maximum variation in pixels for adjustment of the fit centre with respect to the position of detection, measured as the distance between the latter and the brightest local (withing 3-pixel) pixel in the fitting window.} \\   
\parbox{\sizecola}{DOF} &  \parbox{\sizecolb}{I4} & \parbox{\sizecolc}{$-$} &  \parbox{12cm}{Degrees of freedom of the source Gaussian fit.} \\ \hline
\multicolumn{4}{c}{\textbf{Quality Flags}} \\  \hline
\parbox{\sizecola}{CHI2} & \parbox{\sizecolb}{F12.5} & \parbox{\sizecolc}{$-$} & \parbox{12cm}{$\chi ^2$ determined by the fitting engine.} \\
\parbox{\sizecola}{CHI2\_OPP \\ \\ } & \parbox{\sizecolb}{F12.5 \\ \\} & \parbox{\sizecolc}{$-$ \\ \\} & \parbox{12cm}{Estimator of the fidelity between the fit and the data computed as $\phi$= (O(i) - F(i))$^2$ / F(i), where O(i) is the observed data in the i pixel of the fitting window and F(i) is the fitted value in the same position} \\   
\parbox{\sizecola}{FIT\_STATUS \\ \\ \\ } & \parbox{\sizecolb}{I1 \\  \\ \\} & \parbox{\sizecolc}{$-$ \\ \\  \\} & \parbox{12cm}{Flag returned from the fitting engine. Possible values of the flag are: 0 - Fit convergence failed; 1 - Convergence reached; 2 - Convergence reached despite the initial accurary requested to fitting engine was set too low; 3 - Maximum number of iterations in the fitting process reached; 4 - Problems in Fitting due to the initial guess.} \\   
\parbox{\sizecola}{GUESS\_FLAG \\ \\ \\ \\ \\ \\} & \parbox{\sizecolb}{A3 \\ \\ \\ \\ \\ \\} & \parbox{\sizecolc}{$-$ \\ \\ \\ \\ \\ \\} & \parbox{12cm}{Flag on quality of guessed source parameters as determined at the detection stage. The form of the flag is GN where G is a letter defined as: A - Optimal number of positions to estimate the size; B - Sufficient number of positions to estimate the size; C - Low number of positions to estimate the size; and N is a number defining the quality of inital guess size: 0 - Initial Estimate failed; 1 - Good initial estimate for sizes; 2 - One of the two guessed sizes was initially estimate as smaller of PSF; 3 - Initial estimates of source sizes were larger than 3 times the PSF} \\   
\parbox{\sizecola}{GROUP\_FLAG \\ \\ \\ \\ \\ \\} & \parbox{\sizecolb}{I5 \\ \\ \\ \\ \\ \\} & \parbox{\sizecolc}{$-$ \\ \\ \\ \\ \\ \\} & \parbox{12cm}{Flag on quality of guessed source parameters as determined at the detection stage. The form of the flag is GN where G is a letter defined as: A - Optimal number of positions to estimate the size; B - Sufficient number of positions to estimate the size; C - Low number of positions to estimate the size; and N is a number defining the quality of inital guess size: 0 - Initial Estimate failed; 1 - Good initial estimate for sizes; 2 - One of the two guessed sizes was initially estimate as smaller of PSF; 3 - Initial estimates of source sizes were larger than 3 times the PSF} \\   
\parbox{\sizecola}{CONSTRAINTS \\ \\ \\} & \parbox{\sizecolb}{I1 \\ \\ \\} & \parbox{\sizecolc}{$-$ \\ \\ \\ } & \parbox{12cm}{Flag indicating the number of parameters that reached the tolerance limits allowed to the fit process. Values as 4 indicate that the source flux has higher unreliability since either the centre and its sizes have reached the maximum (or the minimum) allowed for the fit engine} \\   
\parbox{\sizecola}{SHIFT\_FLAG \\ } & \parbox{\sizecolb}{F9.3 \\} & \parbox{\sizecolc}{arcsec  \\ } & \parbox{12cm}{Amount of shift of the source peak position from its original detection position, due to Gaussian fitting.} \\   
\hline
   \end{tabular}
\end{table*}

\addtocounter{table}{-1}
\begin{table*}[h]
   \centering
    \caption{Continued}
   \begin{tabular}{llll} \hline\hline
      Field name    & Format & Units & Description \\ \hline
\multicolumn{4}{c}{\textbf{Basic extraction information}} \\ 
\hline
\parbox{\sizecola}{RDETP2DX \\ \\} & \parbox{\sizecolb}{F9.3 \\ \\} & \parbox{\sizecolc}{$-$ \\ \\ } & \parbox{12cm}{Ratio between the 2$^{nd}$ derivative value along $x$ direction expected by the fitted model of the source and the 2$^{nd}$ derivative derivative measureat the detection stage. Values closer to 1 indicate higher reliability of the source.} \\   
\parbox{\sizecola}{RDETP2DY \\ \\} & \parbox{\sizecolb}{F9.3 \\ \\} & \parbox{\sizecolc}{$-$ \\ \\ } & \parbox{12cm}{Ratio between the 2$^{nd}$ derivative value along $y$ direction expected by the fitted model of the source and the 2$^{nd}$ derivative measureat the detection stage. Values closer to 1 indicate higher reliability of the source.} \\   
\parbox{\sizecola}{RDETP2DX45 \\ \\} & \parbox{\sizecolb}{F9.3 \\ \\} & \parbox{\sizecolc}{$-$ \\ \\ } & \parbox{12cm}{Ratio between the 2$^{nd}$ derivative value along the bisector of the $xy$ direction expected by the fitted model of the source and the 2$^{nd}$ derivative measureat the detection stage. Values closer to 1 indicate higher reliability of the source.} \\   
\parbox{\sizecola}{RDETP2DY45 \\ \\} & \parbox{\sizecolb}{F9.3 \\ \\} & \parbox{\sizecolc}{$-$ \\ \\ } & \parbox{12cm}{Ratio between the 2$^{nd}$ derivative value along the bisector of the $yx$ direction expected by the fitted model of the source and the 2$^{nd}$ derivative measureat the detection stage. Values closer to 1 indicate higher reliability of the source.} \\   
\parbox{\sizecola}{OVERLAP\_FLAG \\ \\ \\ \\ \\ \\} & \parbox{\sizecolb}{F9.3 \\ \\ \\ \\ \\ \\} & \parbox{\sizecolc}{$-$ \\ \\ \\ \\ \\ \\ } & \parbox{12cm}{Flag to indicate whether the source has been detected and extracted in one or more adiacent tiles. H indicates that the source has been detected in the tile named in column ATLAS\_IMAGE; E,W indicate that the source is detected only in the eastern or western adiacent tile, respectively (east is higher galactic longitude); if the source has been detected in both H and E or W, then the name of the adiacent tile is also listed (e.g. H\_l060). In those cases, the entry in the catalogue is the one with the highest SNR.} \\   
\parbox{\sizecola}{OVFLUX\_FLAG \\ \\ \\ \\ \\ \\} & \parbox{\sizecolb}{I \\ \\ \\ \\ \\ \\} & \parbox{\sizecolc}{$-$ \\ \\ \\ \\ \\ \\ } & \parbox{12cm}{Flag to indicate which flux values were adopted if detected and extracted in two adiacent tiles. 0 indicates that the soruce has been detected only in once and therefore all fluxes refer to that detection. -1 indicates that the two fluxes differ by more than 15\%; the one listed is that with the highest SNR. 1 indicates that both integrated fluxes lie within 15\%, the one in the catalogue is that with highest SNR. 2 indicates that the integrated fluxes differ by more than 15\% but FPEAK are within 15\%; the one listed is that with highest SNR.} \\   
\hline
   \end{tabular}
\end{table*}

\FloatBarrier

\section{Saturated pixels in \higal\ maps}
\label{saturation}

Table \ref{sat_tab} reports the location of the clusters of saturated pixels in the \higal\ mapped area. The longitudes and latitudes in cols. 1-2 represent the centroid position of the cluster at the shortest wavelength where the saturation conditions exists. The subsequent 8 columns report for each band from 160 to 500 \um, the number of saturated pixels for each location and the radius of the circularized area of the saturated pixels cluster in arcseconds. The last column reports the sources from the IRAS Point Source Catalogue or from the RMS Source Catalogue that are located within 1\amin\ (for IRAS sources) and 40\asec\ (for MSX sources), from the pixels cluster centroid.

\onecolumn

\begin{landscape}
\begin{longtable}{cccrccrccrccrccrll}
\caption{\label{sat_tab} Clumps or clusters of saturated pixels, crossmatched with IRAS point sources, and RMS sources. IRAS sources $<60\arcsec$, RMS sources $<40\arcsec$ from cluster barycentre. }\\
\hline\hline
           &           & &&\multicolumn{2}{c}{red}&&\multicolumn{2}{c}{PSW}&&\multicolumn{2}{c}{PMW}&&\multicolumn{2}{c}{PLW} \\
\cline{5-6}
\cline{8-9}
\cline{11-12}
\cline{14-15}
 Gal. Lon. & Gal. Lat. & Map && Pix. & $R_e$ && Pix. & $R_e$ &&Pix. & $R_e$&&Pix. & $R_e$&& IRAS & RMS \\
($\degr$)& ($\degr$) & &&  & (\arcsec) &&  & (\arcsec)&&& (\arcsec)&&& (\arcsec)&&  & \\

\hline
\endfirsthead
\caption{continued.}\\
\hline\hline
           &           & &&\multicolumn{2}{c}{red}&&\multicolumn{2}{c}{PSW}&&\multicolumn{2}{c}{PMW}&&\multicolumn{2}{c}{PLW} \\
\cline{5-6}
\cline{8-9}
\cline{11-12}
\cline{14-15}
Gal. Lon. & Gal. Lat. & Map && Pix. & $R_e$ && Pix. & $R_e$ &&Pix. & $R_e$&&Pix. & $R_e$&& IRAS & RMS \\
($\degr$)& ($\degr$) & &&  & (\arcsec) &&  & (\arcsec)&&& (\arcsec)&&& (\arcsec)&&  & \\
\hline
\endhead
\hline
\endfoot
291.26944 & -0.71532 & l290 && - & - && 67 & 27 && 1 & 4 && - & - && 11097-6102 & \parbox[t]{6.72cm}{-} \\
291.57932 & -0.43251 & l290 && - & - && 7 & 8 && - & - && - & - && - & \parbox[t]{6.72cm}{G291.5765-00.4310} \\
291.26944 & -0.71508 & l292 && - & - && 67 & 27 && 3 & 7 && - & - && 11097-6102 & \parbox[t]{6.72cm}{-} \\
291.57941 & -0.43112 & l292 && - & - && 4 & 6 && - & - && - & - && - & \parbox[t]{6.72cm}{G291.5765-00.4310} \\
301.13687 & -0.22566 & l301 && - & - && 18 & 14 && 3 & 7 && - & - && 12326-6245 & \parbox[t]{6.72cm}{G301.1364-00.2249} \\
305.20917 & 0.20578 & l305 && - & - && 8 & 9 && - & - && - & - && 13079-6218 & \parbox[t]{6.72cm}{G305.2017+00.2072A, G305.2017+00.2072B} \\
305.35797 & 0.20391 & l305 && - & - && 6 & 8 && - & - && - & - && 13092-6218 & \parbox[t]{6.72cm}{-} \\
305.36215 & 0.15114 & l305 && - & - && 3 & 5 && - & - && - & - && - & \parbox[t]{6.72cm}{-} \\
305.36771 & 0.21224 & l305 && - & - && 1 & 3 && - & - && - & - && - & \parbox[t]{6.72cm}{G305.3676+00.2095, G305.3779+00.2108} \\
305.79990 & -0.24385 & l305 && - & - && 3 & 5 && - & - && - & - && 13134-6242 & \parbox[t]{6.72cm}{G305.7991-00.2461A, G305.7991-00.2461B} \\
309.92188 & 0.47956 & l310 && - & - && 2 & 4 && - & - && - & - && 13471-6120 & \parbox[t]{6.72cm}{G309.9206+00.4790A, G309.9206+00.4790B} \\
311.62744 & 0.29010 & l312 && - & - && 5 & 7 && - & - && - & - && 14013-6105 & \parbox[t]{6.72cm}{G311.6264+00.2897} \\
314.21927 & 0.27282 & l314 && - & - && 1 & 3 && - & - && - & - && 14214-6017 & \parbox[t]{6.72cm}{G314.2204+00.2726} \\
316.81213 & -0.05743 & l316 && - & - && 3 & 5 && - & - && - & - && 14416-5937 & \parbox[t]{6.72cm}{G316.8112-00.0566} \\
318.04984 & 0.08687 & l319 && - & - && 1 & 3 && - & - && - & - && 14498-5856 & \parbox[t]{6.72cm}{G318.0489+00.0854A, G318.0489+00.0854B, G318.0489+00.0854C} \\
318.94812 & -0.19645 & l319 && - & - && 1 & 3 && - & - && - & - && - & \parbox[t]{6.72cm}{G318.9480-00.1969A, G318.9480-00.1969B} \\
322.15808 & 0.63623 & l321 && - & - && 23 & 16 && 4 & 9 && - & - && - & \parbox[t]{6.72cm}{-} \\
322.16376 & 0.62281 & l321 && - & - && 13 & 12 && - & - && - & - && - & \parbox[t]{6.72cm}{-} \\
322.15823 & 0.63590 & l323 && - & - && 21 & 15 && - & - && - & - && - & \parbox[t]{6.72cm}{-} \\
322.16400 & 0.62258 & l323 && - & - && 7 & 8 && - & - && - & - && - & \parbox[t]{6.72cm}{-} \\
323.74069 & -0.26362 & l323 && - & - && 5 & 7 && - & - && - & - && 15278-5620 & \parbox[t]{6.72cm}{G323.7399-00.2617A, G323.7399-00.2617B, G323.7410-00.2552A, G323.7410-00.2552B, G323.7410-00.2552C} \\
324.20078 & 0.12143 & l323 && - & - && 6 & 8 && - & - && - & - && 15290-5546 & \parbox[t]{6.72cm}{G324.1997+00.1192} \\
324.20093 & 0.12056 & l325 && - & - && 5 & 7 && - & - && - & - && 15290-5546 & \parbox[t]{6.72cm}{G324.1997+00.1192} \\
326.47467 & 0.70227 & l325 && - & - && 4 & 6 && - & - && - & - && 15394-5358 & \parbox[t]{6.72cm}{G326.4755+00.6947} \\
326.65701 & 0.59368 & l325 && - & - && 5 & 7 && - & - && - & - && 15408-5356 & \parbox[t]{6.72cm}{-} \\
326.72165 & 0.61432 & l325 && - & - && 1 & 3 && - & - && - & - && 15411-5352 & \parbox[t]{6.72cm}{G326.7249+00.6159A, G326.7249+00.6159B} \\
326.65836 & 0.59550 & l327 && - & - && 1 & 3 && - & - && - & - && 15408-5356 & \parbox[t]{6.72cm}{-} \\
326.67001 & 0.55551 & l327 && - & - && 1 & 3 && - & - && - & - && - & \parbox[t]{6.72cm}{G326.6687+00.5495} \\
326.72278 & 0.61440 & l327 && - & - && 3 & 5 && - & - && - & - && 15411-5352 & \parbox[t]{6.72cm}{G326.7249+00.6159A, G326.7249+00.6159B} \\
327.29385 & -0.57797 & l327 && - & - && 38 & 20 && 17 & 18 && 3 & 11 && - & \parbox[t]{6.72cm}{G327.2852-00.5735} \\
327.30011 & -0.54923 & l327 && - & - && 30 & 18 && - & - && - & - && 15492-5426 & \parbox[t]{6.72cm}{-} \\
327.30661 & -0.54101 & l327 && - & - && 1 & 3 && - & - && - & - && 15492-5426 & \parbox[t]{6.72cm}{-} \\
327.40240 & 0.44477 & l327 && - & - && 4 & 6 && - & - && - & - && 15454-5335 & \parbox[t]{6.72cm}{G327.4014+00.4454} \\
328.23657 & -0.54683 & l327 && - & - && 2 & 4 && - & - && - & - && - & \parbox[t]{6.72cm}{-} \\
328.25488 & -0.53147 & l327 && - & - && 7 & 8 && - & - && - & - && 15541-5349 & \parbox[t]{6.72cm}{G328.2523-00.5320A, G328.2523-00.5320B, G328.2523-00.5320C} \\
328.30728 & 0.43178 & l327 && - & - && 7 & 8 && - & - && - & - && 15502-5302 & \parbox[t]{6.72cm}{G328.3067+00.4308} \\
328.56671 & -0.53409 & l327 && - & - && 9 & 10 && - & - && - & - && 15557-5337 & \parbox[t]{6.72cm}{G328.5669-00.5327} \\
328.57483 & -0.53013 & l327 && - & - && 2 & 4 && - & - && - & - && 15557-5337 & \parbox[t]{6.72cm}{G328.5669-00.5327} \\
328.80792 & 0.63265 & l327 && 2 & 3 && 32 & 19 && 8 & 12 && - & - && 15520-5234 & \parbox[t]{6.72cm}{G328.8074+00.6324} \\
328.57144 & -0.53254 & l330 && - & - && 16 & 13 && - & - && - & - && 15557-5337 & \parbox[t]{6.72cm}{G328.5669-00.5327} \\
328.80875 & 0.63335 & l330 && - & - && 29 & 18 && 7 & 11 && - & - && 15520-5234 & \parbox[t]{6.72cm}{G328.8074+00.6324} \\
329.02948 & -0.20541 & l330 && - & - && 3 & 5 && - & - && - & - && 15566-5304 & \parbox[t]{6.72cm}{-} \\
329.18417 & -0.31345 & l330 && - & - && 2 & 4 && - & - && - & - && 15579-5303 & \parbox[t]{6.72cm}{-} \\
329.33844 & 0.14789 & l330 && - & - && 12 & 11 && - & - && - & - && 15567-5236 & \parbox[t]{6.72cm}{G329.3371+00.1469} \\
329.40497 & -0.45928 & l330 && - & - && 1 & 3 && - & - && - & - && 15596-5301 & \parbox[t]{6.72cm}{G329.4055-00.4574} \\
330.87857 & -0.36690 & l330 && - & - && 18 & 14 && 3 & 7 && - & - && 16065-5158 & \parbox[t]{6.72cm}{G330.8708-00.3715A, G330.8708-00.3715B, G330.8845-00.3721} \\
330.95459 & -0.18134 & l330 && 7 & 6 && 32 & 19 && 13 & 16 && - & - && 16060-5146 & \parbox[t]{6.72cm}{G330.9544-00.1817} \\
330.87854 & -0.36711 & l332 && - & - && 19 & 14 && 2 & 6 && - & - && 16065-5158 & \parbox[t]{6.72cm}{G330.8708-00.3715A, G330.8708-00.3715B, G330.8845-00.3721} \\
330.95401 & -0.18147 & l332 && - & - && 33 & 19 && 12 & 15 && - & - && 16060-5146 & \parbox[t]{6.72cm}{G330.9544-00.1817} \\
331.13147 & -0.24311 & l332 && - & - && 7 & 8 && - & - && - & - && 16071-5142 & \parbox[t]{6.72cm}{G331.1282-00.2436} \\
331.27811 & -0.18765 & l332 && - & - && 1 & 3 && - & - && - & - && 16076-5134 & \parbox[t]{6.72cm}{G331.2759-00.1891B} \\
331.51184 & -0.10189 & l332 && - & - && 15 & 13 && - & - && - & - && - & \parbox[t]{6.72cm}{G331.5131-00.1020, G331.5180-00.0947A, G331.5180-00.0947B} \\
331.55585 & -0.12044 & l332 && - & - && 3 & 5 && - & - && - & - && 16086-5119 & \parbox[t]{6.72cm}{G331.5582-00.1206} \\
332.09387 & -0.42099 & l332 && - & - && 2 & 4 && - & - && - & - && 16124-5110 & \parbox[t]{6.72cm}{G332.0939-00.4206} \\
332.82730 & -0.54833 & l332 && 5 & 5 && 33 & 19 && 10 & 14 && - & - && 16164-5046 & \parbox[t]{6.72cm}{G332.8256-00.5498A, G332.8256-00.5498B} \\
332.96378 & -0.67754 & l332 && - & - && 2 & 4 && - & - && - & - && 16175-5045 & \parbox[t]{6.72cm}{G332.9636-00.6800} \\
333.06830 & -0.44629 & l332 && - & - && 9 & 10 && - & - && - & - && - & \parbox[t]{6.72cm}{G333.0682-00.4461} \\
333.12161 & -0.43240 & l332 && - & - && 9 & 10 && - & - && - & - && 16172-5028 & \parbox[t]{6.72cm}{G333.1256-00.4367, G333.1306-00.4275} \\
333.13354 & -0.43052 & l332 && - & - && 85 & 31 && 14 & 16 && - & - && 16172-5028 & \parbox[t]{6.72cm}{G333.1306-00.4275, G333.1256-00.4367} \\
333.28455 & -0.38667 & l332 && - & - && 37 & 20 && 1 & 4 && - & - && - & \parbox[t]{6.72cm}{G333.2880-00.3907} \\
333.29950 & -0.35258 & l332 && - & - && 1 & 3 && - & - && - & - && 16177-5018 & \parbox[t]{6.72cm}{-} \\
332.82697 & -0.54870 & l334 && - & - && 31 & 18 && 9 & 13 && 1 & 6 && 16164-5046 & \parbox[t]{6.72cm}{G332.8256-00.5498A, G332.8256-00.5498B} \\
332.96283 & -0.67861 & l334 && - & - && 2 & 4 && - & - && - & - && 16175-5045 & \parbox[t]{6.72cm}{G332.9636-00.6800} \\
333.06778 & -0.44717 & l334 && - & - && 16 & 13 && - & - && - & - && - & \parbox[t]{6.72cm}{G333.0682-00.4461} \\
333.12073 & -0.43279 & l334 && - & - && 7 & 8 && - & - && - & - && 16172-5028 & \parbox[t]{6.72cm}{G333.1256-00.4367} \\
333.12357 & -0.42208 & l334 && - & - && 1 & 3 && - & - && - & - && 16172-5028 & \parbox[t]{6.72cm}{G333.1306-00.4275} \\
333.13281 & -0.43099 & l334 && - & - && 85 & 31 && 18 & 19 && 1 & 6 && 16172-5028 & \parbox[t]{6.72cm}{G333.1306-00.4275, G333.1256-00.4367} \\
333.28406 & -0.38753 & l334 && - & - && 36 & 20 && 1 & 4 && - & - && - & \parbox[t]{6.72cm}{G333.2880-00.3907} \\
333.46677 & -0.16383 & l334 && - & - && 3 & 5 && - & - && - & - && 16175-5002 & \parbox[t]{6.72cm}{-} \\
333.60181 & -0.21256 & l334 && 8 & 7 && 93 & 32 && 18 & 19 && - & - && 16183-4958 & \parbox[t]{6.72cm}{G333.6032-00.2184} \\
333.60635 & -0.21475 & l334 && 2 & 3 && 93 & 32 && 18 & 19 && - & - && 16183-4958 & \parbox[t]{6.72cm}{G333.6032-00.2184} \\
333.61011 & -0.21475 & l334 && 2 & 3 && 93 & 32 && 18 & 19 && - & - && 16183-4958 & \parbox[t]{6.72cm}{G333.6032-00.2184} \\
335.58456 & -0.28968 & l336 && - & - && 5 & 7 && - & - && - & - && 16272-4837 & \parbox[t]{6.72cm}{-} \\
335.78864 & 0.17434 & l336 && - & - && 3 & 5 && - & - && - & - && - & \parbox[t]{6.72cm}{-} \\
336.01752 & -0.82527 & l336 && - & - && 1 & 3 && - & - && - & - && 16313-4840 & \parbox[t]{6.72cm}{-} \\
336.99533 & -0.02706 & l336 && - & - && 7 & 8 && - & - && - & - && 16318-4724 & \parbox[t]{6.72cm}{G336.9920-00.0244} \\
337.12134 & -0.17316 & l336 && - & - && 3 & 5 && - & - && - & - && 16330-4725 & \parbox[t]{6.72cm}{-} \\
337.40457 & -0.40154 & l336 && - & - && 22 & 15 && 4 & 9 && - & - && 16351-4722 & \parbox[t]{6.72cm}{G337.4032-00.4037, G337.4050-00.4071A, G337.4050-00.4071B, G337.4050-00.4071C} \\
337.70480 & -0.05338 & l336 && - & - && 10 & 10 && - & - && - & - && 16348-4654 & \parbox[t]{6.72cm}{G337.7051-00.0575B} \\
337.71286 & 0.08847 & l336 && - & - && 3 & 5 && - & - && - & - && 16343-4648 & \parbox[t]{6.72cm}{G337.7091+00.0932A, G337.7091+00.0932B, G337.7091+00.0932C} \\
337.40515 & -0.40158 & l338 && - & - && 24 & 16 && 4 & 9 && - & - && 16351-4722 & \parbox[t]{6.72cm}{G337.4032-00.4037, G337.4050-00.4071A, G337.4050-00.4071B} \\
337.70502 & -0.05350 & l338 && - & - && 6 & 8 && - & - && 1 & 6 && 16348-4654 & \parbox[t]{6.72cm}{G337.7051-00.0575B} \\
337.71362 & 0.08732 & l338 && - & - && 2 & 4 && - & - && - & - && 16343-4648 & \parbox[t]{6.72cm}{G337.7091+00.0932A, G337.7091+00.0932B, G337.7091+00.0932C} \\
337.91513 & -0.47671 & l338 && - & - && 26 & 17 && 5 & 10 && - & - && 16374-4701 & \parbox[t]{6.72cm}{-} \\
337.92151 & -0.45594 & l338 && - & - && 4 & 6 && - & - && - & - && 16374-4701 & \parbox[t]{6.72cm}{G337.9266-00.4588} \\
338.07437 & 0.00983 & l338 && - & - && 1 & 3 && - & - && - & - && 16359-4635 & \parbox[t]{6.72cm}{G338.0715+00.0126A, G338.0715+00.0126B, G338.0715+00.0126C} \\
338.92065 & 0.55064 & l338 && - & - && 1 & 3 && 30 & 24 && - & - && - & \parbox[t]{6.72cm}{G338.9196+00.5495} \\
340.05484 & -0.24336 & l341 && - & - && 16 & 13 && - & - && - & - && 16445-4516 & \parbox[t]{6.72cm}{G340.0543-00.2437A, G340.0543-00.2437B, G340.0543-00.2437C} \\
340.97015 & -1.02083 & l341 && - & - && 2 & 4 && - & - && - & - && 16513-4504 & \parbox[t]{6.72cm}{-} \\
343.12711 & -0.06283 & l343 && - & - && 22 & 15 && 2 & 6 && - & - && 16547-4247 & \parbox[t]{6.72cm}{G343.1261-00.0623} \\
343.75632 & -0.16335 & l343 && - & - && 2 & 4 && - & - && - & - && 16572-4221 & \parbox[t]{6.72cm}{-} \\
344.22766 & -0.56794 & l343 && - & - && 10 & 10 && 1 & 4 && - & - && - & \parbox[t]{6.72cm}{-} \\
344.22128 & -0.59246 & l343 && - & - && 1 & 3 && - & - && - & - && 17006-4215 & \parbox[t]{6.72cm}{G344.2207-00.5953} \\
344.22733 & -0.56887 & l345 && - & - && 7 & 8 && 1 & 4 && - & - && - & \parbox[t]{6.72cm}{-} \\
345.00293 & -0.22400 & l345 && - & - && 15 & 13 && 1 & 4 && - & - && 17016-4124 & \parbox[t]{6.72cm}{G345.0034-00.2240A, G345.0034-00.2240B} \\
345.40417 & -0.94387 & l345 && - & - && 2 & 4 && - & - && - & - && 17059-4132 & \parbox[t]{6.72cm}{-} \\
345.40503 & -0.94054 & l345 && - & - && 1 & 3 && - & - && - & - && 17059-4132 & \parbox[t]{6.72cm}{-} \\
345.40759 & -0.95251 & l345 && - & - && 11 & 11 && - & - && - & - && 17059-4132 & \parbox[t]{6.72cm}{-} \\
345.48730 & 0.31525 & l345 && - & - && 27 & 17 && 2 & 6 && - & - && 17009-4042 & \parbox[t]{6.72cm}{G345.4881+00.3148} \\
345.50470 & 0.34849 & l345 && - & - && 10 & 10 && - & - && - & - && 17008-4040 & \parbox[t]{6.72cm}{G345.5043+00.3480} \\
345.64835 & 0.01016 & l345 && - & - && 2 & 4 && - & - && - & - && 17028-4045 & \parbox[t]{6.72cm}{G345.6495+00.0084} \\
347.62949 & 0.14830 & l347 && - & - && 1 & 3 && - & - && - & - && - & \parbox[t]{6.72cm}{-} \\
348.18448 & 0.48243 & l347 && - & - && 2 & 4 && - & - && - & - && - & \parbox[t]{6.72cm}{-} \\
348.54904 & -0.97985 & l349 && - & - && 8 & 9 && - & - && - & - && - & \parbox[t]{6.72cm}{-} \\
348.69720 & -1.02824 & l349 && - & - && 18 & 14 && 1 & 4 && - & - && - & \parbox[t]{6.72cm}{G348.6972-01.0263} \\
348.70135 & -1.04115 & l349 && - & - && 11 & 11 && - & - && - & - && - & \parbox[t]{6.72cm}{-} \\
348.72681 & -1.03991 & l349 && - & - && 36 & 20 && 8 & 12 && - & - && 17167-3854 & \parbox[t]{6.72cm}{G348.7250-01.0435, G348.7342-01.0359B} \\
349.09204 & 0.10544 & l349 && - & - && 2 & 4 && - & - && - & - && 17130-3756 & \parbox[t]{6.72cm}{-} \\
350.01199 & -1.34354 & l349 && - & - && 1 & 3 && - & - && - & - && 17216-3801 & \parbox[t]{6.72cm}{-} \\
350.10364 & 0.08211 & l349 && - & - && 2 & 4 && - & - && - & - && 17160-3707 & \parbox[t]{6.72cm}{-} \\
350.11032 & 0.08795 & l349 && - & - && 1 & 3 && - & - && - & - && 17160-3707 & \parbox[t]{6.72cm}{-} \\
350.11282 & 0.09461 & l349 && - & - && 2 & 4 && - & - && - & - && 17160-3707 & \parbox[t]{6.72cm}{-} \\
350.50192 & 0.95693 & l349 && - & - && 2 & 4 && - & - && - & - && 17136-3617 & \parbox[t]{6.72cm}{-} \\
350.50858 & 0.95776 & l349 && - & - && 3 & 5 && - & - && - & - && 17136-3617 & \parbox[t]{6.72cm}{-} \\
351.15857 & 0.70003 & l349 && - & - && 67 & 27 && 14 & 16 && 1 & 6 && 17165-3554 & \parbox[t]{6.72cm}{-} \\
351.24503 & 0.66776 & l349 && - & - && 67 & 27 && 8 & 12 && - & - && - & \parbox[t]{6.72cm}{-} \\
351.25070 & 0.65370 & l349 && - & - && 22 & 15 && - & - && - & - && - & \parbox[t]{6.72cm}{-} \\
351.15891 & 0.69968 & l352 && - & - && 58 & 25 && 12 & 15 && - & - && 17165-3554 & \parbox[t]{6.72cm}{-} \\
351.24637 & 0.66349 & l352 && - & - && 84 & 31 && 7 & 11 && - & - && - & \parbox[t]{6.72cm}{-} \\
351.41672 & 0.64628 & l352 && 5 & 5 && 109 & 35 && 22 & 21 && 1 & 6 && 17175-3544 & \parbox[t]{6.72cm}{-} \\
351.44254 & 0.65746 & l352 && - & - && 97 & 33 && 34 & 26 && 5 & 14 && - & \parbox[t]{6.72cm}{-} \\
351.58179 & -0.35232 & l352 && - & - && 27 & 17 && 8 & 12 && 1 & 6 && 17220-3609 & \parbox[t]{6.72cm}{-} \\
351.62717 & -1.26058 & l352 && - & - && 1 & 3 && 41 & 28 && 2 & 9 && 17258-3637 & \parbox[t]{6.72cm}{-} \\
351.77597 & -0.53590 & l352 && 1 & 2 && 46 & 22 && 15 & 17 && - & - && 17233-3606 & \parbox[t]{6.72cm}{-} \\
353.19455 & 0.90645 & l352 && - & - && 1 & 3 && - & - && - & - && - & \parbox[t]{6.72cm}{-} \\
353.41052 & -0.35994 & l354 && - & - && 22 & 15 && 5 & 10 && - & - && 17271-3439 & \parbox[t]{6.72cm}{-} \\
0.65762 & -0.04104 & l000 && 3 & 4 && - & - && - & - && - & - && 17441-2822 & \parbox[t]{6.72cm}{-} \\
0.66608 & -0.03508 & l000 && 30 & 13 && 1 & 3 && - & - && - & - && 17441-2822 & \parbox[t]{6.72cm}{-} \\
0.67696 & -0.02762 & l000 && 5 & 5 && - & - && - & - && - & - && 17441-2822 & \parbox[t]{6.72cm}{-} \\
3.43932 & -0.34878 & l004 && - & - && 4 & 6 && - & - && - & - && 17517-2609 & \parbox[t]{6.72cm}{-} \\
5.88571 & -0.39286 & l006 && 3 & 4 && 28 & 17 && 5 & 10 && - & - && 17574-2403 & \parbox[t]{6.72cm}{-} \\
5.90034 & -0.42896 & l006 && - & - && 4 & 6 && - & - && - & - && - & \parbox[t]{6.72cm}{-} \\
8.13747 & 0.22342 & l006 && - & - && 1 & 3 && - & - && - & - && 17599-2148 & \parbox[t]{6.72cm}{-} \\
8.14080 & 0.22092 & l006 && - & - && 2 & 4 && - & - && - & - && 17599-2148 & \parbox[t]{6.72cm}{-} \\
8.66965 & -0.35544 & l008 && - & - && 19 & 14 && 2 & 6 && - & - && 18032-2137 & \parbox[t]{6.72cm}{-} \\
9.62041 & 0.19508 & l008 && - & - && 14 & 12 && 1 & 4 && - & - && 18032-2032 & \parbox[t]{6.72cm}{-} \\
10.29660 & -0.14639 & l011 && - & - && 1 & 3 && - & - && - & - && - & \parbox[t]{6.72cm}{-} \\
10.30060 & -0.14606 & l011 && - & - && 5 & 7 && - & - && - & - && 18060-2005 & \parbox[t]{6.72cm}{-} \\
10.47293 & 0.02776 & l011 && - & - && 16 & 13 && 5 & 10 && - & - && 18056-1952 & \parbox[t]{6.72cm}{G010.4718+00.0256, G010.4718+00.0206} \\
10.62483 & -0.38273 & l011 && 4 & 5 && 42 & 21 && 13 & 16 && - & - && 18075-1956 & \parbox[t]{6.72cm}{G010.6235-00.3834, G010.6260-00.3744, G010.6311-00.3864} \\
11.93718 & -0.61500 & l011 && - & - && 5 & 7 && - & - && - & - && 18110-1854 & \parbox[t]{6.72cm}{G011.9373-00.6165} \\
12.20841 & -0.10072 & l011 && - & - && 5 & 7 && - & - && - & - && 18097-1825A & \parbox[t]{6.72cm}{-} \\
11.93764 & -0.61586 & l013 && - & - && 9 & 10 && - & - && - & - && 18110-1854 & \parbox[t]{6.72cm}{G011.9373-00.6165} \\
12.20885 & -0.10139 & l013 && - & - && 9 & 10 && - & - && - & - && 18097-1825A & \parbox[t]{6.72cm}{-} \\
12.41826 & 0.50535 & l013 && - & - && 1 & 3 && - & - && - & - && 18079-1756 & \parbox[t]{6.72cm}{-} \\
12.80578 & -0.19902 & l013 && 6 & 6 && 154 & 42 && 39 & 28 && 2 & 9 && - & \parbox[t]{6.72cm}{G012.8062-00.1987} \\
12.88957 & 0.48929 & l013 && - & - && 6 & 8 && - & - && - & - && 18089-1732 & \parbox[t]{6.72cm}{G012.8909+00.4938A, G012.8909+00.4938B, G012.8909+00.4938C} \\
12.90789 & -0.25941 & l013 && - & - && 12 & 11 && - & - && - & - && 18117-1753 & \parbox[t]{6.72cm}{G012.9090-00.2607} \\
14.33189 & -0.64444 & l013 && - & - && 7 & 8 && - & - && - & - && 18159-1648 & \parbox[t]{6.72cm}{G014.3313-00.6397} \\
14.33213 & -0.64354 & l015 && - & - && 7 & 8 && - & - && - & - && 18159-1648 & \parbox[t]{6.72cm}{G014.3313-00.6397} \\
15.01485 & -0.70673 & l015 && - & - && 1 & 3 && - & - && - & - && - & \parbox[t]{6.72cm}{-} \\
15.01896 & -0.67179 & l015 && - & - && 315 & 60 && 36 & 27 && 1 & 6 && 18174-1612 & \parbox[t]{6.72cm}{-} \\
17.63897 & 0.15421 & l017 && - & - && 1 & 3 && - & - && - & - && 18196-1331 & \parbox[t]{6.72cm}{G017.6380+00.1566} \\
19.07808 & -0.28641 & l019 && - & - && 1 & 3 && - & - && - & - && 18239-1228 & \parbox[t]{6.72cm}{G019.0741-00.2861} \\
19.60909 & -0.23325 & l019 && - & - && 11 & 11 && - & - && - & - && 18248-1158 & \parbox[t]{6.72cm}{G019.6085-00.2357} \\
20.08053 & -0.13477 & l019 && - & - && 4 & 6 && - & - && - & - && 18253-1130 & \parbox[t]{6.72cm}{G020.0801-00.1360, G020.0722-00.1419} \\
23.20654 & -0.37680 & l024 && - & - && 1 & 3 && - & - && - & - && - & \parbox[t]{6.72cm}{-} \\
24.32809 & 0.14480 & l024 && - & - && 1 & 3 && - & - && - & - && 18324-0737 & \parbox[t]{6.72cm}{-} \\
24.79058 & 0.08432 & l024 && - & - && 14 & 12 && - & - && - & - && - & \parbox[t]{6.72cm}{G024.7891+00.0846} \\
25.64959 & 1.04886 & l026 && - & - && 5 & 7 && - & - && - & - && 18316-0602 & \parbox[t]{6.72cm}{G025.6498+01.0491} \\
25.82551 & -0.17718 & l026 && - & - && 1 & 3 && - & - && - & - && - & \parbox[t]{6.72cm}{-} \\
26.51048 & 0.28280 & l026 && - & - && 1 & 3 && - & - && - & - && 18360-0537 & \parbox[t]{6.72cm}{G026.5107+00.2824A, G026.5107+00.2824B, G026.5107+00.2824C} \\
27.36542 & -0.16550 & l026 && - & - && 1 & 3 && - & - && - & - && 18391-0504 & \parbox[t]{6.72cm}{-} \\
27.36667 & -0.16609 & l028 && - & - && 1 & 3 && - & - && - & - && 18391-0504 & \parbox[t]{6.72cm}{-} \\
28.19999 & -0.04926 & l028 && - & - && 8 & 9 && - & - && - & - && 18403-0417 & \parbox[t]{6.72cm}{G028.2007-00.0494A, G028.2007-00.0494B} \\
29.95521 & -0.01511 & l028 && - & - && 13 & 12 && - & - && - & - && 18434-0242 & \parbox[t]{6.72cm}{G029.9564-00.0174} \\
29.95584 & -0.01561 & l030 && - & - && 10 & 10 && - & - && - & - && 18434-0242 & \parbox[t]{6.72cm}{G029.9564-00.0174} \\
30.70293 & -0.06791 & l030 && - & - && 16 & 13 && 1 & 4 && - & - && - & \parbox[t]{6.72cm}{-} \\
30.71995 & -0.08227 & l030 && - & - && 5 & 7 && - & - && - & - && - & \parbox[t]{6.72cm}{G030.7206-00.0826} \\
30.81767 & -0.05530 & l030 && - & - && 21 & 15 && 4 & 9 && - & - && - & \parbox[t]{6.72cm}{-} \\
30.81573 & -0.05528 & l031 && - & - && 13 & 12 && 4 & 9 && - & - && - & \parbox[t]{6.72cm}{-} \\
31.28092 & 0.06224 & l031 && - & - && 2 & 4 && - & - && - & - && 18456-0129 & \parbox[t]{6.72cm}{G031.2803+00.0615A, G031.2803+00.0615B} \\
31.41228 & 0.30732 & l031 && - & - && 11 & 11 && - & - && - & - && 18449-0115 & \parbox[t]{6.72cm}{G031.4134+00.3092} \\
32.79803 & 0.19142 & l033 && - & - && 10 & 10 && - & - && - & - && 18479-0005 & \parbox[t]{6.72cm}{G032.7977+00.1903} \\
33.91713 & 0.11108 & l033 && - & - && 1 & 3 && - & - && - & - && 18502+0051 & \parbox[t]{6.72cm}{G033.9148+00.1093} \\
34.25706 & 0.15616 & l033 && - & - && 82 & 30 && 18 & 19 && 3 & 11 && 18507+0110 & \parbox[t]{6.72cm}{-} \\
34.41228 & 0.23503 & l033 && - & - && 5 & 7 && - & - && - & - && 18507+0121 & \parbox[t]{6.72cm}{-} \\
34.25696 & 0.15447 & l035 && 3 & 4 && 75 & 29 && 19 & 19 && 2 & 9 && 18507+0110 & \parbox[t]{6.72cm}{-} \\
34.41156 & 0.23619 & l035 && - & - && 1 & 3 && - & - && - & - && 18507+0121 & \parbox[t]{6.72cm}{-} \\
35.19689 & -0.74271 & l035 && - & - && 8 & 9 && - & - && - & - && 18556+0136 & \parbox[t]{6.72cm}{G035.1979-00.7427} \\
43.16719 & 0.01141 & l041 && 29 & 13 && - & - && - & - && - & - && 19078+0901 & \parbox[t]{6.72cm}{G043.1679+00.0095} \\
43.16536 & -0.02749 & l044 && - & - && 7 & 8 && - & - && - & - && - & \parbox[t]{6.72cm}{G043.1650-00.0285} \\
43.16829 & 0.01070 & l044 && 9 & 7 && 86 & 31 && 23 & 21 && 2 & 9 && 19078+0901 & \parbox[t]{6.72cm}{G043.1679+00.0095} \\
43.79532 & -0.12582 & l044 && - & - && 1 & 3 && - & - && - & - && 19095+0930 & \parbox[t]{6.72cm}{G043.7955-00.1275} \\
45.07181 & 0.13207 & l044 && - & - && 4 & 6 && - & - && - & - && 19110+1045 & \parbox[t]{6.72cm}{G045.0711+00.1325} \\
45.12346 & 0.13248 & l044 && - & - && 1 & 3 && - & - && - & - && 19111+1048 & \parbox[t]{6.72cm}{G045.1221+00.1323} \\
45.07174 & 0.13254 & l046 && - & - && 3 & 5 && - & - && - & - && 19110+1045 & \parbox[t]{6.72cm}{G045.0711+00.1325} \\
49.36889 & -0.29932 & l048 && - & - && 1 & 3 && - & - && - & - && 19209+1421 & \parbox[t]{6.72cm}{G049.3697-00.3031} \\
49.48889 & -0.38775 & l048 && 6 & 6 && 313 & 59 && 50 & 31 && 7 & 17 && - & \parbox[t]{6.72cm}{G049.4885-00.3799} \\
49.49014 & -0.36838 & l048 && 1 & 2 && 313 & 59 && 50 & 31 && - & - && 19213+1424 & \parbox[t]{6.72cm}{G049.4903-00.3694} \\
49.48544 & -0.37678 & l048 && - & - && 313 & 59 && 2 & 6 && 1 & 6 && 19213+1424 & \parbox[t]{6.72cm}{G049.4885-00.3799, G049.4903-00.3694} \\
49.48867 & -0.38734 & l050 && 9 & 7 && 326 & 61 && 51 & 32 && 10 & 20 && 19213+1424 & \parbox[t]{6.72cm}{G049.4885-00.3799} \\
49.48973 & -0.36881 & l050 && 9 & 7 && 326 & 61 && 51 & 32 && 10 & 20 && 19213+1424 & \parbox[t]{6.72cm}{G049.4903-00.3694} \\
49.48466 & -0.37687 & l050 && - & - && 326 & 61 && 4 & 9 && - & - && 19213+1424 & \parbox[t]{6.72cm}{G049.4885-00.3799, G049.4903-00.3694} \\
49.48466 & -0.37687 & l050 && - & - && 326 & 61 && 2 & 6 && - & - && 19213+1424 & \parbox[t]{6.72cm}{G049.4885-00.3799, G049.4903-00.3694} \\
49.48551 & -0.36269 & l050 && - & - && - & - && - & - && 1 & 6 && 19213+1424 & \parbox[t]{6.72cm}{G049.4883-00.3545A, G049.4883-00.3545B, G049.4903-00.3694} \\
61.47805 & 0.09104 & l061 && - & - && 2 & 4 && - & - && - & - && 19446+2505 & \parbox[t]{6.72cm}{G061.4736+00.0908A, G061.4736+00.0908B, G061.4736+00.0908C} \\
\end{longtable}
\end{landscape}

\end{document}